\documentclass{aa}  

\usepackage{graphicx}
\usepackage{natbib}
\usepackage{txfonts}
\newcommand*\diff{\mathop{}\!\mathrm{d}}
\begin{document}

   \title{Kinematic differences between multiple  populations \\ in Galactic globular clusters\thanks{Table 2 is only available at the CDS via anonymous ftp to
    cdsarc.u-strasbg.fr (XXXX) or via http://cdsarc.u-strasbg.fr/viz-bin/cat/J/A+A/XXX/Lzzz}\fnmsep \thanks{Based on observations collected at the European Organisation for Astronomical Research in the Southern Hemisphere, Chile (proposal IDs 094.D-0142, 095.D-0629, 096.D-0175, 097.D-0295, 098.D-0148, 099.D-0019, 100.D-0161, 101.D-0268, 102.D-0270, 103.D-0204, 105.20CR.002)}}


   \author{Sven Martens \inst{1}
          \and Sebastian Kamann \inst{2}
          \and Stefan Dreizler \inst{1}
          \and Fabian Göttgens \inst{1}
          \and Tim-Oliver Husser \inst{1}
          \and Marilyn Latour \inst{1}
          \and Elena Balakina \inst{2}
          \and Davor Krajnović \inst{3}
          \and Renuka Pechetti \inst{2}
          \and Peter M. Weilbacher \inst{3}
          }

   \institute{Institut für Astrophysik und Geophysik, Georg-August-Universität Göttingen, Friedrich-Hund-Platz 1, 37077 Göttingen, Germany\\  \email{sven.martens@uni-goettingen.de}
   \and
   Astrophysics Research Institute, Liverpool John Moores University, 146 Brownlow Hill, Liverpool L3 5RF, UK
   \and
   Leibniz-Institut für Astrophysik Potsdam (AIP), An der Sternwarte 16, 14482 Potsdam, Germany 
   }

   \date{Received MONTH DD, YYYY; accepted MONTH DD, YYYY}

  \abstract
   {}
   {The formation process of multiple populations in globular clusters is still up for debate. These populations are characterized by different light-element abundances. Kinematic differences between the populations are particularly interesting in this respect, because they allow us to distinguish between single-epoch formation scenarios and multi-epoch formation scenarios. We derive rotation and dispersion profiles for 25 globular clusters and aim to find kinematic differences between multiple populations in 21 of them to constrain the formation process.}
   {We split red-giant branch (RGB) stars in each cluster into three populations (P1, P2, P3) for the type-II clusters and two populations (P1 and P2) otherwise using Hubble photometry. We derive the global rotation and dispersion profiles for each cluster by using all stars with radial velocity measurements obtained from MUSE spectroscopy. We also derive these profiles for the individual populations of each cluster. Based on the rotation and dispersion profiles, we calculate the rotation strength in terms of ordered-over-random motion $\left(v/\sigma\right)_\mathrm{HL}$ evaluated at the half-light radius of the cluster. We then consistently analyse all clusters for differences in the rotation strength of their populations.}
   {We detect rotation in all but four clusters. For NGC~104, NGC~1851, NGC~2808, NGC~5286, NGC~5904, NGC~6093, NGC~6388, NGC~6541, NGC~7078 and NGC~7089 we also detect rotation for P1 and/or P2 stars. For NGC~2808, NGC~6093 and NGC~7078 we find differences in $\left(v/\sigma\right)_\mathrm{HL}$ between P1 and P2 that are larger than $1\sigma$. Whereas we find that P2 rotates faster than P1 for NGC~6093 and NGC~7078, the opposite is true for NGC~2808. However, even for these three clusters the differences are still of low significance. We find that the strength of rotation of a cluster generally scales with its median relaxation time. For P1 and P2 the corresponding relation is very weak at best. We observe no correlation between the difference in rotation strength between P1 and P2 and cluster relaxation time. The stellar radial velocities derived from MUSE data that this analysis is based on are made publicly available.}
   {}


   \maketitle
%
\section{Introduction}
\label{sec:intro}
Classically, it is assumed that all stars in a globular cluster form in the same molecular cloud and therefore are identical in age and chemical abundances. The discovery of multiple populations of stars within  globular clusters calls this into question. These populations generally differ in light element abundances \citep{carretta_na-o_2009}, but there is no evidence of age differences larger than $\sim 0.1\,$Gyrs \citep{bastian_multiple_2018, martocchia_search_2018}. Using these abundance differences, the stars of most clusters (type-I clusters) can be separated into at least two populations. One population has a scaled solar metallicity, whereas the other populations are always enriched in some light elements (such as N or Na) and depleted in others (such as C or O). The fraction of enriched to non-enriched stars and the strength of the spread in light element abundances increase with cluster mass \citep[e.g.][]{carretta_properties_2010, milone_hubble_2017}. There is evidence of metallicity spreads within some globular clusters for quite some time \citep[e.g.][]{carretta_abundances_2010}. \citet{milone_hubble_2017} found that for these clusters (type-II clusters) the two stellar populations are themselves split. Type-II clusters exhibit multiple subgiant and red giant branches, likely due to variations in heavy-elements abundances. \citet{pfeffer_accreted_2021} discussed several of these clusters and argued that some of them are actually remnants of nuclear star clusters. The occurrence of multiple populations is not limited to Galactic globular clusters, but they have also been observed in clusters of other galaxies \citep[e.g.][]{mucciarelli_looking_2009, milone_multiple_2009, dalessandro_multiple_2016}. Furthermore, \citet{martocchia_age_2018} found that cluster age might play a role in the onset of multiple populations, because they did not detect light-element variations in clusters younger than $\sim 1.7\,$Gyr. However, by analyzing main sequence stars instead of red giant branch stars \citet{cadelano_expanding_2022} found evidence for multiple populations in the ~1.5 Gyr old star cluster NGC~1783.

\begin{figure*}[t]
    \centering
    \includegraphics[width=0.49\textwidth]{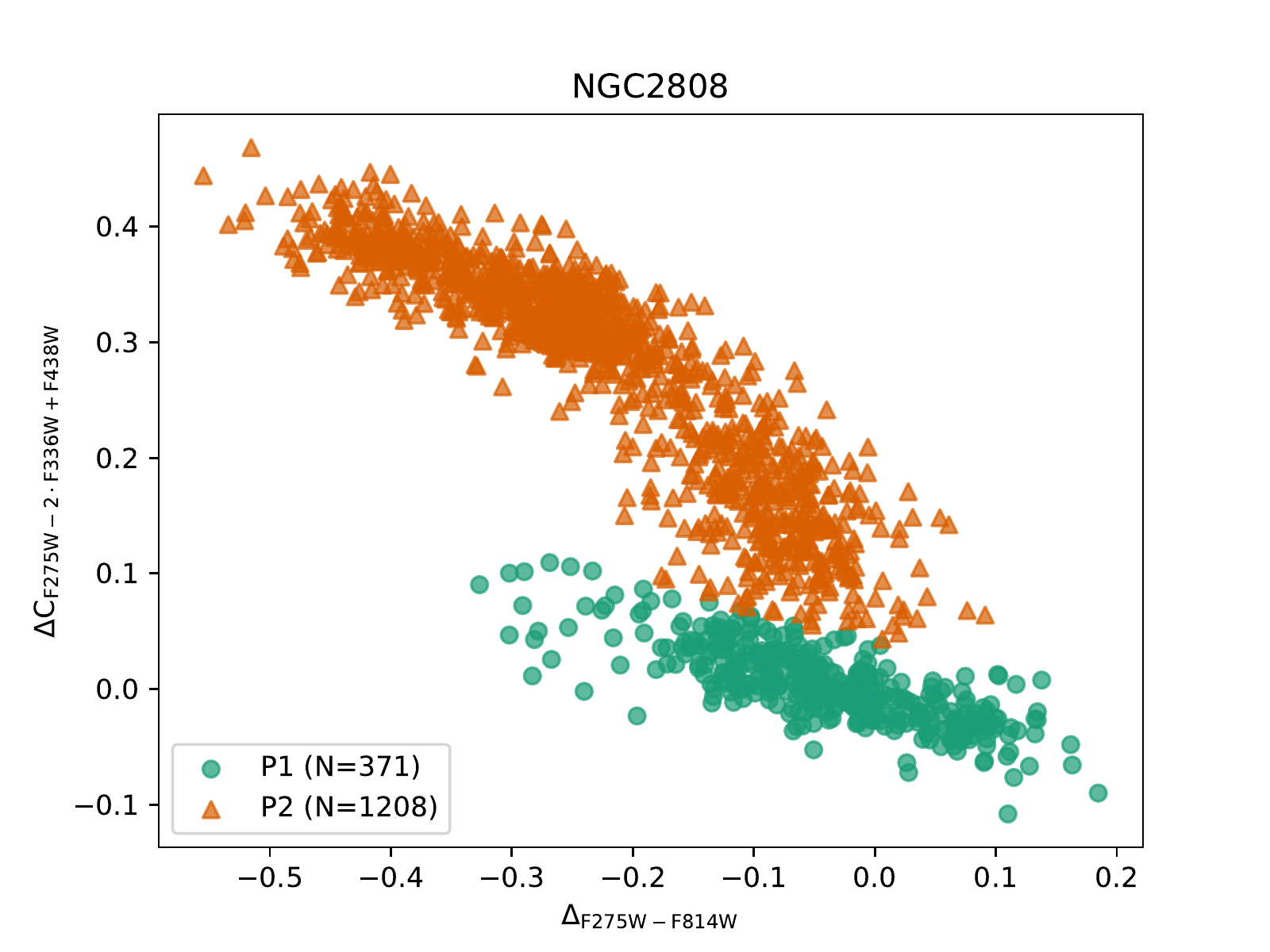}
    \includegraphics[width=0.49\textwidth]{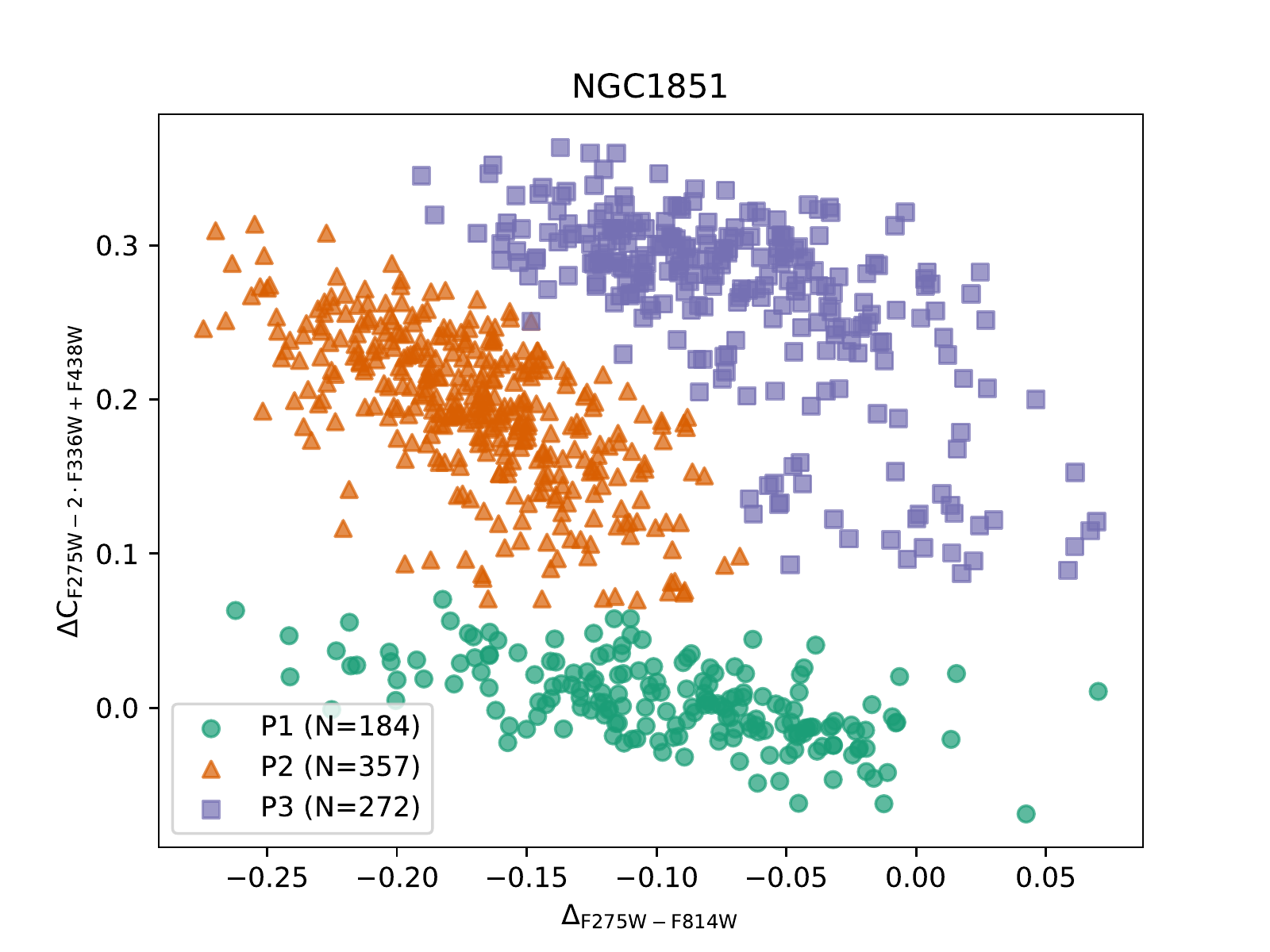}
    \caption{Chromosome maps for NGC 2808 and NGC 1851, where the identified populations of each cluster are labeled with different colors and symbols. For the type-II cluster NGC 1851 there is an additional population compared to NGC 2808.}
    \label{fig:cmap_ngc2808_ngc1851}
\end{figure*}

Several formation scenarios for multiple populations in globular clusters have been put forward, but each scenario comes with its caveats \citep[for a detailed review see][]{bastian_multiple_2018}. There are two types of formation scenarios that we discuss here briefly. On the one hand, multi-epoch formation scenarios propose the formation of a second generation of stars that form from gas polluted by primordial stellar sources, such as asymptotic giant branch stars or fast-rotating massive stars  \citep{cottrell_correlated_1981, renzini_rethinking_2013, decressin_origin_2007, decressin_fast_2007}. On the other hand, single-epoch formation scenarios suggest that some stars accrete material when moving through the cluster center in order to explain the observed spread in light-element abundances \citep{bastian_early_2013, gieles_concurrent_2018}.

The structural and kinematical differences between the multiple populations of globular clusters today could give insights on the formation process of these populations. Even though kinematic differences were imprinted during the birth of each cluster, and they diminish over time due to the interactions between stars during cluster evolution, at least some clusters are still expected to show measurable differences in their present day kinematics \citep{vesperini_dynamical_2013, henault-brunet_multiple_2015, tiongco_kinematical_2019}. Differences in the concentrations of stars between populations at the time of formation would entail differences in radial anisotropies over time \citep[e.g.][]{richer_dynamical_2013, bellini_hubble_2015}. Currently, a majority of globular clusters shows a higher concentration of stars from the enriched population compared to the pristine population \citep[e.g.][]{dalessandro_family_2019}. 

\citet{henault-brunet_multiple_2015} showed that multi-epoch and single-epoch formation scenarios result in very similar radial anisotropy profiles, as they all assume that the second populations forms centrally concentrated relative to the first population. Therefore, anisotropy cannot be used to distinguish between these two types of scenarios. However, these two formation scenarios would entail different initial conditions for the kinematics of these populations of stars. For multi-epoch formation scenarios, the rotation velocity is expected to be lower for the non-enriched population compared to the N-enriched population. On the contrary, for single-epoch formation scenarios, the non-enriched population is expected to have a higher rotation velocity than the N-enriched population \citep{henault-brunet_multiple_2015}.

Studies have been carried out to look for these differences in several globular clusters. In particular, \citet{cordero_differences_2017} were the first to find rotational differences between multiple populations in a globular cluster for NGC~6205. Furthermore, no kinematic differences between populations have been found for NGC~6121 \citep{cordoni_three-component_2020}, NGC~6838 \citep{cordoni_three-component_2020}, NGC~6352 \citep{libralato_hubble_2019}, NGC~6205 and NGC~7078 \citep{szigeti_rotation_2021}. For NGC~104, \citet{milone_gaia_2018} and \citet{cordoni_three-component_2020} found no differences in the rotation pattern between populations, but the latter did find that the enriched population exhibits stronger anisotropy than the non-enriched population. \citet{cordoni_three-component_2020} also found significant differences in the phases of the rotation curves between populations for NGC~5904. However, \citet{szigeti_rotation_2021} could not find significant differences in the rotation curves of this cluster, but for NGC~5272 they found that the enriched population is rotating faster than the non-enriched population. \citet{bellini_hubble_2015} found differences in the radial anisotropy of NGC~2808 based on proper motion data, but there is no analysis on radial rotation and dispersion profiles yet for this cluster. For NGC~6362, \citet{dalessandro_3d_2021} found that the enriched population is rotating faster than the non-enriched population. \citet{kamann_peculiar_2020} found similar results for NGC~6093 in that one of the enriched populations they identified is rotating significantly stronger than the non-enriched population. The overall lack of uniformity and agreement in these results emphasizes the need to further study the kinematic differences in multiple populations of globular clusters.

In this work, we are following the approach described by \citet{kamann_peculiar_2020} to systematically study the kinematics for 25 galactic globular clusters and looking at differences between populations in 21 of them. We used radial velocities derived from MUSE spectroscopy in combination with radial velocities from \citet{baumgardt_catalogue_2018} to extend the radial coverage of each globular cluster, as described in Section \ref{sec:data}. In Section \ref{sec:methods} we describe our approach to split the stars of each cluster into two or three populations based on photometric data. We use the stellar radial velocities to create radial rotation and dispersion profiles and derive parameters to characterize the global dynamics of each cluster and of its stellar populations. We present the results of this analysis in Section \ref{sec:results} and conclude in Section \ref{sec:conclusion}.


\section{Data}
\label{sec:data}
The globular clusters analyzed in this study were observed with the Multi Unit Spectroscopic Explorer  \citep[MUSE,][]{bacon_muse_2010}. MUSE is an integral field spectrograph mounted at UT4 of the ESO Very Large Telescope (VLT) that has been in operation since 2014. It features a wide-field mode with a field of view of $1'\times 1'$ at a sampling of $0.2''$ per pixel. Since 2019 MUSE also possesses a narrow-field mode that covers a field of view of $7.5''\times 7.5''$ at a sampling of $0.025"$ per pixel. Both modes cover a spectral range of $4750\,\AA - 9350\,\AA$ with a corresponding spectral resolution (R) of $1770$ at $4750\,\AA$ and $3590$ at $9350\,\AA$. This analysis is based on the MUSE Galactic globular cluster survey, presented in \citet{kamann_stellar_2018}. We are using all available wide-field mode data for each globular cluster.

Basic data reduction is carried out using the official MUSE pipeline \citep{weilbacher_data_2020}. The process of extracting spectra of single stars from the resulting datacube is described in detail by \citet{kamann_stellar_2018}. In short, the program \texttt{PAMPELMUSE} by \citet{kamann_resolving_2013} is used in combination with a reference source catalog derived from HST photometry \citep{sarajedini_acs_2007, anderson_acs_2008} to determine the position of each resolved star in the MUSE data and to fit the MUSE PSF as a function of wavelength to retrieve stellar spectra.

As described by \citet{husser_muse_2016}, the line-of-sight velocity of each star is derived from its spectrum using cross-correlation and a full spectral fitting approach. By cross-correlating each spectrum against a set of template spectra, the velocity $v_\mathrm{los, cc}$ is derived. The value of $v_\mathrm{los, cc}$ is then used as an initial guess for the full spectral fitting method, described in detail in \citet{husser_muse_2016}. Using a Levenberg-Marquardt algorithm, the observed stellar spectra are fitted against the \textit{Göttingen Spectral Library} \citep{husser_new_2013} to derive the stellar metallicity [M/H], effective temperature $T_\mathrm{eff}$, and radial velocity $v_\mathrm{los, fit}$.
\begin{table}[t]
\renewcommand{\arraystretch}{1.0}
\caption{Overview of the number of stars per cluster and population.}
\label{tab:clusters_num}
\centering
\begin{tabular}{l l l l l l}
\hline\hline \noalign{\smallskip}
Cluster &$N_\mathrm{all}$ &$N_\mathrm{MUSE}$ &$N_\mathrm{P1}$ &$N_\mathrm{P2}$ &$N_\mathrm{P3}$\\ \noalign{\smallskip} \hline
NGC 104 & $32486$ & $28588$ & $343$ & $1250$ & --\\
NGC 362 & $8937$ & $8244$ & $234$ & $599$ & $22$\\
NGC 1851 & $15182$ & $13250$ & $184$ & $357$ & $272$\\
NGC 1904 & $5523$ & $4373$ & -- & -- & --\\
NGC 2808 & $15621$ & $13760$ & $371$ & $1208$ & --\\
NGC 3201 & $5206$ & $3917$ & $40$ & $53$ & --\\
NGC 5286 & $7373$ & $6472$ & $264$ & $371$ & $123$\\
NGC 5904 & $19096$ & $18297$ & $177$ & $587$ & --\\
NGC 6093 & $11419$ & $10879$ & $367$ & $582$ & --\\
NGC 6218 & $8053$ & $6156$ & $88$ & $123$ & --\\
NGC 6254 & $15418$ & $14744$ & $135$ & $243$ & --\\
NGC 6266 & $15527$ & $14255$ & -- & -- & --\\
NGC 6293 & $4455$ & $3081$ & -- & -- & --\\
NGC 6388 & $14711$ & $12528$ & $726$ & $1431$ & $404$\\
NGC 6397 & $11728$ & $8681$ & $19$ & $50$ & --\\
NGC 6441 & $13226$ & $11494$ & $1005$ & $1793$ & --\\
NGC 6522 & $6333$ & $2641$ & -- & -- & --\\
NGC 6541 & $12560$ & $11029$ & $339$ & $352$ & --\\
NGC 6624 & $7815$ & $6012$ & $130$ & $332$ & --\\
NGC 6656 & $17230$ & $11544$ & $122$ & $166$ & $113$\\
NGC 6681 & $5797$ & $4749$ & $38$ & $249$ & --\\
NGC 6752 & $15525$ & $13721$ & $104$ & $267$ & --\\
NGC 7078 & $13770$ & $12899$ & $390$ & $697$ & --\\
NGC 7089 & $12727$ & $12167$ & $264$ & $1086$ & $36$\\
NGC 7099 & $10046$ & $7705$ & $81$ & $182$ & --\\
\hline
\end{tabular}
\tablefoot{$N_\mathrm{all}$ describes the numer of stars including data from \citet{baumgardt_catalogue_2018}, whereas $N_\mathrm{MUSE}$ is based solely on MUSE data.}
\end{table}
\\
To retain a reliable data set, we are using a set of filters on the line-of-sight velocities derived from MUSE spectra. We are employing the reliability parameter $R_\mathrm{total} > 0.8$ described in \citet{giesers_stellar_2019}. It ensures that the value of $v_\mathrm{los, cc}$ is trustworthy and consistent with $v_\mathrm{los, fit}$ and that the analyzed spectrum has a signal-to-noise ratio of $S/N > 5$. Additionally, we are removing stars that show temporal variations in their line-of-sight velocities (e.g. binaries, pulsating stars) because in such cases the measured velocities do not trace the gravitational potentials of the host clusters. We use the method described by \citet{giesers_stellar_2019} to derive the probability $p_\mathrm{var}$ that any star that was observed multiple times shows temporal variance in velocity. We exclude all stars with $p_\mathrm{var} > 0.5$. To derive one value for the line-of-sight velocity for each star, we average over MUSE velocity measurements for stars that have been observed multiple times. The stellar coordinates and averaged radial velocities used for this analysis are listed in Table 2 which is only available in electronic form at CDS and on the Göttingen Research Online repository, at \url{XXX}.

To increase the radial coverage of each cluster, we use stellar velocities from \citet{baumgardt_catalogue_2018} in addition to the MUSE data. We match the \citet{baumgardt_catalogue_2018} data to HST photometry from \citet{anderson_acs_2008}, but the radial velocities extend much further out than the HST data, so that most radial velocities are not matched, but simply added to our data set. Before combining, the respective systematic cluster velocity is subtracted from the stellar velocities to minimize systematic differences between data sets. If a star is included in the samples of both MUSE and \citet{baumgardt_catalogue_2018}, we average the stellar velocities from both sources. As shown by \citet{kamann_stellar_2018}, the stellar radial velocities from MUSE and \citet{baumgardt_catalogue_2018} agree in regions where they overlap. In the outer regions the completeness in terms of fraction of stars with a radial velocity measure is significantly smaller than in the center since the \citeauthor{baumgardt_catalogue_2018} sample only consists of giant branch stars. Because of energy equipartition, which causes mass segregation, it is possible that this affects the velocity dispersion in the outer regions. However, using the formula by \citet{bianchini_novel_2016} we estimated that the difference between our MUSE sample and the \citeauthor{baumgardt_catalogue_2018} sample in dispersion is $\lesssim 0.1\,$ km/s based on the average masses of stars in either sample, which is fully within the uncertainties of our measurements.

\section{Methods}
\label{sec:methods}
\subsection{Population Split}
\label{subsec:pop_split}
The separation into multiple populations is based on the chromosome map of each globular cluster. A chromosome map, which was first introduced by \citet{milone_hubble_2017}, is a pseudo-color color diagram using a combination of the HST filters F275W, F336W, F438W and F814W that splits the stars of a cluster into its populations. We use photometry from the HST UV Globular Cluster Survey (HUGS) from \citet{piotto_hubble_2015} and \citet{nardiello_hubble_2018} to create these maps, as explained in \citet{latour_stellar_2019}. We note that only RGB stars are included in the population analysis because the chromosome maps are tailored to distinguish populations at this evolutionary phase. 

We need a consistent population separation because we want to compare the kinematics of equivalent populations between clusters. For type-I clusters, we use the fact that these clusters always consist of one population with a scaled-solar abundance (P1) and at least one additional population (P2) that differs in abundances to P1. Therefore, we follow the classification by \citet{milone_hubble_2017} and divide the stars of type-I clusters into two groups using the chromosome map. The left panel of Fig. \ref{fig:cmap_ngc2808_ngc1851} shows our chromosome map and identified populations P1 and P2 for NGC~2808. Stars from P1 are always found in the lower part of the chromosome map around (0, 0) coordinate and, depending on the cluster, partly extend horizontally, whereas P2 stars are located above P1 stars and extend diagonally toward the top left.

Type-II clusters contain stars that are enhanced in some particular heavy-elements, such as barium and lanthanum, possibly iron as well, compared to P1 and P2 stars \citep{marino_iron_2015}. These metal-enhanced stars are found on the reddest part of the RGB in the CMD \citep[see, e.g.][]{milone_hubble_2017}. For the Type-II clusters, we consider these stars as a third population (P3). The right panel in Fig. \ref{fig:cmap_ngc2808_ngc1851} shows our split into three populations for the type-II cluster NGC~1851. The position of P1 and P2 for NGC~1851 are similar to NGC~2808, but the third population has a distinct position on the upper-right region of the chromosome map.

We analyze 25 globular clusters, six of which are considered type-II clusters. As briefly mentioned in Section \ref{sec:intro}, some type-II clusters might actually be the remnants of nuclear star clusters. However, \citet{pfeffer_accreted_2021} conclude that none of the type-II clusters in our sample are remnants of nuclear star clusters. Therefore, we treat them as globular clusters. A list of all clusters with the number of stars used in the analysis of the global kinematics, and the number of stars per population, are shown in Table \ref{tab:clusters_num}. For NGC~1904, NGC~6266, NGC~6293 and NGC~6522 the necessary HUGS photometric data to separate their RGB stars into populations are not available. In these cases, we only derive the global kinematic properties.

\subsection{Kinematics}
\label{subsec:kinematics}
To study the kinematic differences between populations, we first use the radial velocities of all stars in the cluster to derive its global kinematics. Then we repeat the same procedure, including only the RGB stars assigned to specific populations. From the radial velocity data we create kinematic profiles and finally derive the ratio of ordered-over-random motion $\left(v/\sigma\right)_\mathrm{HL}$ and a proxy for the spin parameter $\lambda_\mathrm{R, HL}$. Both are evaluated at the half-light radius to characterize the strength of rotation for all clusters and each of their populations. We use a very similar approach as described in \citet{kamann_peculiar_2020}, but we repeat the important assumptions and formulas below.

To create kinematic profiles, we need to derive the rotational velocity $v_\mathrm{rot}$ and the line-of-sight velocity dispersion $\sigma_\mathrm{los}$ of each cluster as a function of radius. To derive these parameters, we employ the maximum-likelihood approach described in \citet{kamann_stellar_2018}. We assume that the line-of-sight stellar velocities at any position  $(x, y)$ may be approximated by a Gaussian with mean $v_\mathrm{los}$ and standard deviation $\sigma_\mathrm{los}$. To account for the position of each star within the cluster with respect to the rotation axis of that cluster, we parameterize the line-of-sight velocities according to $v_\mathrm{los}(r,  \theta) = v_\mathrm{rot}(r)\sin(\theta-\theta_0)$, where $r=\sqrt{x^2+y^2}$ is the distance from the cluster center, $\theta_0$ is the angle of the cluster rotation axis, and $\theta=\text{atan2}(y, x)$ is the position angle measured counter-clockwise from north to east.

As described in \citet{kamann_peculiar_2020}, we use parametric and non-parametric models to analyze the stellar rotation velocities. For the non-parametric approach, we simply bin velocities radially and derive the rotation velocity and velocity dispersion for each bin. The parametric rotation profile $v_\mathrm{rot}$ we employ is characteristic for systems that have undergone violent relaxation \citep{lynden-bell_statistical_1967,gott_dynamics_1973}:
\begin{align}
\label{eq:vrot}
    v(r) = v_\mathrm{sys} + v_\mathrm{rot}(r)= v_\mathrm{sys} + 2 v_\mathrm{max} \frac{r_\mathrm{peak} r}{r_\mathrm{peak}^2+r^2},
\end{align}
where $v_\mathrm{sys}$ is the systematic velocity, $v_\mathrm{max}$ is the maximum rotation velocity reached at radial distance $r_\mathrm{peak}$. The parametric dispersion profile we use is a \citet{plummer_problem_1911} profile
\begin{align}
\label{eq:sigma}
    \sigma_\mathrm{los}=\frac{\sigma_\mathrm{max}}{\left(1+\left(\frac{r}{a_0}\right)^2\right)^\frac{1}{4}}.
\end{align}
\\
We differ in our approach to handling non-member stars compared to \citet{kamann_peculiar_2020}. They used cluster membership probabilities that were derived from stellar metallicities and line-of-sight velocities, as described by \citet{kamann_muse_2016}. We cannot take the same approach, because we do not have metallicities for the additional \citeauthor{baumgardt_catalogue_2018} stars. We modified the method of membership determination in order to have consistent membership probabilities for stars from both data sets. In particular, we introduce a prior on the membership probability for each star $p_i$ that is related to the stellar surface density of the cluster $\rho(r_i)$ at the radial distance $r_i$ of that star from the cluster center according to
\begin{align}
    p_i(r)=\frac{\rho(r_i)}{\rho(r_i) + f_\mathrm{fg}},
\end{align}
where $f_\mathrm{fg}$ measures the fractional contribution of foreground sources to the observed source density. To describe the stellar surface density of each cluster, we use the \texttt{LIMEPY} models described in \citet{gieles_family_2015}, with parameters for each cluster as determined by \citet{de_boer_globular_2019}. For NGC~6441 and NGC~6522 \citet{de_boer_globular_2019} do not provide any parameters for these models, so we chose to use King models \citep{king_structure_1966} with the necessary parameters of central concentration and core radius taken from \citet[][2010 edition]{harris_catalog_1996}. We modified the likelihood function $\mathcal{L}_\mathrm{i}$ of each star $i$ to not solely be based on the likelihood of the rotation and dispersion model $\mathcal{L}_\mathrm{cl, i}$, but to include the membership probability $p_i$ as follows:
\begin{align}
    \mathcal{L}_\mathrm{i} = p_i  \mathcal{L}_\mathrm{cl,i} + (1-p_i)  \mathcal{L}_\mathrm{fg,i}
\end{align}
where $\mathcal{L}_\mathrm{fg, i}$ is the likelihood that star $i$ is part of a foreground population. This foreground population is built from single stars included in a Besançon model \citep{robin_synthetic_2003} at the position of each cluster. The foreground likelihood for each star $\mathcal{L}_\mathrm{fg, i}$ is then defined as the superposition of Gaussian kernels for $M$ simulated stars with line-of-sight velocities $v_\mathrm{fg, j}$:
\begin{align}
    \mathcal{L}_\mathrm{fg, i}=\frac{1}{M}\sum_{j=1}^M \exp\left(-\frac{\left(v_\mathrm{fg, j} - v_\mathrm{los, i}\right)^2}{2\cdot v_\mathrm{err, i}^2}\right),
\end{align}
where $v_\mathrm{err, i}$ is the uncertainty of the measured line-of-sight velocities $v_\mathrm{los, i}$.

\begin{figure*}[t]
    \centering
    \includegraphics[width=\textwidth]{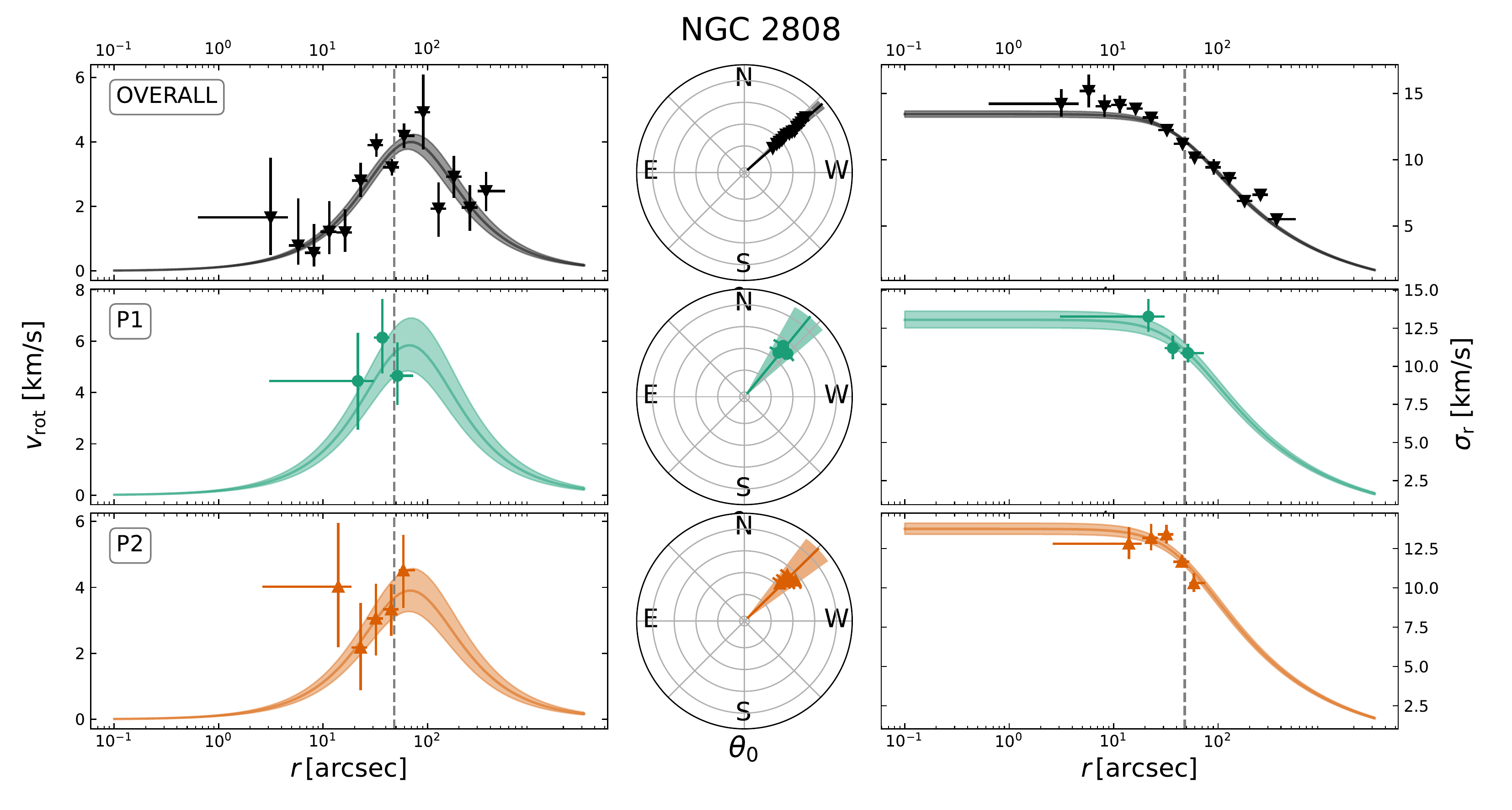}
    \caption{Rotation and dispersion profiles for NGC 2808 and each of its populations. The rotation profiles for each population are shown on the left, whereas the dispersion profiles are shown on the right. In the center, the angle of rotation is shown. The continuous profiles (solid lines) and binned profiles (symbols) shown here are each determined as fits on single stars, and the shaded area represents the $1\sigma$ uncertainty of each continuous profile and rotation angle. The dotted vertical line in each radial profile illustrates the half-light radius of NGC~2808.}
    \label{fig:profiles_ngc2808}
\end{figure*}

To maximize the likelihood of our model given the data, we use the Python package \texttt{emcee} from \citet{foreman-mackey_emcee_2013}, which is an implementation of the invariant Markov chain Monte Carlo (MCMC) ensemble sampler by \citet{goodman_ensemble_2010}. Thinning the samples of each parameter by 16 ensures that the final set of values for each parameter is again uncorrelated. We calculate the best fit parameters as the median of each distribution of thinned samples. In accordance with the standard deviation of a Gaussian, the lower and upper uncertainties of each fitted parameter are calculated using the 16th and 84th percentile of the corresponding distribution. However, in some cases, the distribution of the maximum rotation velocity $v_\mathrm{max}$ may have a peak at zero and drop towards zero for larger values of $v_\mathrm{max}$. In those cases we give an upper limit on the rotation velocity that corresponds to the 95th percentile of the distribution and the rotation angle cannot be calculated. The actual fitting procedure for each cluster is performed as follows:
\begin{enumerate}
    \item The parametric models are fitted simultaneously to the line-of-sight velocities for the overall cluster including the \citeauthor{baumgardt_catalogue_2018} data, with the systematic velocity $v_\mathrm{sys}$, rotation amplitude $v_\mathrm{max}$, rotation angle $\theta_0$,  $r_\mathrm{peak}$, $\sigma_\mathrm{max}$, $a_0$ and $f_\mathrm{fg}$ as free parameters. The corresponding priors for each parameter are presented in Table \ref{tab:prior_param} with the subscript 'o'. The priors on $r_\mathrm{peak}$ and $a_0$ were chosen to exclude unphysical solutions for either small or large values of these parameters.
    
    \item The parametric models are fitted independently for each population of that cluster with $v_\mathrm{sys}$, $v_\mathrm{max}$, $\theta_0$,  $r_\mathrm{peak}$, $\sigma_\mathrm{max}$, $a_0$ as free parameters. The fraction of foreground stars $f_\mathrm{fg}$ from the parametric fit of the overall cluster is used to derive membership probabilities for each star that are kept fixed. The priors of all fitted parameters are listed in Table \ref{tab:prior_param} with the subscript 'p'. The radial extent of the separation in multiple populations is limited by the availability of HUGS photometry from \citet{piotto_hubble_2015} and \citet{nardiello_hubble_2018}, which is why the radial coverage of the velocity and dispersion profiles for each population is limited compared to the overall cluster. Therefore, we chose to apply strict priors on $r_\mathrm{peak}$ and $a_0$ based on the distributions of samples from the parametric fit of the overall cluster. We chose to take a similar approach on $v_\mathrm{sys}$, since the systematic velocity should not vary between populations. Furthermore, we applied a soft prior in $v_\mathrm{max}$, that is based on the distributions of samples from the parametric fit of the overall cluster for $v_\mathrm{max}$ and $\sigma_\mathrm{max}$, and the escape velocity $v_\mathrm{esc}$ of that cluster, where we used values for the escape velocities from \citet{baumgardt_catalogue_2018}.
    
    \item The non-parametric model is fitted to the overall cluster and each of its populations with $v_\mathrm{sys}$, $v_\mathrm{max}$, $\theta_\mathrm{0}$ and $\sigma_\mathrm{los}$ as free parameters per radial bin. The priors for each parameter are listed in Table \ref{tab:prior_nonparam}. Again, we chose to limit the systematic velocity $v_\mathrm{sys}$ based on the results from the corresponding parametric fit. Additionally, we applied a prior on $\theta_\mathrm{0}$ based on the value of $\theta_\mathrm{0}$ from the corresponding parametric fit to ensure that the non-parametric profiles are consistent with the parametric profiles. We note that this approach introduces bias against depicting changes in the rotation axis with radius.
    
\end{enumerate}

As described above, we only use the additional radial velocities from \citet{baumgardt_catalogue_2018} in the analysis of the overall cluster, but not its populations, because we are unable to split those stars into populations. Nonetheless, the inclusion of this data is still important, since the strict priors on $r_\mathrm{peak}$ and $a_0$ for the population fit are solely based on the fit of the whole cluster. Without the addition of that data, we would not be able to reliably derive these parameters for most clusters, as a result of the smaller radial range. 

To quantify the effect of rotation, we calculate the ratio of ordered-over-random $\left(v/\sigma\right)_\mathrm{HL}$ motion for each population in all clusters. Classically, this ratio is defined as the ratio between the maximum rotation velocity to the central velocity dispersion. However, because of the weaknesses of this approach mentioned by \citet{binney_rotation_2005}, we follow the definition of $\left(v/\sigma\right)_\mathrm{HL}$ by \citet{cappellari_sauron_2007}:
\begin{align}
\label{eq:vsigma}
    \left(\frac{v}{\sigma}\right)_\mathrm{HL} &= 
    \frac{\langle v^2\rangle}{\langle {\sigma_\mathrm{r}}^2 \rangle} = 
    \frac{\int_0^{r_\mathrm{HL}} \rho(r) \frac{1}{2} v_\mathrm{rot}(r)^2 \, r\, \diff r}
    {\int_0^{r_\mathrm{HL}} \rho(r) \sigma_\mathrm{los}^2 \, r \, \diff r},
\end{align}
where $r_\mathrm{HL}$ is the half-light radius of each cluster. \citet{emsellem_sauron_2007} highlighted a potential shortcoming of using $\left(v/\sigma\right)_\mathrm{HL}$ to characterize velocity fields, in that structurally different velocity fields can result in very similar values of $\left(v/\sigma\right)_\mathrm{HL}$. To address this issue, they introduced $\lambda_\mathrm{R, HL}$ as an alternative, which is a proxy for the spin parameter of the velocity field:
\begin{align}
\label{eq:lambda}
    \lambda_\mathrm{R, HL} &=
    \frac{\langle r|v|\rangle}{\langle r\sqrt{{v^2 + \sigma_\mathrm{r}}^2} \rangle} = \frac{\int_0^{r_\mathrm{HL}} \rho(r) \frac{2}{\pi} |v_\mathrm{rot}(r)| \, r^2\, \diff r}
    {\int_0^{r_\mathrm{HL}} \rho(r) \sqrt{\sigma_\mathrm{los}^2 + \frac{1}{2} v_\mathrm{rot}(r)^2} \, r \, \diff r}.
\end{align}
We calculate both parameters based on the rotation and dispersion profiles described in Eq.~\ref{eq:vrot} and Eq.~\ref{eq:sigma} for each cluster and its populations to quantify kinematical differences between clusters and populations. For these calculations, we use the values of the half-light radius $r_\mathrm{HL}$ from \citet[][2010 edition]{harris_catalog_1996}. For each cluster and all of its populations, we use the global density profiles $\rho(r)$. For some clusters, the radial rotation profiles do not extend beyond the half-light radius. This could introduce some bias to the values of $\left(v/\sigma\right)_\mathrm{HL}$ and $\lambda_\mathrm{R, HL}$. However, as described earlier, we apply a strict prior on the radial scales of the rotation and dispersion profiles for each population based on the overall profile. Therefore, when we calculate $\left(v/\sigma\right)_\mathrm{HL}$ and $\lambda_\mathrm{R, HL}$ based on the MCMC results of the radial rotation and dispersion profiles, we expect that any bias that may occur is correctly reflected in our uncertainties of these parameters. The kinematical model we employ to derive the rotation and dispersion profiles is not sensitive to structural differences between velocity fields of different clusters. Therefore, we expect that $\left(v/\sigma\right)_\mathrm{HL}$ and $\lambda_\mathrm{R, HL}$ are qualitatively the same for each cluster with this model.

\section{Results}
\label{sec:results}

\subsection{Global Kinematics}
\label{subsec:global_kinematics}
Figure \ref{fig:profiles_ngc2808} shows the radial rotation and dispersion profiles for NGC~2808. In the top panel of this figure, the global profiles for the cluster are presented, where the outermost radial velocity data points are from the stars in the \citet{baumgardt_catalogue_2018} catalog. In the lower panels of this figure, the corresponding profiles are shown for each population. The continuous profiles (solid lines) and binned profiles (symbols) shown here are each determined as fits on single stars, and the shaded area represents the $1\sigma$ uncertainty of each continuous profile and rotation angle. The dashed line indicates the value of the half-light radius of this cluster \citep[][2010 edition]{harris_catalog_1996}. The profiles for the other 24 globular clusters and their corresponding chromosome maps are displayed in Figures \ref{fig:profiles_ngc104} to \ref{fig:profiles_ngc7099} in the appendix. The binned profiles highlight again that the radial extent of our data is limited to the center of each cluster. This could bias our ability to detect differences in kinematics between populations in the outer regions of the cluster. However, based on the work of \citet{henault-brunet_multiple_2015} we expect to find the largest differences between populations around the half-light radius of each cluster. Even closer to the center, differences should still be detectable. Nevertheless, the extension of our work to the outskirts of the clusters appears as a promising opportunity for future studies. Overall, the binned non-parametric profiles are in good agreement with the continuous parametric profiles. The largest discrepancies between these types of profiles are found close to the center, where the binned profiles indicate a rise in rotation velocity for some clusters (e.g. NGC~1904 and NGC~7089). Since the uncertainties of the binned rotation profiles are also the largest close to the center, it is uncertain whether this is a significant effect. We stress that the binned profiles are only used for visualization purposes and all following analyses are based on the parametric profiles. As mentioned in Sec. \ref{subsec:kinematics} the radial extent for the P1, P2 and P3 profiles is limited, which is revealed by the binned profiles. We used priors on the $r_\mathrm{peak}$ that are based on the parametric fit including all stars in the cluster. We use the parametric rotation and dispersion profiles to derive $\left(v/\sigma\right)_\mathrm{HL}$ and $\lambda_\mathrm{R, HL}$, according to Eq. \ref{eq:vsigma} and Eq. \ref{eq:lambda}, for each cluster and its populations. Both of these values are integrals of these profiles up to the half-light radius of each cluster, which makes both parameters robust against changes in $r_\mathrm{peak}$, as shown by \citet{kamann_peculiar_2020}. The fitted parameters and the values of $\left(v/\sigma\right)_\mathrm{HL}$ and $\lambda_\mathrm{R, HL}$ for the populations of all clusters are listed in Table \ref{tab:results}. Since the value of $v_\mathrm{sys}$ is close to zero for each cluster and its populations, it is not important for the subsequent analysis and is not discussed further. 

\begin{figure}[t]
    \centering
    \includegraphics[width=0.5\textwidth]{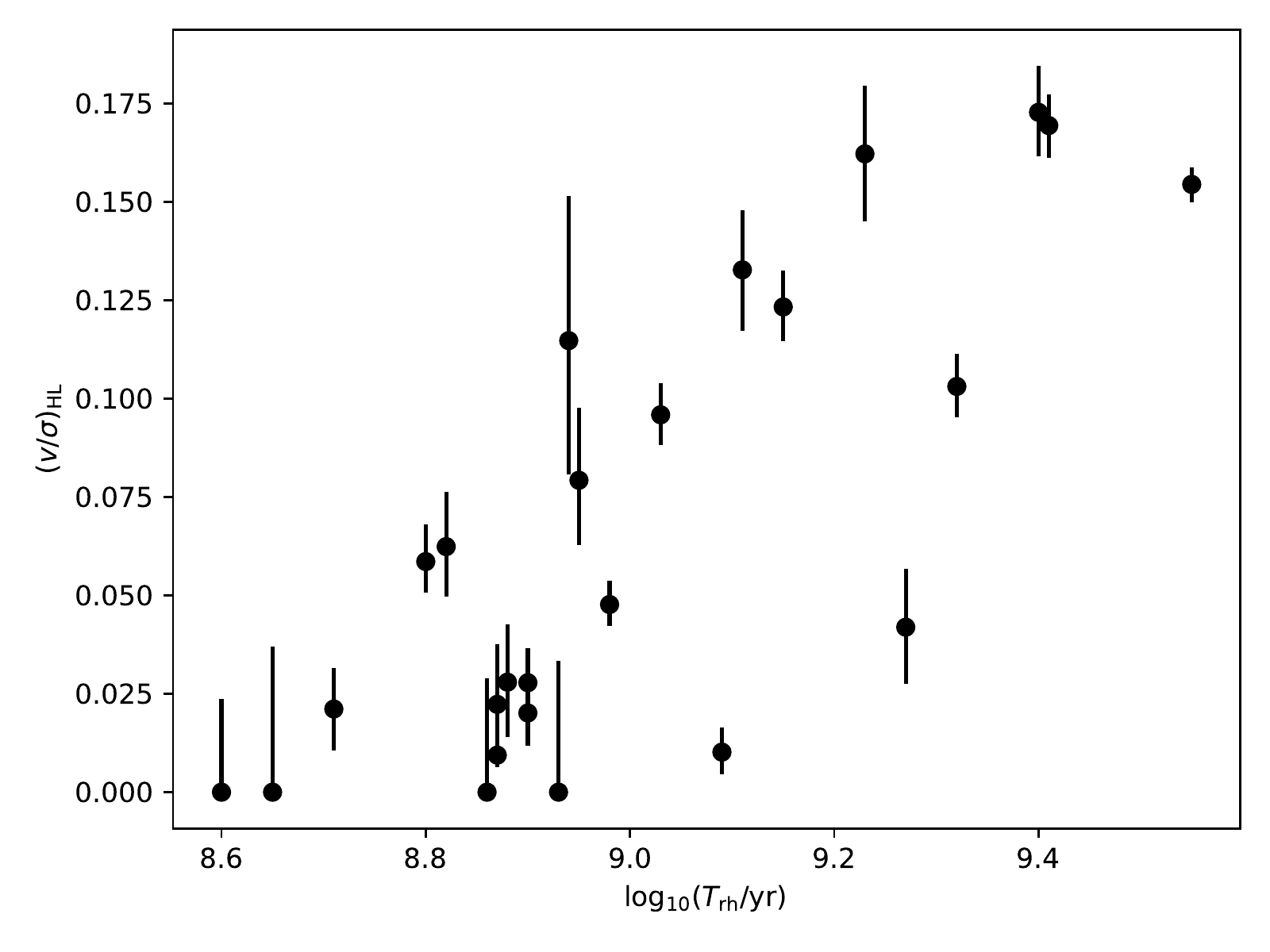}
    \caption{Relation between rotation strength in $\left(v/\sigma\right)_\mathrm{HL}$ and the median relaxation time $T_\mathrm{rh}$ of each cluster taken from \citet[][2010 edition]{harris_catalog_1996}}
    \label{fig:relax}
\end{figure}

In Fig. \ref{fig:relax} we show our values of $\left(v/\sigma\right)_\mathrm{HL}$ as a function of the median relaxation time $T_\mathrm{rh}$ of each cluster. The values for $T_\mathrm{rh}$ are from \citep[][2010 edition]{harris_catalog_1996}, see Tab. \ref{tab:results}. For the global kinematics of the clusters, we find that there is a relationship between $\left(v/\sigma\right)_\mathrm{HL}$ and the median relaxation time, in that for clusters with higher relaxation times we tend to get higher values in $\left(v/\sigma\right)_\mathrm{HL}$. In particular, for NGC~362, NGC~6397, NGC~6522 and NGC~6681 we do not find a significant sign of rotation and all of them have relaxation times of $\log_{10}(T_\mathrm{rh}/\mathrm{Gyr}) < 8.95$. Similar relations between cluster rotation and relaxation time have been found by \citet{kamann_stellar_2018}, \citet{bianchini_internal_2018} and \citet{sollima_eye_2019}. This is to be expected if we assume that globular clusters are imprinted at birth with the angular momentum of their parent molecular clouds. Over time, this angular momentum is dissipated outwards through two-body relaxation. In fact, numerical simulations show that star clusters can be rotating shortly after their birth \citep{mapelli_rotation_2017, bekki_formation_2019} and that the strength of rotation declines over time \citep{lahen_structure_2020}. For several clusters in our sample, the relaxation times provided by \citet{sollima_global_2017} differ from those by \citet[][2010 edition]{harris_catalog_1996}. If we use the values provided by \citet{sollima_global_2017}, we find a similar relation with our values of $\left(v/\sigma\right)_\mathrm{HL}$ but the correlation is weaker.
\\
We find a linear relation between $\left(v/\sigma\right)_\mathrm{HL}$ and $\lambda_\mathrm{R, HL}$ for all clusters and populations. This is shown in Figure \ref{fig:vsigma_vs_lambdar} in the appendix. We find that the constant of proportionality is $\approx 0.8$ for all cases. Since our kinematic model is not sensitive to structural differences in the velocity field of a cluster it is expected that $\left(v/\sigma\right)_\mathrm{HL}$ and $\lambda_\mathrm{R, HL}$ are qualitatively the same. In the following, we only use $\left(v/\sigma\right)_\mathrm{HL}$ to describe the kinematics of clusters and populations.

\subsection{Differences between P1 and P2}
\label{subsec:diff_pops}
\begin{figure*}[t]
    \centering
    \includegraphics[width=\textwidth]{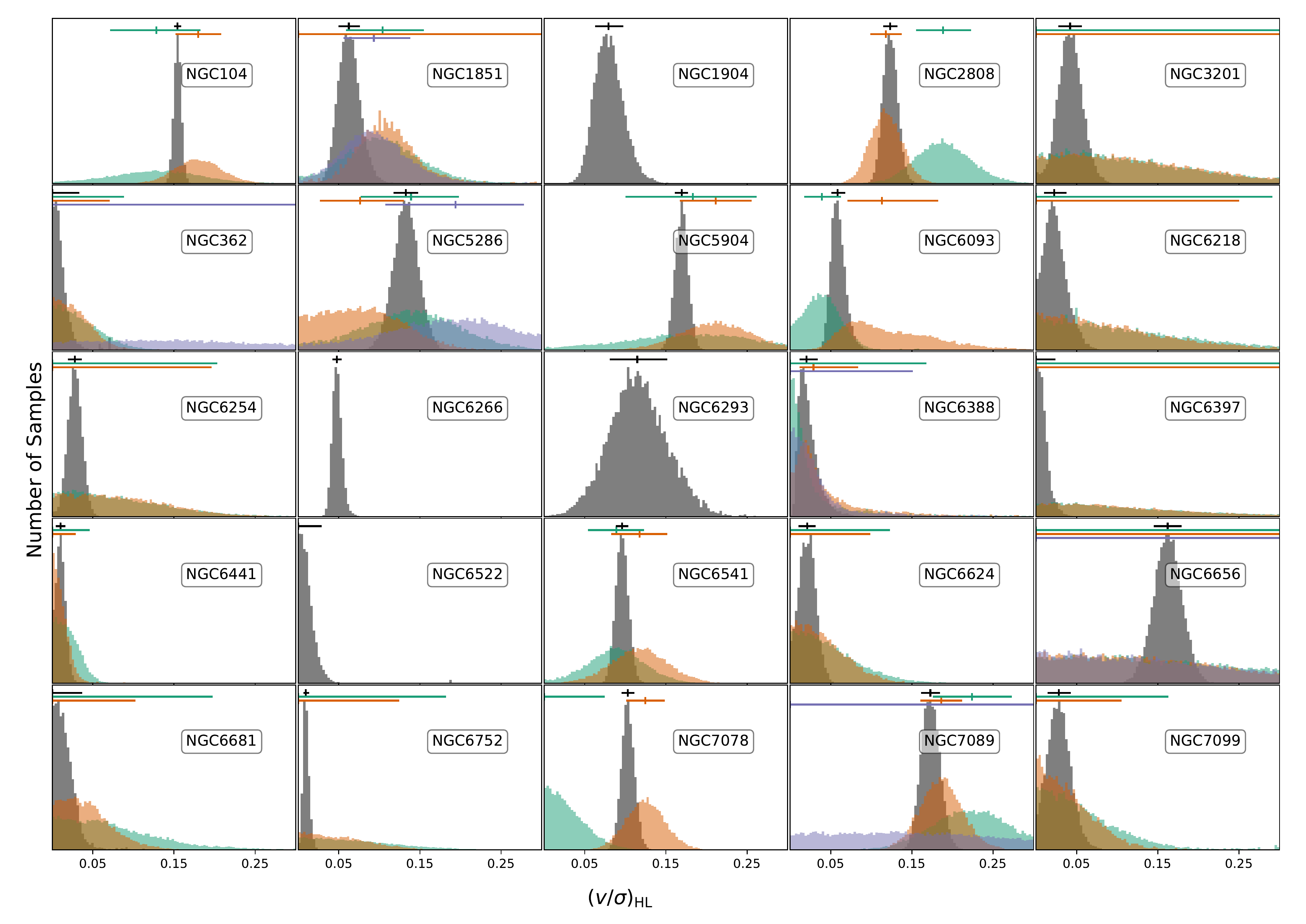}
    \caption{Distributions of samples of $\left(v/\sigma\right)_\mathrm{HL}$, which describes the strength of rotation for each cluster in this analysis. These distributions are shown for the overall cluster (gray) and each of its populations (P1:green, P2:orange, P3:violet). The 16th, 50th and 84th percentile for each distribution are shown on top of the corresponding distribution. For distributions that peak at zero, only the 95th percentile is shown to provide an upper limit on the rotation strength.}
    \label{fig:vsigma_hist}
\end{figure*}

To analyze P1 and P2 for differences in their kinematics, we compare the distributions of $\left(v/\sigma\right)_\mathrm{HL}$ derived from the thinned MCMC samples for $v_\mathrm{max}$, $\theta_0$,  $r_\mathrm{peak}$, $\sigma_\mathrm{max}$ and $a_0$ using Eq. \ref{eq:vrot}, Eq. \ref{eq:sigma} and Eq. \ref{eq:vsigma}. Figure \ref{fig:vsigma_hist} shows these distributions for all populations of each cluster. The distributions for P1 and P2 are shown in green and orange, respectively. For NGC~1904, NGC~6266, NGC~6293, and NGC~6522, only the distribution for all stars, in gray, is plotted because there is no separation into populations for these clusters.

\begin{figure}[t]
    \centering
    \includegraphics[width=0.5\textwidth]{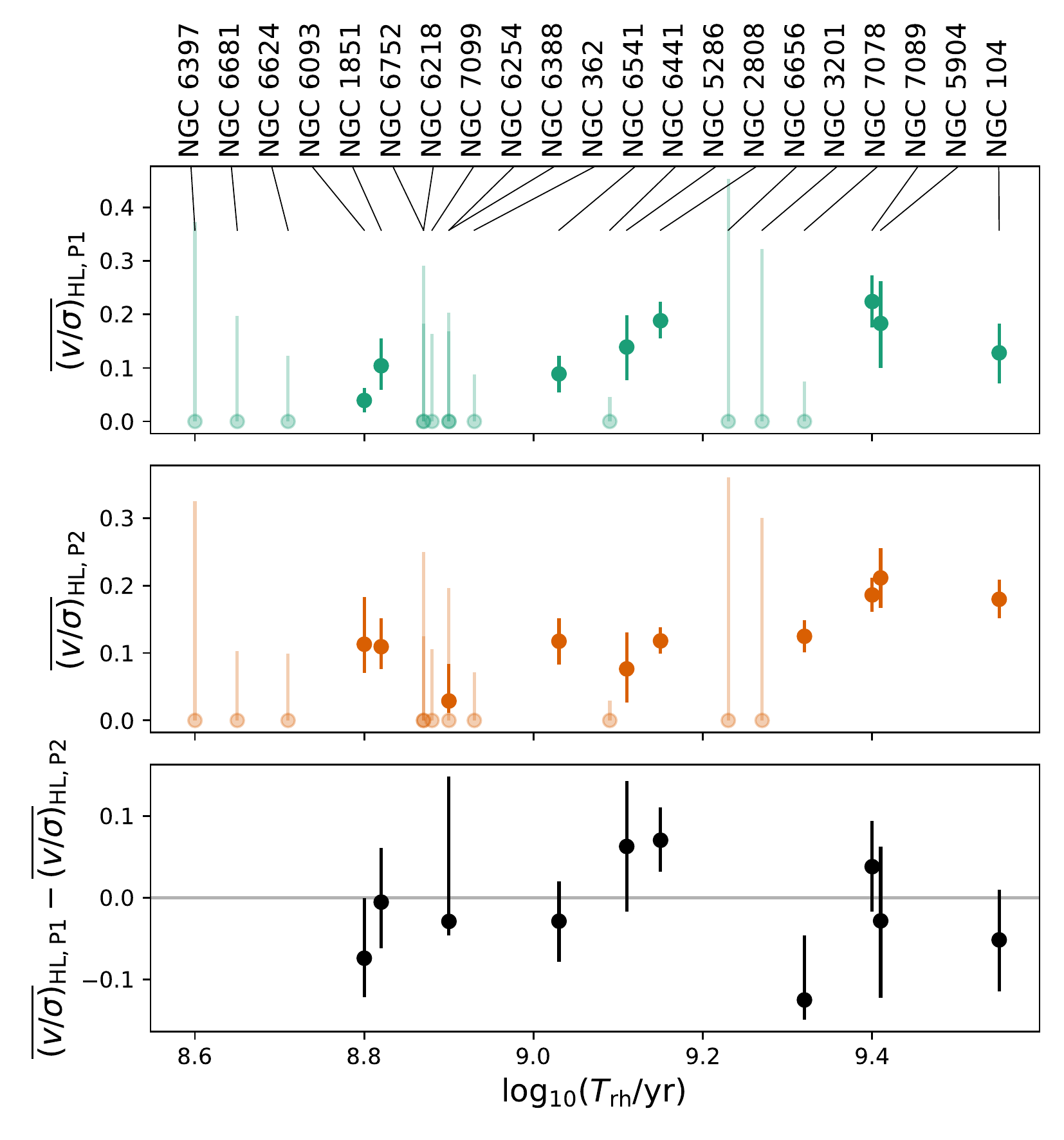}
    \caption{Relation between rotation strength in $\left(v/\sigma\right)_\mathrm{HL}$ of P1 and P2 and the median relaxation time $T_\mathrm{rh}$ for each cluster taken from \citet[][2010 edition]{harris_catalog_1996}, and the difference in rotation strength between P1 and P2 as a function of median relaxation time.}
    \label{fig:relax_per_pop_with_diff}
\end{figure}
\begin{figure*}[ht]
    \centering
    \includegraphics[width=0.49\textwidth]{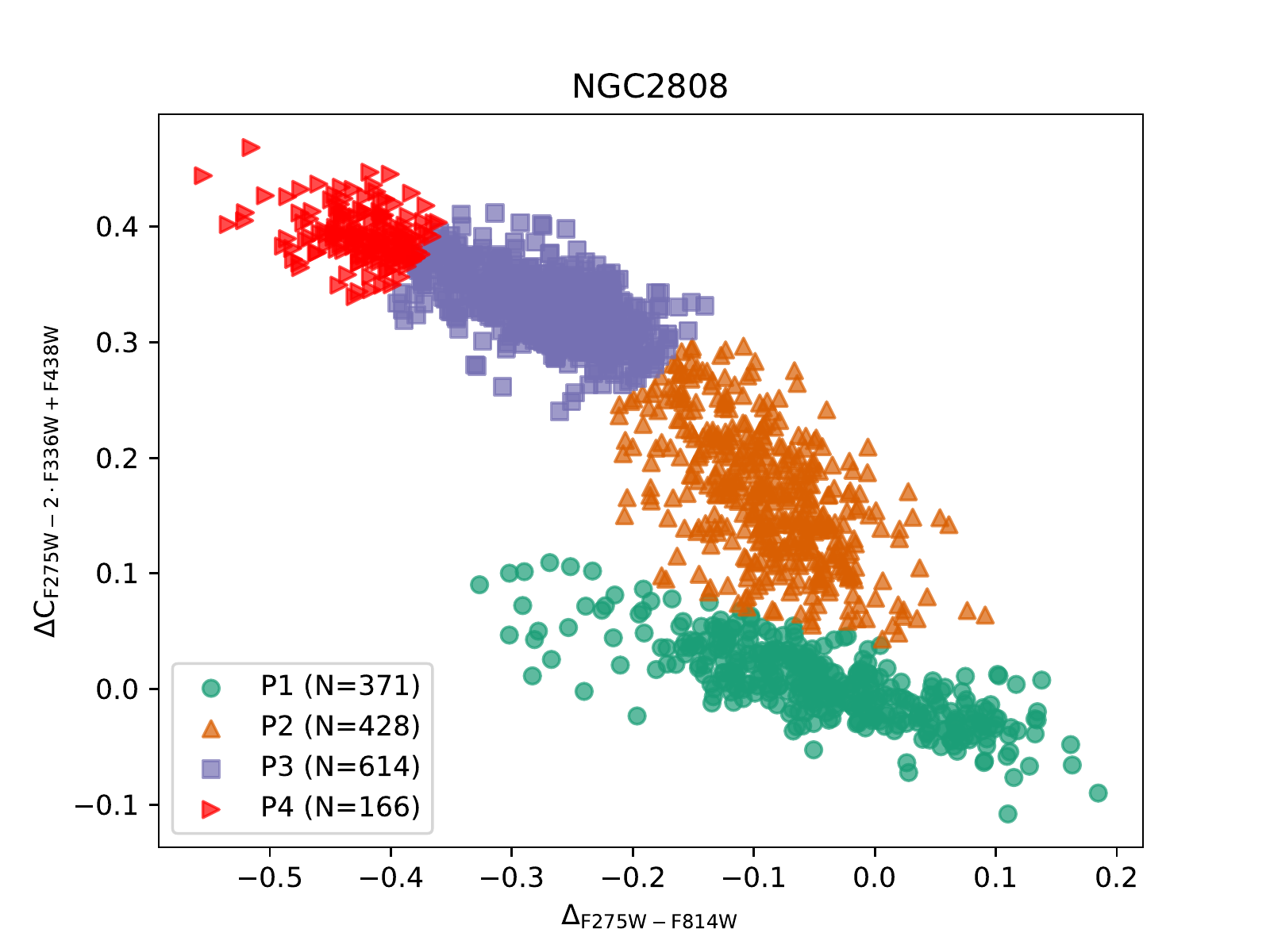}
    \includegraphics[width=0.49\textwidth]{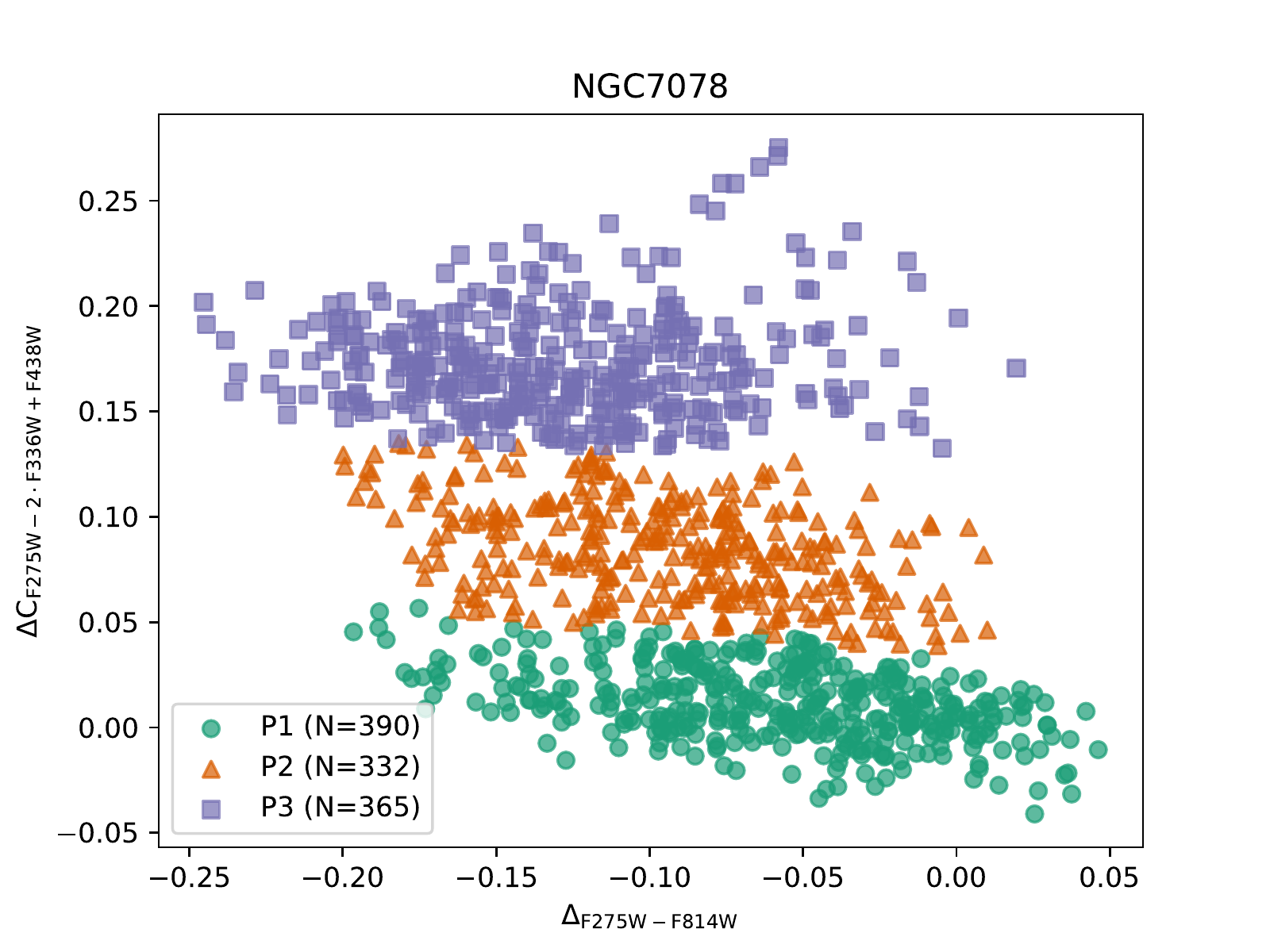}\\
    \includegraphics[width=0.49\textwidth]{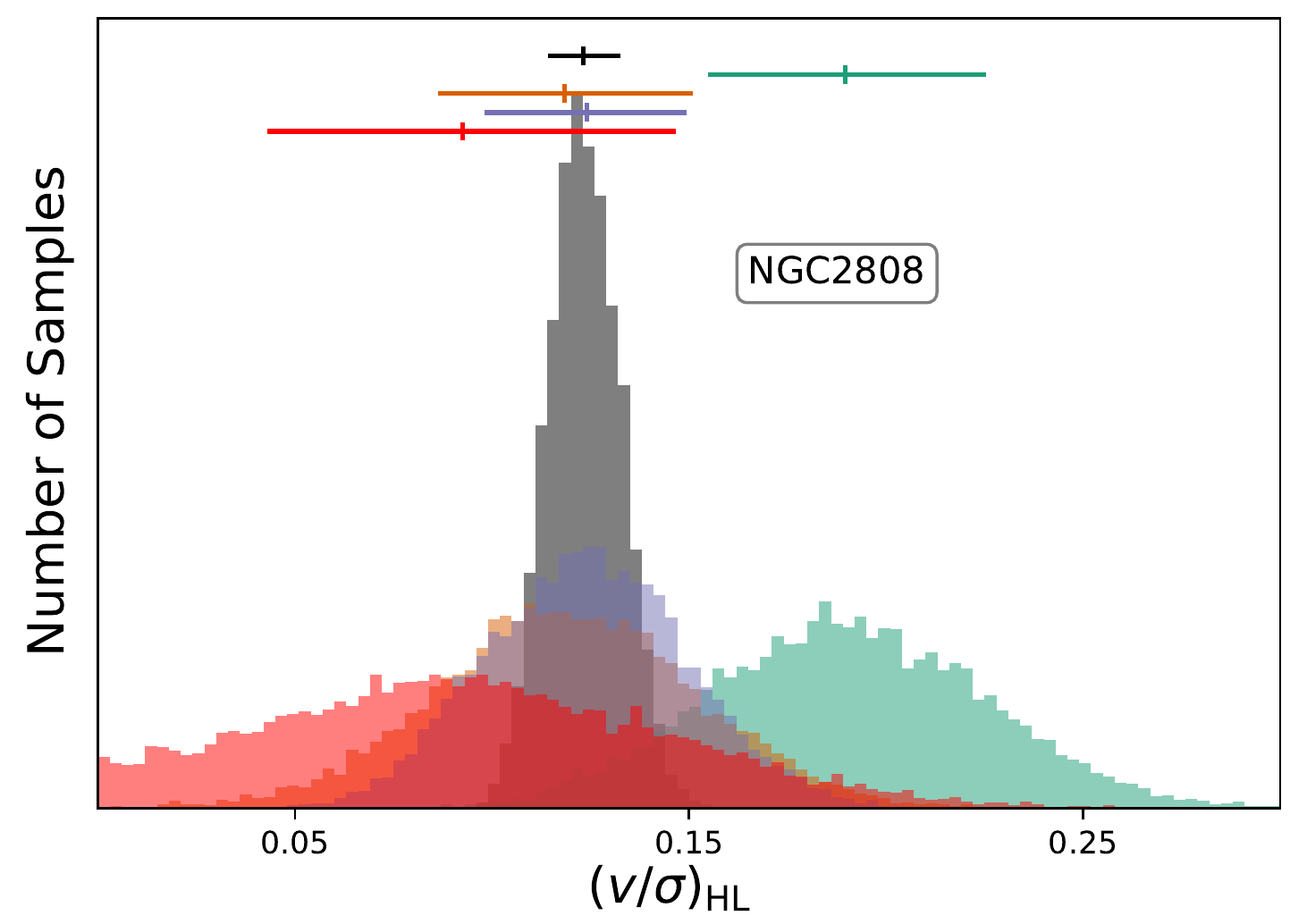}
    \includegraphics[width=0.49\textwidth]{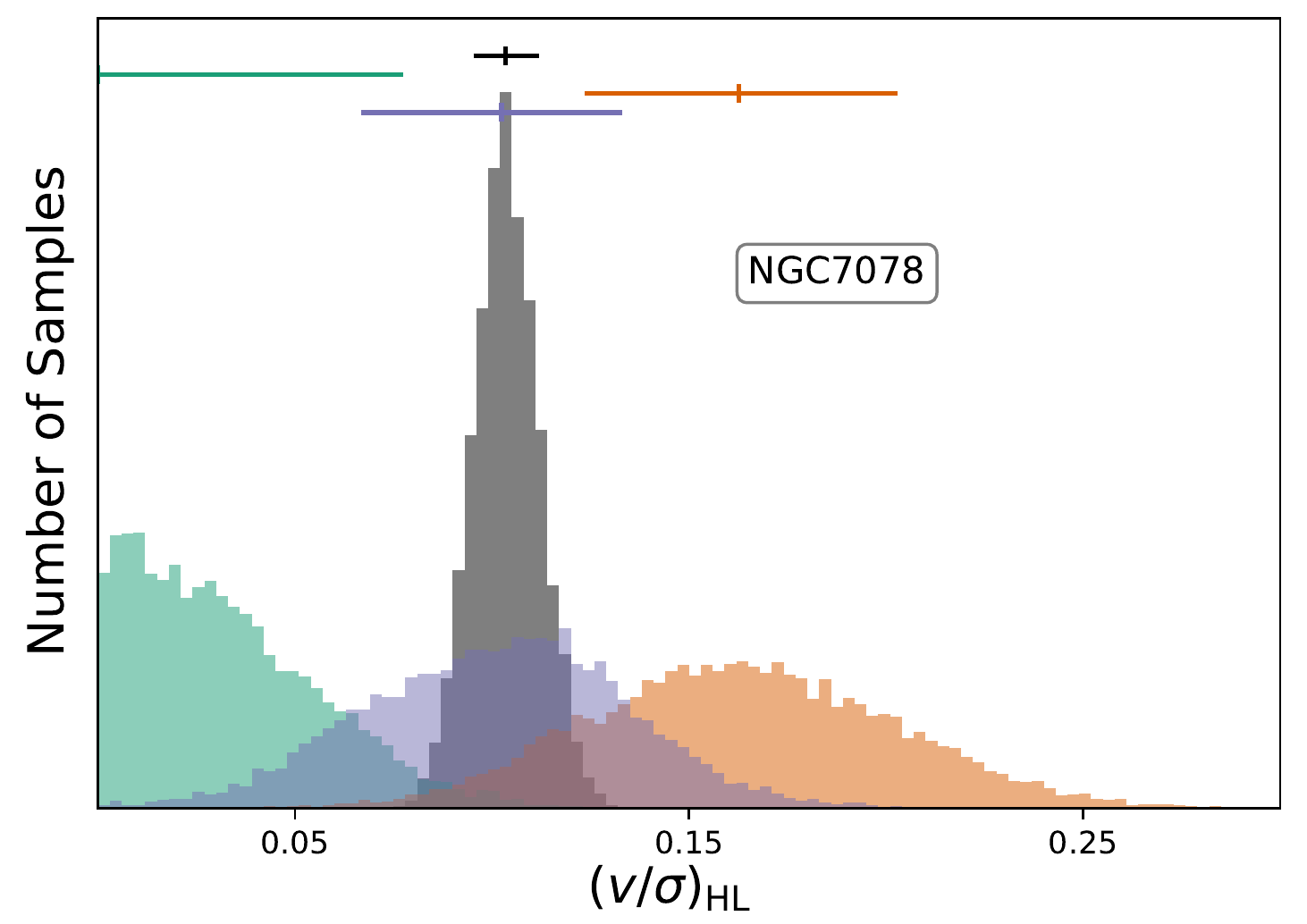}
    \caption{\textbf{Top}: Chromosome maps of NGC~2808 and NGC~7078, with additional population splits. For NGC~2808 the original P2 is split into three subpopulations (P2', P3' and P4'), whereas P2 of NGC~7078 is split into two subpopulations (P2' and P3') according to the respective morphology of the chromosome maps. \textbf{Bottom}: Distributions of samples of $\left(v/\sigma\right)_\mathrm{HL}$ for NGC~2808 and NGC~7078 for the overall cluster and each of their newly identified populations shown atop.}
    \label{fig:more_pops_ngc2808_ngc7078}
\end{figure*}

For NGC~362, NGC~3201, NGC~6218, NGC~6254, NGC~6397, NGC~6624, NGC~6656, NGC~6681, NGC~6752 and NGC~7099 we find that the distributions of $\left(v/\sigma\right)_\mathrm{HL}$ for P1 and P2 shown in Figure \ref{fig:vsigma_hist} are consistent with zero. In these cases, we are unable to detect rotation for either population (see Table \ref{tab:results} and the corresponding rotation profiles in the appendix). Based on our analysis, we find that our ability to detect rotation for any population depends mainly on two factors. First, the uncertainties of our analysis increase substantially for clusters with less than $\sim 200$ stars per population (e.g. NGC~3201 and NGC~6218), resulting in a very broad distribution of $\left(v/\sigma\right)_\mathrm{HL}$. Second, if a cluster is slowly rotating ($\left(v/\sigma\right)_\mathrm{HL} \lesssim 0.05$), it is challenging to constrain the rotation of its populations given our uncertainties. When both factors are present, like in the cases of NGC~6397 and NGC~6752, our analysis only provides broad upper limits on the rotation strength of each population.

For NGC~104, NGC~1851, NGC~2808, NGC~5286, NGC~5904, NGC~6093, NGC~6388, NGC~6541, NGC~7078 and NGC~7089 we are able to detect rotation for P1 or P2. All of these clusters fulfill the condition $\left(v/\sigma\right)_\mathrm{HL} \gtrsim 0.05$. NGC~6656 is the only other cluster in our sample that also fulfills this condition, but we are unable to detect rotation in P1 and P2 because its populations contain fewer than $200$ stars. This shows that the global rotation of the cluster strongly affects the rotation of the individual populations, as expected. Based on the distributions of $\left(v/\sigma\right)_\mathrm{HL}$ shown in Fig. \ref{fig:vsigma_hist}, we find kinematic differences between P1 and P2 that are significant above a $1\sigma$-level for NGC~2808, NGC~6093 and NGC~7078. For NGC~6093 and NGC 7078 we find that P2 rotates faster than P1 at a confidence level of $1.5\sigma$ and $2.2\sigma$ respectively, whereas P2 rotates slower than P1 in NGC~2808 at a confidence level of $1.8\sigma$. For NGC~104, NGC~1851, NGC~5286, NGC~5904, NGC~6388, NGC~6541 and NGC~7089 the strength of rotation of P1 is consistent with that of P2.

Furthermore, we investigated whether the strength of rotation of P1 and P2 or their difference can be related to the relaxation time of the corresponding cluster. In the top and middle panel of Fig. \ref{fig:relax_per_pop_with_diff} the values of $\left(v/\sigma\right)_\mathrm{HL}$ are plotted against the relaxation time of the corresponding cluster for P1 and P2. For both P1 and P2, the strength of rotation only depends weakly on the relaxation time, if there is any correlation at all. Moreover, we do not see a correlation between the difference of rotation strength between P1 and P2 with relaxation time, which is illustrated in the bottom panel of Fig. \ref{fig:relax_per_pop_with_diff}. However, the significance of these results should be taken with a grain of salt, since we only find differences above the $1\sigma$-level for three clusters.

In addition, we looked into possible connections of our kinematic differences with the radial concentration of P1 and P2. In fact, both NGC~6093 and NGC~7078 have been reported to contain a more centrally concentrated P1 compared to P2 according to \citet{dalessandro_peculiar_2018} and \citet{larsen_radial_2015}, respectively. For both clusters, we find that P1 rotates slower than P2. However, \citet{nardiello_hubble_2018} find no difference in concentration between P1 and P2 for NGC~7078. For NGC~104 \citep[e.g.]{milone_multiple_2018, cordoni_three-component_2020} and NGC~2808 \citep{dalessandro_family_2019} P2 was found to be more centrally concentrated than P1. Whereas we do not find significant kinematic differences between P1 and P2 for NGC~104, we find that P1 rotates faster than P2 for NGC~2808. Overall, this could hint at a connection between kinematic differences and the radial concentration of multiple populations in globular clusters, so that a population more centrally concentrated would rotate less. However, since we only find kinematic differences in three clusters and the information on the concentrations is only available for a small subset of our sample of clusters, additional data are needed to investigate this further. Furthermore, the observations for e.g. NGC~5272 \citep{lee_multiple_2021}, NGC~6205 \citep{johnson_oxygen_2012, cordero_differences_2017} and NGC~6362 \citep{dalessandro_family_2019, dalessandro_3d_2021} do not support this trend in our data that more centrally concentrated populations rotate less.

\subsection{Additional Population in Type-II Clusters}
\label{subsec:add_pops}
The distributions of $\left(v/\sigma\right)_\mathrm{HL}$ for type-II clusters are also shown in Figure \ref{fig:vsigma_hist}, where P1, P2 and P3 are shown in green, orange and purple respectively. For four of the six type-II clusters in our sample, we did not detect rotation in P3. For NGC~362, NGC~6656 and NGC~7089 this is most likely due to the low number of stars in P3 as discussed previously. For NGC~6388 P3 is populated well, but the global rotation of the cluster is very low. For NGC~1851 and NGC~5286 we detect rotation in P1, P2 and P3. For NGC~1851 we find that the distribution of $\left(v/\sigma\right)_\mathrm{HL}$ for P3 is very similar to that of P1 and P2, whereas for NGC 5286, there might be a hint of P3 rotating faster than P1 and P2. However, the observed difference in $\left(v/\sigma\right)_\mathrm{HL}$ is still within the $1\sigma$ uncertainty interval, so further data are needed to draw any solid conclusions. In particular, these results also do not give any clear hints on other formation scenarios for type-II clusters.

\begin{figure*}[t]
    \centering
    \includegraphics[width=0.66\textwidth]{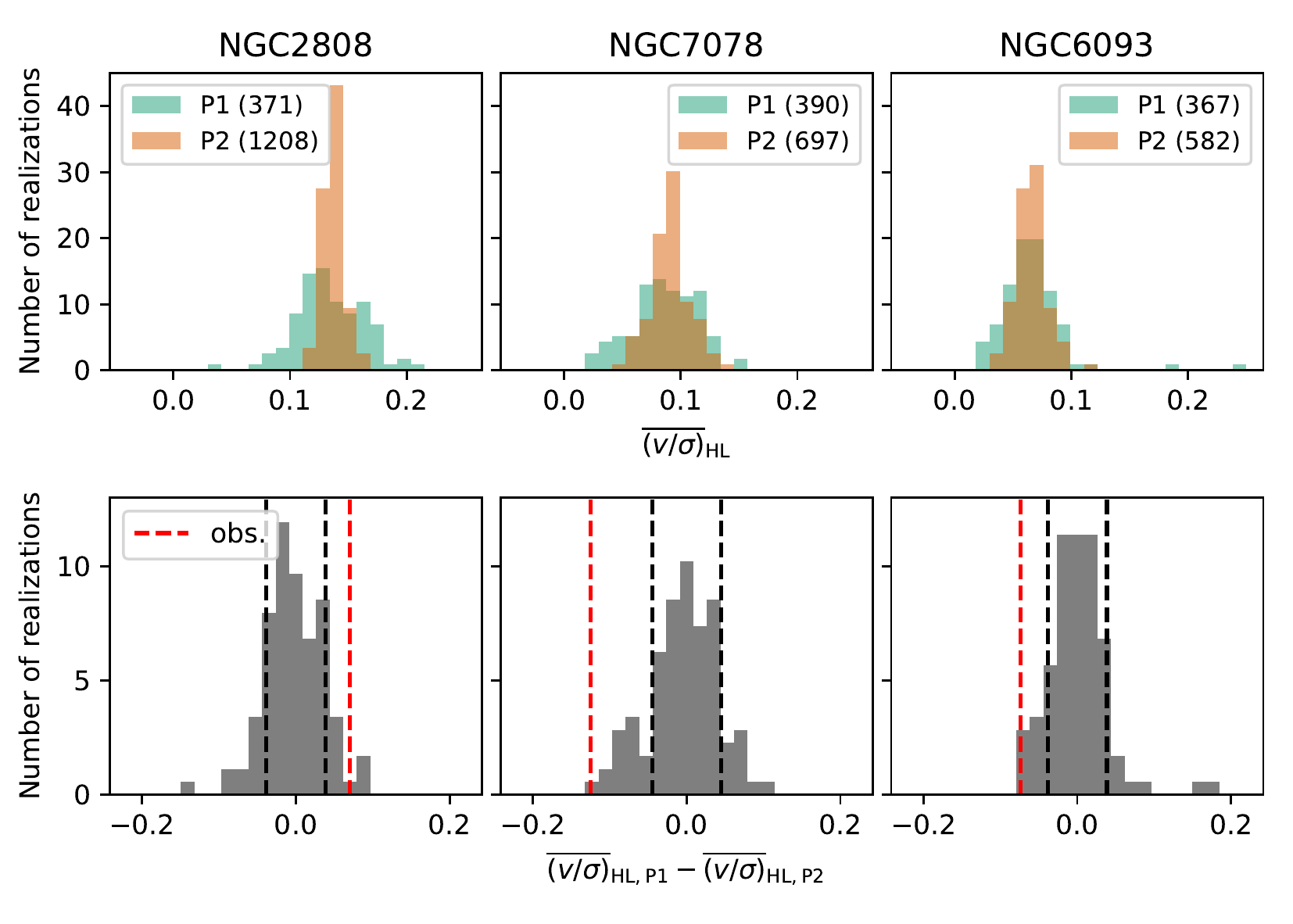}
    \caption{\textbf{Top:} Values of $\overline{\left(v/\sigma\right)}_\mathrm{HL}$ derived from randomly sampling the chromosome maps of NGC~2808, NGC~6093 and NGC~7078. The numbers of stars in the randomly drawn P1 and P2 are the same as the observed populations. \textbf{Bottom:} Differences of the randomly drawn values of $\overline{\left(v/\sigma\right)}_\mathrm{HL}$ between P1 and P2 for NGC~2808, NGC~6093 and NGC~7078. The observed differences for these clusters are shown as the red dotted line, whereas the black dotted line is the standard deviation of the shown distribution.}
    \label{fig:vsigma_random}
\end{figure*}

\subsection{Notes on Individual Clusters}
\label{subsec:individual_clusters}

\subsubsection{NGC~6093}
\label{subsubsec:ngc6093}
For NGC~6093 we observe that $r_\mathrm{peak}$ varies between P1 and P2 (see Table \ref{tab:results}). While the values of $r_\mathrm{peak}$ for the whole cluster and P1 are consistent, the distribution of samples for $r_\mathrm{peak}$ of P2 peaks at a much smaller value, which is also apparent in the radial rotation profile in Fig. \ref{fig:profiles_ngc6093} in the appendix. Furthermore, the distribution of $r_\mathrm{peak}$ for P2 is asymmetric and there is a strong anti-correlation between $r_\mathrm{peak}$ and $v_\mathrm{max}$. This causes the asymmetry of $\left(v/\sigma\right)_\mathrm{HL}$ in Fig. \ref{fig:vsigma_hist} for P2 of this cluster. As discussed in Sec. \ref{subsec:global_kinematics} the value of $\left(v/\sigma\right)_\mathrm{HL}$ is generally robust to changes in $r_\mathrm{peak}$. However, in this case the value of $r_\mathrm{peak}$ for P2 is very close to the center of that cluster, and it seems worthwhile to find out whether this has a significant effect on our results. If we apply a uniform prior on $r_\mathrm{peak}$ for P2, we find that the difference between P1 and P2 in $\left(v/\sigma\right)_\mathrm{HL}$ is even larger and the asymmetry in the distribution of $\left(v/\sigma\right)_\mathrm{HL}$ vanishes. If we fix the value of $r_\mathrm{peak}$ for P2 to that obtained for the whole cluster, the asymmetry also vanishes, but in this case $\left(v/\sigma\right)_\mathrm{HL}$ is consistent with that value for P1. This cluster was already analyzed for kinematic differences between populations based on MUSE data, with a very similar approach to the one presented here by \citet{kamann_peculiar_2020}. They did not find this peculiar behavior of P2 for that cluster, because they used a different population split than the one used here. Notably, they also used individual population density profiles when calculating $\left(v/\sigma\right)_\mathrm{HL}$ for their populations and not the global density profile of that cluster. \citet{kamann_peculiar_2020} split the cluster in three populations, where our P1 is consistent with their primordial population and our P2 includes their intermediate and extreme populations. If we average their values of $\left(v/\sigma\right)_\mathrm{HL}$ for the intermediate and extreme population, the result is consistent with the value of $\left(v/\sigma\right)_\mathrm{HL}$ they obtained for the primordial population. If we consider that \citet{kamann_peculiar_2020} fixed the radial scale of each population, our results are consistent with theirs in that we do not find significant kinematic differences between P1 and P2 if the radial scale is fixed to that of the global profile. However, there is no physical reason for why the radial scale of P2 cannot be different to that of P1.

\subsubsection{NGC~2808 \& NGC~7078}
\label{subsubsec:ngc2808_ngc7078}
For NGC~2808 and NGC~7078 we find differences above the $1\sigma$-level in $\left(v/\sigma\right)_\mathrm{HL}$ and the rotation profiles of their populations, whereas the dispersion profiles do not differ significantly. For NGC~7078 we observe that P2 rotates faster than P1. Qualitatively, this behavior is similar to that of NGC~6093 and NGC~6205, where differences of this type between similar populations have also been reported by \citet{kamann_peculiar_2020} and \citet{cordero_differences_2017} respectively. For NGC~2808 we find the opposite in that P1 rotates faster than P2. 

To investigate the relationship between the populations in the chromosome maps and their kinematic differences further for NGC~2808 and NGC~7078, we decided to reiterate the analysis with different population splits. Incidentally, the structure of the chromosome maps of these clusters allows us to distinguish between four populations (P1, P2', P3' and P4') for NGC~2808 \citep{milone_hubble_2015, latour_stellar_2019} and three populations (P1, P2' and P3') for NGC~7078 \citep{nardiello_hubble_2018-1}. Compared to \citet{milone_hubble_2015} our P2', P3' and P4' in NGC~2808 are equivalent to their populations C, D and E, while our P1' is their populations A and B combined. For NGC~7078, our P2' is equivalent to population B by \citet{nardiello_hubble_2018-1}, whereas our P1' corresponds to their populations A and D and our population P3' is equal to their populations C and E. We do not split P1' for both clusters and P3' for NGC~7078 any further to ensure that there are still enough stars per population to get meaningful results from our analysis. These populations according to our splitting are shown in the chromosome maps in the upper panels of Fig. \ref{fig:more_pops_ngc2808_ngc7078}, whereas the distributions of $\left(v/\sigma\right)_\mathrm{HL}$ for these populations are depicted in the lower panels of that figure. For NGC~2808 we find that P2', P3' and P4' do not differ in their rotation significantly, but the difference between these three populations and P1' is larger than $1\sigma$. For NGC~7078 the value of $\left(v/\sigma\right)_\mathrm{HL}$ of P3' stars is consistent with that of P1' and P2' stars, but the difference between P1' stars and P2' is larger than $1\sigma$. Therefore, we do not find a general trend of $\left(v/\sigma\right)_\mathrm{HL}$ along these populations.

NGC~7078 has been analyzed for kinematic differences in populations by \citet{szigeti_rotation_2021}. They used high precision radial velocity data from the SDSS-IV APOGEE-2 survey for 138 stars in NGC~7078 to measure its rotation amplitude as a function of position angle for the whole cluster and two populations that were identified based on single element abundance changes. They derived the angular rotation profile of NGC~7078 by splitting the cluster into two halves through the cluster center. This line of separation is rotated in small angular increments, and the difference in mean radial velocity $\Delta V$ between both halves is calculated at each step. If the cluster is rotating, the relation between velocity difference $\Delta V$ and position angle relative to the rotation axis $\alpha$ is $\Delta V = 2 v_\mathrm{rot} \cdot \sin\left(\alpha\right)$. The global rotation amplitude and rotation angle that they find agree with our values for $v_\mathrm{max}$ and $\theta_0$ for the whole cluster. Furthermore, they found no kinematic difference between their two populations, which is in contrast to our results. Their method may only be applied to a fully sampled area that is symmetric with respect to the position angle. To ensure that our data comply with those restrictions, we filter our stars to make the covered area of the cluster circular. If we apply their method to our filtered data for P1 and P2, we find rotation amplitudes and rotation angles that are consistent with our radial rotation curves for P1 and P2. The corresponding angular rotation profiles and the spatial coverage of the cluster are shown in Fig. \ref{fig:ngc7078_theta_binning_1}. In particular, we still find kinematic differences between P1 and P2. One striking difference between their data and ours is that we have 390 and 697 stars in P1 and P2, compared to their 33 and 49 stars in those populations. To investigate this further, we decreased the number of stars by randomly sampling from P1 and P2 and then applied their method again. Figure \ref{fig:ngc7078_theta_binning_3} shows three of these angular rotation profiles and the spatial coverage of the cluster. We find that with so few stars, this method results in a wide range of fundamentally different profiles, presumably because the method is very sensitive to single stars. Probably the uncertainties of the differential rotation profiles are underestimated because the strong correlations between different values in those profiles are generally not taken into account. Together with our much larger sample of analyzed stars, this could be the cause for the discrepancies between our results and those from \citet{szigeti_rotation_2021}.

NGC~2808 has not been analyzed for differences in its radial rotation and dispersion profiles yet, but \citet{bellini_hubble_2015} found differences in the radial anisotropy profiles for this cluster, by using proper motion data. Given that this cluster shows kinematic differences using both radial velocities and proper motion data, it would be very interesting to combine both data sets and analyze the 3D stellar velocities of this cluster for differences between multiple populations.

\subsubsection{NGC~104 \& NGC~5904}
\label{subsubsec:ngc104_ngc5904}

Neither \citet{milone_gaia_2018}, nor \citet{cordoni_three-component_2020}, find differences in the rotation amplitude between the populations of NGC~104, which agrees with our results. For NGC~5904 \citet{cordoni_three-component_2020} do not observe any differences in rotation amplitude, but they do find differences in the phase of their rotation curves. These phase differences translate to a differing angle of rotation. We do not find any significant differences in rotation amplitudes for NGC~5904 either, but we also do not find a significant difference in the angle of rotation between P1 and P2. 

\subsection{Random Sampling of the Chromosome Map}
\label{subsec:random_sampling}
To further analyze the solidity of our results for NGC~2808, NGC~6093 and NGC~7078 and to evaluate our uncertainty estimations, we randomly sampled the chromosome maps of these clusters to create random populations P1 and P2. To ensure comparability to the original population split, we used the same number of stars per population as in the original separations (see Table \ref{tab:clusters_num}). To achieve statistically relevant results, we created 100 random population pairs for each cluster. We calculated the rotation and dispersion profiles and derived $\left(v/\sigma\right)_\mathrm{HL}$ for each of these random populations, as described in Sec. \ref{subsec:kinematics}. As a result, we calculate 100 values of $\left(v/\sigma\right)_\mathrm{HL}$ for each population of each cluster, which are shown in the upper panels of Fig. \ref{fig:vsigma_random}. It is apparent that the distributions for P1 and P2 for each cluster have the same median value. This shows that, on average, we do not find kinematic differences when randomly sampling the chromosome map. We also find that the distributions of $\overline{\left(v/\sigma\right)}_\mathrm{HL}$ obtained with the random sampling are broader for smaller numbers of stars per population, as expected. In the lower panels of Fig. \ref{fig:vsigma_random} we show the differences of $\overline{\left(v/\sigma\right)}_\mathrm{HL}$ between P1 and P2 obtained from the random sampling. As expected, these distributions are centered around zero. We also indicated the observed differences for NGC~2808, NGC~6093 and NGC~7078 in red. The observed differences lie outside the $1\sigma$-regions of the distributions, which supports the solidity of our uncertainty estimations and indicates that the kinematic differences that we find for NGC~2808, NGC~6093 and NGC~7078 are truly connected to the populations defined in the chromosome maps.

\section{Conclusion \& Outlook}
\label{sec:conclusion}
We created and analyzed the rotation and dispersion profiles of 25 Galactic globular clusters in the search for kinematic differences between different populations within each cluster. Based on these kinematic profiles, we derived the rotation strength in terms of the ratio of ordered-over-random motion $\left(v/\sigma\right)_\mathrm{HL}$, evaluated at the half-light radius, for each cluster and its populations to quantify kinematic differences. For NGC~362, NGC~6397, NGC~6522 and NGC~6681 we find no significant global rotation when using all stars in these clusters. For NGC~104, NGC~1851, NGC~2808, NGC~5286, NGC5904, NGC~6093, NGC~6388, NGC~6541, NGC~7078 and NGC~7089 we are able to detect rotation in at least one of their populations. For three clusters we find differences above the $1\sigma$ level: For NGC~6093 and NGC~7078 we find that P2 stars rotate faster than P1 stars, whereas we find the opposite for NGC~2808, where P1 stars rotate faster than P2 stars. 

Our results do not give a clear hint on the formation scenario of multiple populations in globular clusters. We find support for both multi-epoch and single-epoch formation scenarios in our data. For multi-epoch formation scenarios, we expect to find that P2 rotates faster than P1, which matches our results for NGC~6093 and NGC~7078. Assuming a single-epoch formation scenario, it follows that P1 rotates faster than P2, which is what we find for NGC~2808. However, the kinematic differences that we find are still relatively uncertain at confidence levels of $\lesssim 2\sigma$, so further data are needed for a definitive answer. We find further support for both scenarios if we consider clusters with kinematical variations between P1 and P2 below the $1\sigma$-level. While we find that the rotation strength of each cluster is positively correlated with median relaxation time, the correlation between relaxation time and the rotation strength of P1 or P2 is weak at best, and we do not see a correlation of the kinematical difference between P1 and P2 with relaxation time. Based on our analysis, neither of the two types of formation scenarios for multiple populations in globular clusters is favored. \citet{bastian_multiple_2018} discussed that none of the formation scenarios put forward to date is able to explain all the observations. It is also possible that the formation of multiple populations does not affect the rotation of either population in a way that we are able to observe. However, it would be very interesting to see what primordial differences between P1 and P2 are consistent with the differences between those populations that we find.

To get a better understanding of the formation scenarios of multiple populations and their kinematics in general, there are several problems worth addressing. For many clusters in our sample, we were unable to detect rotation for P1 or P2. However, since most clusters are rotating overall, we suspect that at least one population should be rotating as well for most clusters. As mentioned in Sec. \ref{subsec:global_kinematics} we are generally unable to detect rotation in P1 or P2 if the overall cluster is already rotating slowly with $\left(v/\sigma\right)_\mathrm{HL} \lesssim 0.05$ or when the number of stars per population is of the order of 200 stars or below. Especially the results for NGC~6656 and NGC~3201 could be improved significantly by increasing the number of stars per population, since these clusters are rotating fast enough to overcome that limitation. One way to tackle this issue is to determine the population tags directly from the spectra. An alternative would be to add additional photometric data to be able to assign more stars to their respective population. This would be especially useful outside the core of each cluster, where radial velocity measurements of stars are available \citep[e.g.][]{baumgardt_catalogue_2018}, but these stars have not been separated into populations yet. Leitinger et al. (in preparation) are currently working on using ground-based photometry to split the populations and derive density profiles for each population in globular clusters. When we include their additional population tags for NGC~7078, we observed small changes in the distribution of $\left(v/\sigma\right)_\mathrm{HL}$, but nothing significant since the number of stars per population is already comparatively large for that cluster. Nonetheless, it seems worthwhile to pursue that approach, since it also increases the radial range of the population data. It is possible that this could increase the accuracy of measurements of the rotation in P1 and P2 substantially. Another possibility for a future work would be to use density profiles per population to derive the rotation strength. Using data from Leitinger et al. (in preparation) we checked whether our results for NGC~7078 change if we use their density profiles for P1 and P2, but we still find that P2 rotates faster than P1 with a difference larger than $1\sigma$. However, if the populations have the same rotation and dispersion curves, but different concentrations, then we would expect to find kinematic differences between them. This is because the observed kinematics at a given projected radius correspond to a different intrinsic radius relative to the cluster center. Ultimately, one needs more sophisticated models to understand all the details.

\begin{acknowledgements}
SM, FG, ML, PMW and SD acknowledge funding from the Deutsche Forschungsgemeinschaft (grant LA 4383/4-1, DR 281/35-1 and KA 4537/2-1) and by the BMBF from the ErUM program through grants 05A14MGA, 05A17MGA, 05A20MGA and 05A20BAB. SK and RP gratefully acknowledge funding from UKRI in the form of a Future Leaders Fellowship (grant no. MR/T022868/1).
\end{acknowledgements}
\bibliographystyle{aa}
\bibliography{database}
\begin{appendix}
\section{Additional Tables and Figures}
\begin{table}[h]
\renewcommand{\arraystretch}{1.5}
\caption{Priors of the parametric fit described in Sec. \ref{subsec:kinematics}.}             
\label{tab:prior_param}      
\centering                          
\begin{tabular}{c l}         
\hline\hline                 
\noalign{\smallskip}
Parameter & Prior \\
\noalign{\smallskip}
\hline                        
$v_\mathrm{sys, o}$ & $\mathcal{U}(\text{-10 km/s, 10 km/s})$ \\
$v_\mathrm{sys, p}$ & $\mathcal{N}(\mu(v_\mathrm{sys, o}), \sigma(v_\mathrm{sys, o}))$\\
$v_\mathrm{max, o}$ & $\mathcal{U}(0, v_\text{esc})$\\
$v_\mathrm{max, p}$ & $ \propto
            \begin{cases}
            \mathcal{N}(\mu(v_\mathrm{max, o}), \mu(\sigma_\mathrm{max, o})) & \text{if $0 < v_\text{max, p} < v_\text{esc}$}\\
            0 & \text{else}\\
        \end{cases}
    $\\
$\theta_0$ & $\mathcal{U}(-\pi, \pi)$\\
$r_\mathrm{peak, o}$ & $ \propto
            \begin{cases}
            0 & \text{if $r_\mathrm{peak, o}<r_\mathrm{HL} / 30$}\\
            1 & \text{if $r_\mathrm{HL} / 30<r_\mathrm{peak, o}<5\, r_\mathrm{HL}$}\\
            \mathcal{N}(5\, r_\mathrm{HL}, r_\mathrm{HL}) & \text{else},
        \end{cases}
    $\\
$r_\mathrm{peak, p}$ & $ \propto
            \begin{cases}
            \mathcal{N}(\mu(r_\mathrm{peak, o}), \sigma(r_\mathrm{peak, o})) & \text{if $r_\mathrm{peak, p}> 0$}\\
            0 & \text{else}\\
        \end{cases}
    $\\
$\sigma_\mathrm{max}$ & $\mathcal{U}(0, \infty)$ \\
$a_\text{0, o}$ & same prior as $r_\mathrm{peak, o}$\\
$a_\text{0, p}$ & $ \propto
            \begin{cases}
            \mathcal{N}(\mu(a_\mathrm{0, o}), \sigma(a_\mathrm{0, o})) & \text{if $a_\mathrm{0, p}> 0$}\\
            0 & \text{else}\\
        \end{cases}
    $\\
$f_\mathrm{fg}$ & $\mathcal{U}(0, 1)$ \\
\hline                        
\end{tabular}
\tablefoot{The subscript 'o' denotes priors for fit of the overall population, whereas the subscript 'p' describes priors for each population fit.}
\end{table}

\begin{table}[h]
\renewcommand{\arraystretch}{1.5}
\caption{Priors of the non-parametric fit described in Sec. \ref{subsec:kinematics}.}             
\label{tab:prior_nonparam}      
\centering                          
\begin{tabular}{c l}         
\hline\hline                 
\noalign{\smallskip}
Parameter & Prior \\
\noalign{\smallskip}
\hline
$v_\text{sys}$ &  $\mathcal{N}(\mu(v_\mathrm{sys}^*), \sigma(v_\mathrm{sys}^*))$\\
$v_\text{max}$ &  $\mathcal{U}(0, \infty)$\\
$\theta_\text{0}$ &  $\mathcal{N}(\mu(\theta_\mathrm{0}^*), \sigma(\theta_\mathrm{0}^*))$\\
$\sigma_\text{max}$ & $\mathcal{U}(0, \infty)$\\
\hline                        
\end{tabular}
\tablefoot{The superscript '*' denotes that these parameters are distributions of samples from the parametric fit.}
\end{table}

\longtab[3]{
\renewcommand{\arraystretch}{1.425}
\begin{longtable}{l l l l l l l l l l}
\label{tab:results}\\
\caption[width=2\textwidth]{Median and upper limits of the parameter distributions for the parametric model of each cluster.}\\
\hline\hline \noalign{\smallskip}
Cluster & Population & $\sigma_\mathrm{max}$ [km/s]& $v_\mathrm{max}$ [km/s]& $\theta_0$ [rad]&$r_\mathrm{peak}$ ["]& $a$ ["]& $\left(v/\sigma\right)_\mathrm{HL}$ &$\lambda_\mathrm{R, HL}$ & $\log_{10}(T_\mathrm{rh}/\mathrm{yr})$ \\ \noalign{\smallskip} \hline
\endfirsthead

\caption{continued.}\\
\hline\hline \noalign{\smallskip}
Cluster & Population & $\sigma_\mathrm{max}$ [km/s]& $v_\mathrm{max}$ [km/s]& $\theta_0$ [rad]&$r_\mathrm{peak}$ ["]& $a$ ["]& $\left(v/\sigma\right)_\mathrm{HL}$ &$\lambda_\mathrm{R, HL}$ & $\log_{10}(T_\mathrm{rh}/\mathrm{yr})$ \\ \noalign{\smallskip} \hline
\endhead
\endfoot
\endlastfoot
NGC 104 & Overall & $12.37_{-0.09}^{+0.09}$ & $4.75_{-0.14}^{+0.14}$ & $-2.34_{-0.03}^{+0.03}$ & $165_{-10}^{+10}$ & $142_{-5}^{+6}$ & $0.154_{-0.005}^{+0.004}$ & $0.117_{-0.004}^{+0.004}$ & $9.55$\\
& P1 & $12.4_{-0.5}^{+0.5}$ & $3.9_{-1.7}^{+1.7}$ & $-2.5_{-0.6}^{+0.5}$ & $165_{-10}^{+10}$ & $141_{-6}^{+5}$ & $0.13_{-0.06}^{+0.05}$ & $0.10_{-0.04}^{+0.04}$ & --\\
& P2 & $12.33_{-0.25}^{+0.26}$ & $5.5_{-0.9}^{+0.9}$ & $-2.45_{-0.19}^{+0.18}$ & $166_{-10}^{+10}$ & $141_{-6}^{+6}$ & $0.180_{-0.028}^{+0.029}$ & $0.136_{-0.020}^{+0.021}$ & --\\
\hline
NGC 362 & Overall & $8.67_{-0.23}^{+0.24}$ & $ < 1.0$ & -- & $140_{-80}^{+110}$ & $47_{-6}^{+6}$ & $ < 0.03$ & $ < 0.028$ & $8.93$\\
& P1 & $8.1_{-0.4}^{+0.5}$ & $ < 5$ & -- & $190_{-100}^{+90}$ & $46_{-6}^{+6}$ & $ < 0.09$ & $ < 0.07$ & --\\
& P2 & $8.3_{-0.3}^{+0.3}$ & $ < 4$ & -- & $190_{-100}^{+100}$ & $45_{-6}^{+6}$ & $ < 0.07$ & $ < 0.06$ & --\\
& P3 & $7.5_{-1.2}^{+1.5}$ & $ < 15$ & -- & $160_{-80}^{+90}$ & $47_{-6}^{+6}$ & $ < 0.4$ & $ < 0.3$ & --\\
\hline
NGC 1851 & Overall & $10.34_{-0.25}^{+0.29}$ & $1.42_{-0.16}^{+0.17}$ & $1.45_{-0.09}^{+0.09}$ & $26_{-6}^{+8}$ & $27.3_{-2.2}^{+2.4}$ & $0.062_{-0.013}^{+0.014}$ & $0.050_{-0.011}^{+0.012}$ & $8.82$\\
& P1 & $10.1_{-0.6}^{+0.6}$ & $2.4_{-1.1}^{+1.0}$ & $1.6_{-0.4}^{+0.4}$ & $28_{-8}^{+7}$ & $27.8_{-2.3}^{+2.4}$ & $0.10_{-0.05}^{+0.05}$ & $0.08_{-0.04}^{+0.04}$ & --\\
& P2 & $8.7_{-0.4}^{+0.4}$ & $2.3_{-0.6}^{+0.8}$ & $1.8_{-0.3}^{+0.3}$ & $29_{-9}^{+8}$ & $28.3_{-2.2}^{+2.4}$ & $0.11_{-0.03}^{+0.04}$ & $0.086_{-0.027}^{+0.03}$ & --\\
& P3 & $10.0_{-0.5}^{+0.5}$ & $2.1_{-0.8}^{+0.9}$ & $1.8_{-0.4}^{+0.4}$ & $27_{-8}^{+8}$ & $26.4_{-2.4}^{+2.5}$ & $0.09_{-0.04}^{+0.04}$ & $0.07_{-0.03}^{+0.04}$ & --\\
\hline
NGC 1904 & Overall & $7.02_{-0.28}^{+0.3}$ & $1.50_{-0.20}^{+0.21}$ & $0.26_{-0.13}^{+0.14}$ & $65_{-16}^{+18}$ & $30_{-4}^{+5}$ & $0.079_{-0.016}^{+0.018}$ & $0.064_{-0.014}^{+0.016}$ & $8.95$\\
\hline
NGC 2808 & Overall & $13.44_{-0.23}^{+0.23}$ & $4.02_{-0.22}^{+0.23}$ & $0.72_{-0.05}^{+0.05}$ & $69_{-8}^{+9}$ & $50_{-4}^{+4}$ & $0.123_{-0.009}^{+0.009}$ & $0.100_{-0.007}^{+0.008}$ & $9.15$\\
& P1 & $13.1_{-0.5}^{+0.6}$ & $5.9_{-1.0}^{+1.1}$ & $0.88_{-0.18}^{+0.18}$ & $68_{-9}^{+9}$ & $50_{-4}^{+4}$ & $0.19_{-0.03}^{+0.03}$ & $0.151_{-0.026}^{+0.027}$ & --\\
& P2 & $13.8_{-0.3}^{+0.4}$ & $3.9_{-0.6}^{+0.7}$ & $0.78_{-0.15}^{+0.15}$ & $69_{-9}^{+9}$ & $49_{-4}^{+4}$ & $0.118_{-0.019}^{+0.020}$ & $0.095_{-0.015}^{+0.016}$ & --\\
\hline
NGC 3201 & Overall & $4.68_{-0.07}^{+0.07}$ & $1.4_{-0.5}^{+0.5}$ & $1.8_{-0.4}^{+0.4}$ & $950_{-210}^{+190}$ & $970_{-100}^{+90}$ & $0.042_{-0.014}^{+0.015}$ & $0.033_{-0.011}^{+0.012}$ & $9.27$\\
& P1 & $4.7_{-0.5}^{+0.6}$ & $ < 10$ & -- & $960_{-220}^{+200}$ & $960_{-100}^{+110}$ & $ < 0.3$ & $ < 0.24$ & --\\
& P2 & $4.4_{-0.4}^{+0.5}$ & $ < 9$ & -- & $950_{-200}^{+210}$ & $970_{-100}^{+110}$ & $ < 0.3$ & $ < 0.23$ & --\\
\hline
NGC 5286 & Overall & $10.01_{-0.22}^{+0.23}$ & $4.4_{-1.0}^{+1.1}$ & $0.21_{-0.10}^{+0.10}$ & $110_{-30}^{+40}$ & $42_{-6}^{+8}$ & $0.133_{-0.015}^{+0.015}$ & $0.107_{-0.013}^{+0.013}$ & $9.11$\\
& P1 & $10.5_{-0.5}^{+0.6}$ & $4.9_{-2.3}^{+2.8}$ & $0.2_{-0.5}^{+0.5}$ & $110_{-40}^{+30}$ & $43_{-7}^{+8}$ & $0.14_{-0.06}^{+0.06}$ & $0.11_{-0.05}^{+0.05}$ & --\\
& P2 & $10.4_{-0.5}^{+0.5}$ & $2.7_{-1.7}^{+2.3}$ & $0.2_{-0.7}^{+0.8}$ & $120_{-40}^{+40}$ & $40_{-8}^{+8}$ & $0.08_{-0.05}^{+0.05}$ & $0.06_{-0.04}^{+0.04}$ & --\\
& P3 & $10.0_{-0.7}^{+0.8}$ & $7_{-3}^{+4}$ & $0.0_{-0.5}^{+0.5}$ & $110_{-30}^{+40}$ & $40_{-8}^{+8}$ & $0.19_{-0.09}^{+0.08}$ & $0.15_{-0.07}^{+0.06}$ & --\\
\hline
NGC 5904 & Overall & $7.80_{-0.10}^{+0.08}$ & $3.11_{-0.17}^{+0.19}$ & $0.82_{-0.06}^{+0.06}$ & $126_{-14}^{+16}$ & $176_{-21}^{+30}$ & $0.169_{-0.008}^{+0.008}$ & $0.135_{-0.007}^{+0.007}$ & $9.41$\\
& P1 & $7.2_{-0.4}^{+0.4}$ & $3.1_{-1.4}^{+1.3}$ & $0.3_{-0.4}^{+0.5}$ & $127_{-16}^{+16}$ & $180_{-30}^{+30}$ & $0.18_{-0.08}^{+0.08}$ & $0.15_{-0.07}^{+0.06}$ & --\\
& P2 & $7.05_{-0.21}^{+0.23}$ & $3.5_{-0.8}^{+0.8}$ & $0.45_{-0.22}^{+0.22}$ & $127_{-16}^{+16}$ & $170_{-30}^{+30}$ & $0.21_{-0.04}^{+0.04}$ & $0.17_{-0.03}^{+0.03}$ & --\\
\hline
NGC 6093 & Overall & $11.6_{-0.4}^{+0.4}$ & $2.3_{-0.3}^{+0.6}$ & $-0.47_{-0.09}^{+0.09}$ & $78_{-21}^{+30}$ & $21.8_{-2.3}^{+2.4}$ & $0.059_{-0.008}^{+0.010}$ & $0.046_{-0.006}^{+0.008}$ & $8.8$\\
& P1 & $12.3_{-0.6}^{+0.7}$ & $1.8_{-1.1}^{+1.3}$ & $-0.5_{-0.7}^{+0.7}$ & $90_{-30}^{+30}$ & $21.8_{-2.3}^{+2.4}$ & $0.039_{-0.022}^{+0.023}$ & $0.031_{-0.017}^{+0.018}$ & --\\
& P2 & $11.9_{-0.6}^{+0.6}$ & $3.0_{-0.7}^{+0.8}$ & $-0.55_{-0.22}^{+0.22}$ & $35_{-18}^{+50}$ & $19.8_{-2.4}^{+2.3}$ & $0.11_{-0.04}^{+0.07}$ & $0.09_{-0.04}^{+0.06}$ & --\\
\hline
NGC 6218 & Overall & $6.27_{-0.13}^{+0.14}$ & $0.45_{-0.25}^{+0.3}$ & $2.0_{-0.6}^{+0.6}$ & $350_{-220}^{+230}$ & $58_{-6}^{+7}$ & $0.022_{-0.013}^{+0.015}$ & $0.019_{-0.011}^{+0.013}$ & $8.87$\\
& P1 & $5.9_{-0.4}^{+0.5}$ & $ < 9$ & -- & $440_{-200}^{+210}$ & $59_{-6}^{+7}$ & $ < 0.29$ & $ < 0.23$ & --\\
& P2 & $5.2_{-0.3}^{+0.4}$ & $ < 6$ & -- & $460_{-190}^{+200}$ & $58_{-6}^{+6}$ & $ < 0.25$ & $ < 0.20$ & --\\
\hline
NGC 6254 & Overall & $6.34_{-0.08}^{+0.09}$ & $0.8_{-0.3}^{+0.4}$ & $2.37_{-0.28}^{+0.3}$ & $410_{-170}^{+200}$ & $113_{-11}^{+13}$ & $0.028_{-0.009}^{+0.009}$ & $0.022_{-0.007}^{+0.007}$ & $8.9$\\
& P1 & $6.5_{-0.4}^{+0.4}$ & $ < 8$ & -- & $470_{-170}^{+180}$ & $114_{-13}^{+13}$ & $ < 0.20$ & $ < 0.16$ & --\\
& P2 & $5.83_{-0.26}^{+0.30}$ & $ < 7$ & -- & $470_{-170}^{+180}$ & $114_{-12}^{+12}$ & $ < 0.20$ & $ < 0.15$ & --\\
\hline
NGC 6266 & Overall & $15.34_{-0.27}^{+0.30}$ & $3.8_{-1.5}^{+1.6}$ & $-2.60_{-0.12}^{+0.11}$ & $200_{-90}^{+90}$ & $39_{-3}^{+3}$ & $0.048_{-0.005}^{+0.006}$ & $0.037_{-0.004}^{+0.005}$ & $8.98$\\
\hline
NGC 6293 & Overall & $6.8_{-0.5}^{+0.5}$ & $3.5_{-2.1}^{+3}$ & $0.43_{-0.29}^{+0.29}$ & $170_{-150}^{+120}$ & $18_{-5}^{+9}$ & $0.11_{-0.03}^{+0.04}$ & $0.087_{-0.026}^{+0.03}$ & $8.94$\\
\hline
NGC 6388 & Overall & $17.8_{-0.3}^{+0.4}$ & $1.06_{-0.26}^{+0.29}$ & $0.84_{-0.24}^{+0.26}$ & $44_{-22}^{+50}$ & $33.7_{-2.3}^{+2.6}$ & $0.020_{-0.008}^{+0.014}$ & $0.016_{-0.007}^{+0.012}$ & $8.9$\\
& P1 & $17.4_{-0.6}^{+0.6}$ & $ < 20$ & -- & $40_{-40}^{+60}$ & $34.6_{-2.4}^{+2.6}$ & $ < 0.17$ & $ < 0.12$ & --\\
& P2 & $18.6_{-0.5}^{+0.5}$ & $1.6_{-0.9}^{+1.5}$ & $0.2_{-0.5}^{+0.6}$ & $37_{-29}^{+50}$ & $33.5_{-2.4}^{+2.6}$ & $0.029_{-0.017}^{+0.05}$ & $0.023_{-0.014}^{+0.05}$ & --\\
& P3 & $18.9_{-0.7}^{+0.9}$ & $ < 11$ & -- & $60_{-50}^{+50}$ & $33.4_{-2.6}^{+2.6}$ & $ < 0.15$ & $ < 0.12$ & --\\
\hline
NGC 6397 & Overall & $6.30_{-0.12}^{+0.12}$ & $ < 0.7$ & -- & $600_{-400}^{+300}$ & $83_{-7}^{+7}$ & $ < 0.024$ & $ < 0.019$ & $8.6$\\
& P1 & $6.2_{-1.0}^{+1.2}$ & $ < 12$ & -- & $700_{-300}^{+400}$ & $84_{-7}^{+7}$ & $ < 0.4$ & $ < 0.27$ & --\\
& P2 & $5.6_{-0.5}^{+0.6}$ & $ < 11$ & -- & $700_{-300}^{+300}$ & $83_{-7}^{+7}$ & $ < 0.3$ & $ < 0.23$ & --\\
\hline
NGC 6441 & Overall & $18.9_{-0.4}^{+0.4}$ & $1.3_{-0.8}^{+1.0}$ & $3.5_{-0.5}^{+0.6}$ & $150_{-70}^{+50}$ & $31.4_{-3.0}^{+3}$ & $0.010_{-0.006}^{+0.006}$ & $0.008_{-0.004}^{+0.005}$ & $9.09$\\
& P1 & $19.4_{-0.7}^{+0.8}$ & $ < 9$ & -- & $180_{-70}^{+60}$ & $32_{-3}^{+3}$ & $ < 0.05$ & $ < 0.04$ & --\\
& P2 & $18.3_{-0.6}^{+0.6}$ & $ < 5$ & -- & $170_{-70}^{+70}$ & $32_{-3}^{+3}$ & $ < 0.029$ & $ < 0.022$ & --\\
\hline
NGC 6522 & Overall & $13.7_{-1.5}^{+2.4}$ & $ < 7$ & -- & $260_{-130}^{+90}$ & $11_{-4}^{+4}$ & $ < 0.029$ & $ < 0.021$ & $8.86$\\
\hline
NGC 6541 & Overall & $9.8_{-0.3}^{+0.3}$ & $3.4_{-0.5}^{+0.6}$ & $-0.19_{-0.07}^{+0.07}$ & $126_{-27}^{+30}$ & $28_{-3}^{+3}$ & $0.096_{-0.008}^{+0.008}$ & $0.073_{-0.006}^{+0.007}$ & $9.03$\\
& P1 & $8.8_{-0.4}^{+0.4}$ & $2.9_{-1.2}^{+1.3}$ & $-0.4_{-0.4}^{+0.4}$ & $135_{-29}^{+30}$ & $29_{-3}^{+3}$ & $0.09_{-0.03}^{+0.03}$ & $0.067_{-0.026}^{+0.026}$ & --\\
& P2 & $9.3_{-0.4}^{+0.5}$ & $4.0_{-1.3}^{+1.5}$ & $-0.07_{-0.28}^{+0.27}$ & $130_{-30}^{+30}$ & $29_{-3}^{+3}$ & $0.12_{-0.03}^{+0.03}$ & $0.089_{-0.026}^{+0.026}$ & --\\
\hline
NGC 6624 & Overall & $8.48_{-0.24}^{+0.24}$ & $1.2_{-0.7}^{+0.9}$ & $0.4_{-0.5}^{+0.5}$ & $210_{-100}^{+70}$ & $26_{-3}^{+4}$ & $0.021_{-0.011}^{+0.010}$ & $0.015_{-0.008}^{+0.008}$ & $8.71$\\
& P1 & $8.1_{-0.6}^{+0.6}$ & $ < 9$ & -- & $230_{-80}^{+90}$ & $25_{-4}^{+4}$ & $ < 0.12$ & $ < 0.09$ & --\\
& P2 & $7.7_{-0.4}^{+0.4}$ & $ < 7$ & -- & $240_{-90}^{+90}$ & $24_{-4}^{+4}$ & $ < 0.10$ & $ < 0.07$ & --\\
\hline
NGC 6656 & Overall & $9.44_{-0.10}^{+0.10}$ & $3.0_{-0.4}^{+0.4}$ & $0.25_{-0.10}^{+0.10}$ & $230_{-40}^{+50}$ & $150_{-13}^{+14}$ & $0.162_{-0.017}^{+0.017}$ & $0.134_{-0.014}^{+0.014}$ & $9.23$\\
& P1 & $8.6_{-0.5}^{+0.6}$ & $ < 8$ & -- & $240_{-40}^{+40}$ & $150_{-14}^{+15}$ & $ < 0.5$ & $ < 0.3$ & --\\
& P2 & $9.0_{-0.5}^{+0.5}$ & $ < 7$ & -- & $240_{-40}^{+50}$ & $150_{-15}^{+14}$ & $ < 0.4$ & $ < 0.28$ & --\\
& P3 & $8.6_{-0.5}^{+0.6}$ & $ < 7$ & -- & $240_{-50}^{+40}$ & $151_{-15}^{+14}$ & $ < 0.4$ & $ < 0.3$ & --\\
\hline
NGC 6681 & Overall & $7.9_{-0.4}^{+0.5}$ & $ < 2.2$ & -- & $170_{-100}^{+70}$ & $18_{-3}^{+4}$ & $ < 0.04$ & $ < 0.028$ & $8.65$\\
& P1 & $8.2_{-1.0}^{+1.3}$ & $ < 12$ & -- & $200_{-90}^{+90}$ & $16_{-4}^{+4}$ & $ < 0.20$ & $ < 0.14$ & --\\
& P2 & $6.9_{-0.5}^{+0.6}$ & $ < 7$ & -- & $210_{-90}^{+90}$ & $18_{-4}^{+4}$ & $ < 0.10$ & $ < 0.07$ & --\\
\hline
NGC 6752 & Overall & $8.39_{-0.11}^{+0.11}$ & $0.61_{-0.19}^{+0.19}$ & $-1.89_{-0.3}^{+0.30}$ & $450_{-150}^{+170}$ & $92_{-5}^{+6}$ & $0.009_{-0.003}^{+0.004}$ & $0.0067_{-0.0022}^{+0.0030}$ & $8.87$\\
& P1 & $7.9_{-0.5}^{+0.6}$ & $ < 12$ & -- & $480_{-170}^{+160}$ & $91_{-6}^{+6}$ & $ < 0.18$ & $ < 0.13$ & --\\
& P2 & $7.8_{-0.4}^{+0.4}$ & $ < 9$ & -- & $490_{-150}^{+160}$ & $92_{-6}^{+5}$ & $ < 0.12$ & $ < 0.09$ & --\\
\hline
NGC 7078 & Overall & $12.24_{-0.30}^{+0.3}$ & $3.21_{-0.26}^{+0.27}$ & $-2.26_{-0.06}^{+0.06}$ & $77_{-10}^{+11}$ & $26.6_{-2.2}^{+2.3}$ & $0.103_{-0.008}^{+0.008}$ & $0.080_{-0.006}^{+0.007}$ & $9.32$\\
& P1 & $11.6_{-0.5}^{+0.5}$ & $ < 2.2$ & -- & $78_{-11}^{+11}$ & $26.4_{-2.3}^{+2.1}$ & $ < 0.07$ & $ < 0.06$ & --\\
& P2 & $12.8_{-0.4}^{+0.5}$ & $4.1_{-0.8}^{+0.9}$ & $-2.41_{-0.19}^{+0.19}$ & $78_{-11}^{+11}$ & $26.2_{-2.1}^{+2.0}$ & $0.125_{-0.024}^{+0.024}$ & $0.097_{-0.018}^{+0.018}$ & --\\
\hline
NGC 7089 & Overall & $10.91_{-0.22}^{+0.24}$ & $3.90_{-0.27}^{+0.3}$ & $2.20_{-0.05}^{+0.05}$ & $66_{-12}^{+13}$ & $47_{-4}^{+5}$ & $0.173_{-0.011}^{+0.012}$ & $0.140_{-0.010}^{+0.011}$ & $9.4$\\
& P1 & $11.0_{-0.5}^{+0.6}$ & $5.1_{-1.2}^{+1.1}$ & $2.11_{-0.21}^{+0.22}$ & $66_{-12}^{+13}$ & $46_{-5}^{+5}$ & $0.22_{-0.05}^{+0.05}$ & $0.18_{-0.04}^{+0.04}$ & --\\
& P2 & $11.3_{-0.3}^{+0.3}$ & $4.1_{-0.6}^{+0.6}$ & $2.26_{-0.14}^{+0.13}$ & $58_{-12}^{+13}$ & $45_{-4}^{+5}$ & $0.186_{-0.026}^{+0.026}$ & $0.152_{-0.021}^{+0.022}$ & --\\
& P3 & $13.5_{-1.7}^{+2.1}$ & $ < 11$ & -- & $66_{-13}^{+12}$ & $47_{-5}^{+5}$ & $ < 0.4$ & $ < 0.3$ & --\\
\hline
NGC 7099 & Overall & $5.86_{-0.15}^{+0.16}$ & $0.53_{-0.18}^{+0.17}$ & $-0.8_{-0.4}^{+0.3}$ & $70_{-30}^{+80}$ & $53_{-7}^{+8}$ & $0.028_{-0.014}^{+0.015}$ & $0.021_{-0.011}^{+0.012}$ & $8.88$\\
& P1 & $5.3_{-0.4}^{+0.5}$ & $ < 4$ & -- & $120_{-80}^{+80}$ & $53_{-8}^{+8}$ & $ < 0.16$ & $ < 0.12$ & --\\
& P2 & $5.9_{-0.3}^{+0.4}$ & $ < 4$ & -- & $130_{-80}^{+80}$ & $50_{-8}^{+8}$ & $ < 0.11$ & $ < 0.08$ & --\\
\hline
\hline
\end{longtable}
\centering
\tablefoot{The two measures of the rotation strength $\left(v/\sigma\right)_\mathrm{HL}$ and $\lambda_\mathrm{R, HL}$ are derived from the parameter distribution of the parametric model, as described in Sec. \ref{subsec:kinematics}. The median relaxation times $T_\mathrm{rh}$ for each cluster are from \citet[][2010 edition]{harris_catalog_1996}.}
}

\begin{figure*}[b]
    \centering
    \includegraphics[width=\textwidth]{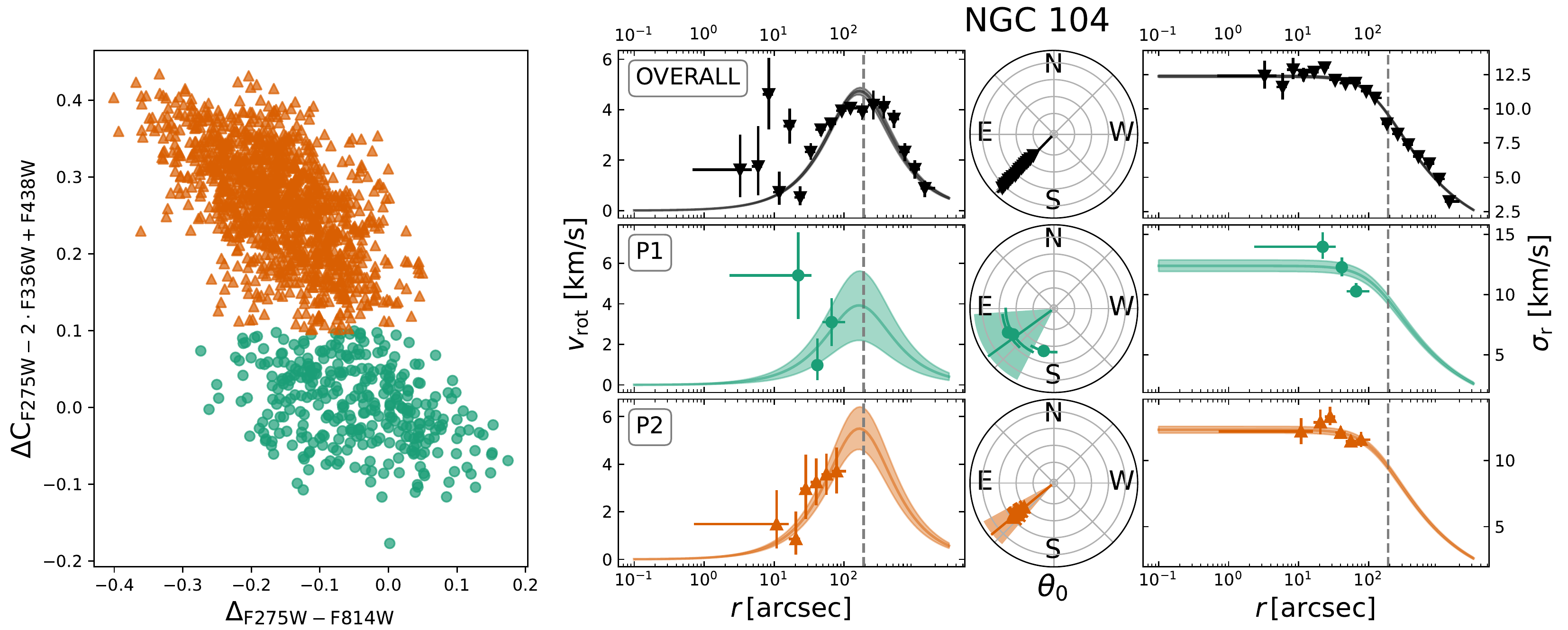}
    \caption{Chromosome maps, rotation and dispersion profiles for NGC~104 and each of its populations. The rotation profiles for each population are shown on the left, whereas the dispersion profiles are shown on the right. In the center, the angle of rotation is shown.}
    \label{fig:profiles_ngc104}
\end{figure*}
\begin{figure*}[b]
    \centering
    \includegraphics[width=\textwidth]{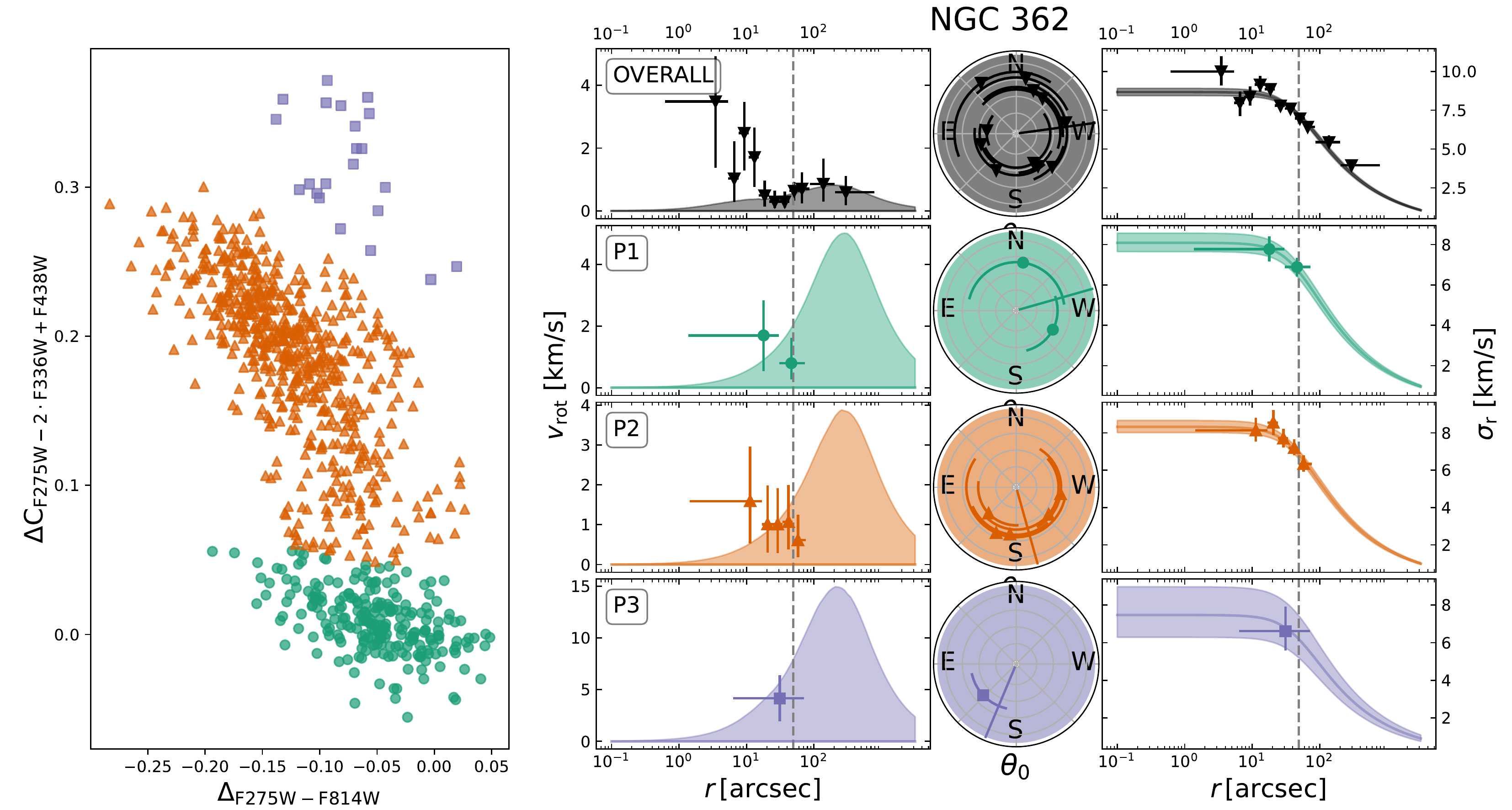}
    \caption{Continuation of Fig. \ref{fig:profiles_ngc104} for NGC~362.}
    \label{fig:profiles_ngc362}
\end{figure*}
\begin{figure*}[b]
    \centering
    \includegraphics[width=\textwidth]{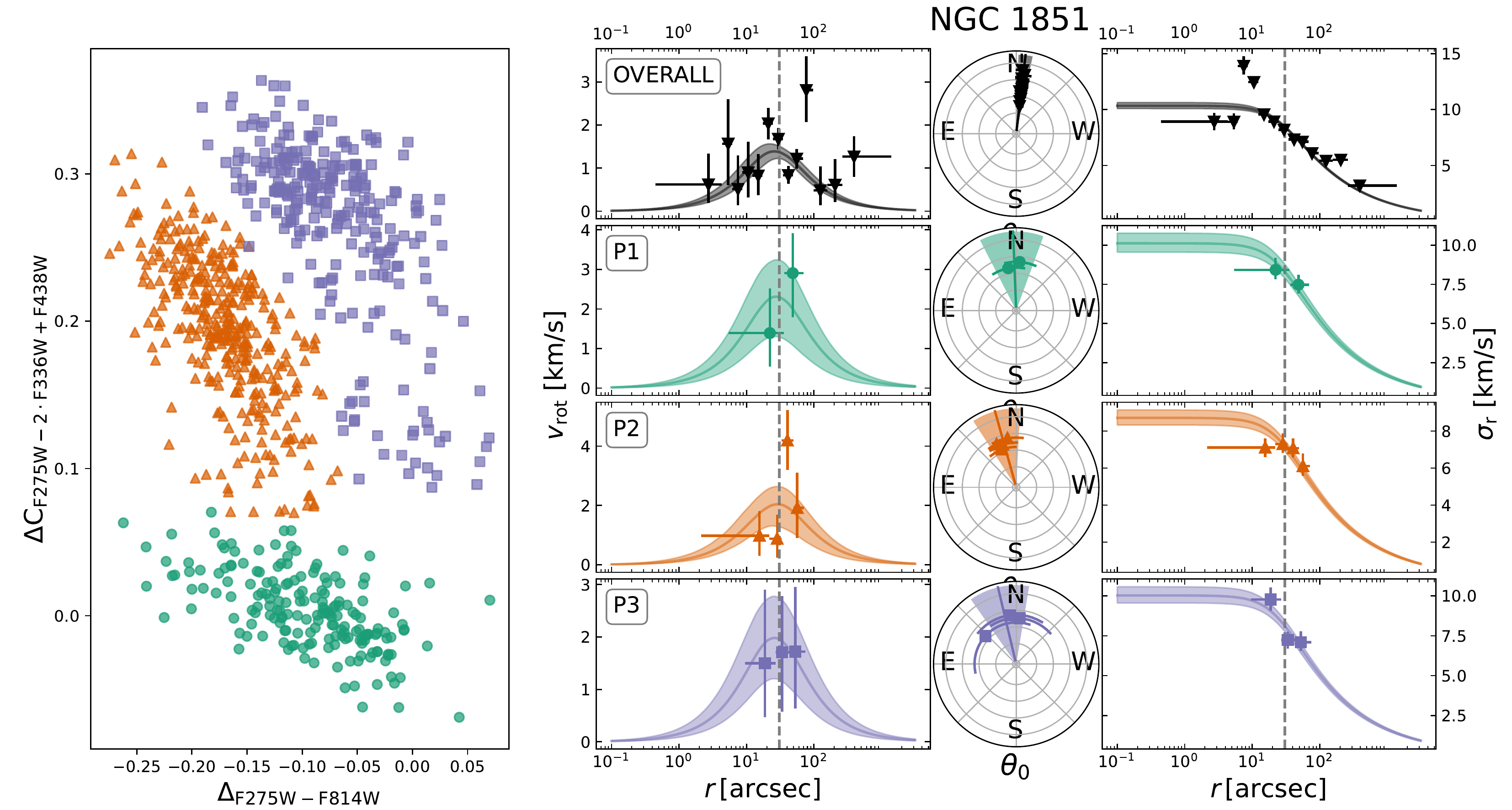}
    \caption{Continuation of Fig. \ref{fig:profiles_ngc104} for NGC~1851.}
    \label{fig:profiles_ngc1851}
\end{figure*}
\begin{figure*}[b]
    \centering
    \includegraphics[width=\textwidth]{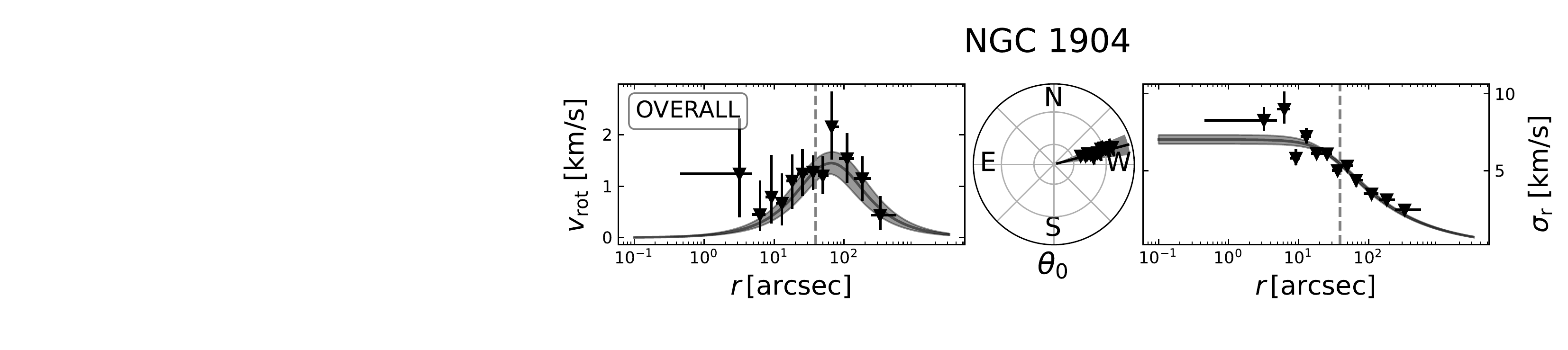}
    \caption{Continuation of Fig. \ref{fig:profiles_ngc104} for NGC~1904.}
    \label{fig:profiles_ngc1904}
\end{figure*}
\begin{figure*}[b]
    \centering
    \includegraphics[width=\textwidth]{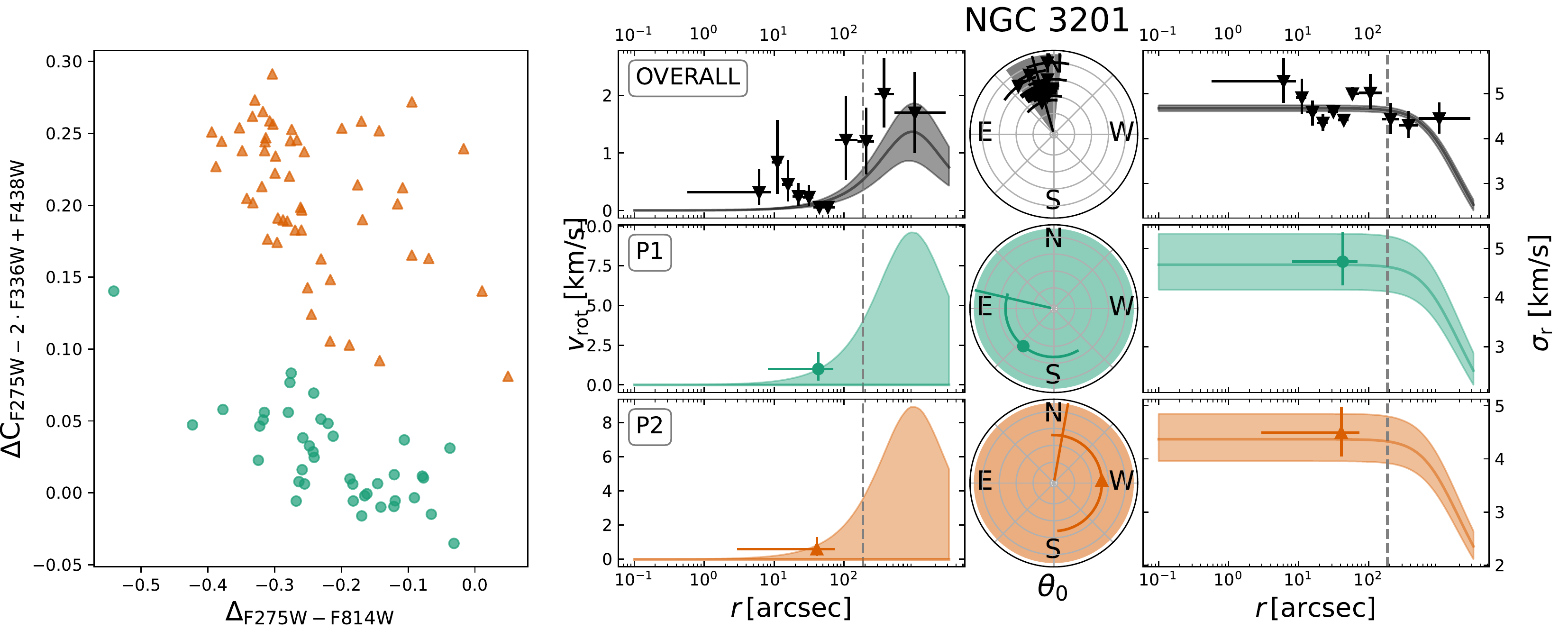}
    \caption{Continuation of Fig. \ref{fig:profiles_ngc104} for NGC~3201.}
    \label{fig:profiles_ngc3201}
\end{figure*}
\begin{figure*}[b]
    \centering
    \includegraphics[width=\textwidth]{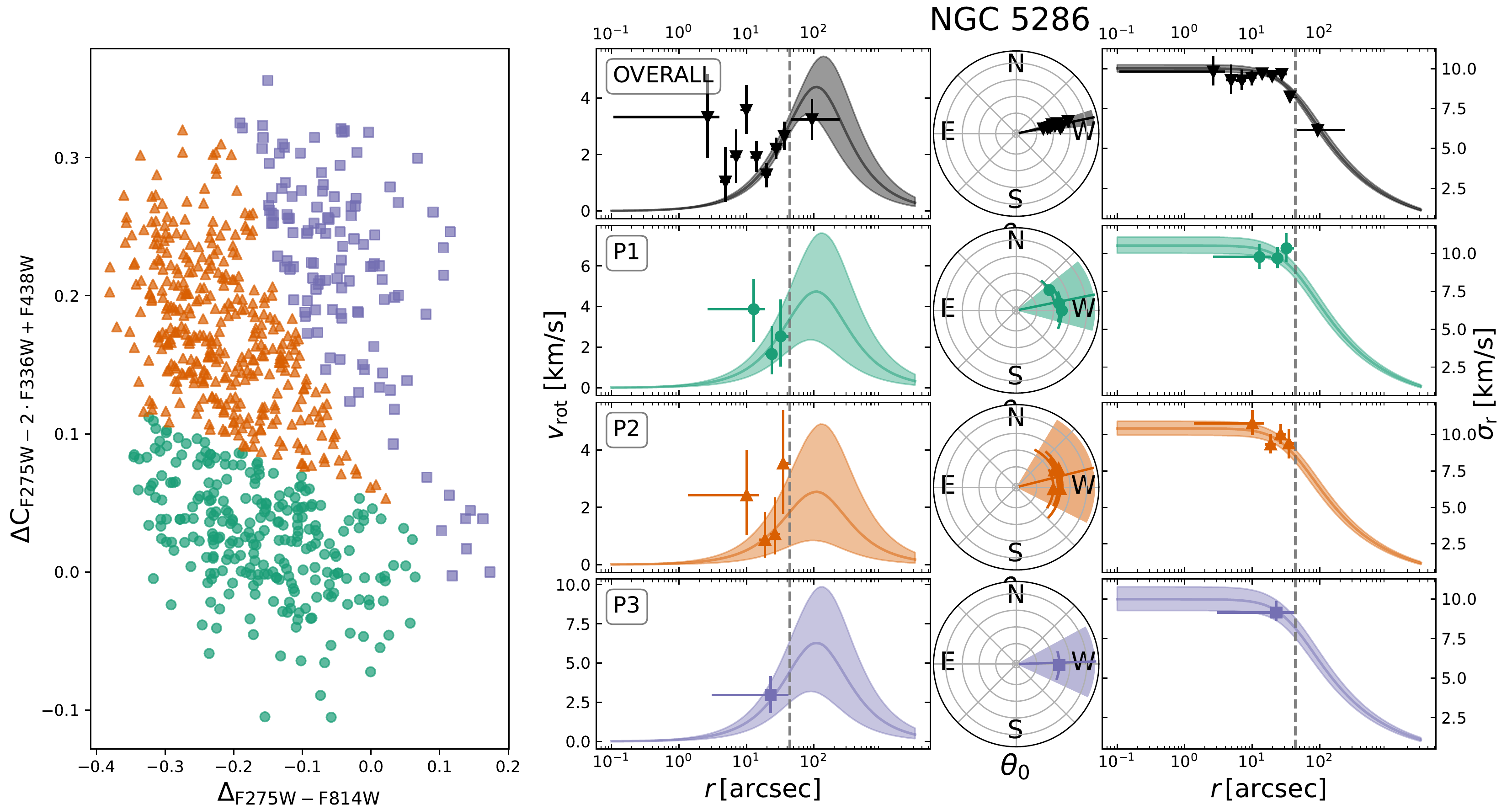}
    \caption{Continuation of Fig. \ref{fig:profiles_ngc104} for NGC~5286.}
    \label{fig:profiles_ngc5286}
\end{figure*}
\begin{figure*}[b]
    \centering
    \includegraphics[width=\textwidth]{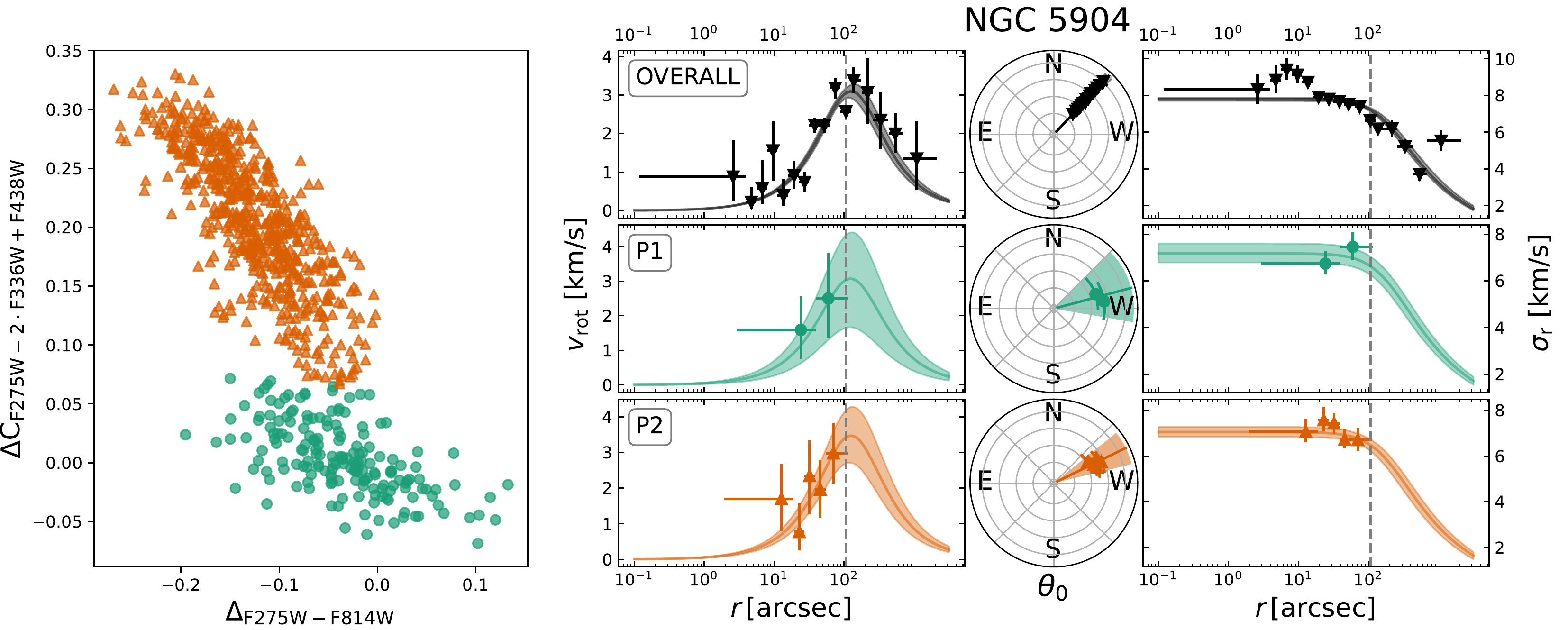}
    \caption{Continuation of Fig. \ref{fig:profiles_ngc104} for NGC~5904.}
    \label{fig:profiles_ngc5904}
\end{figure*}
\begin{figure*}[b]
    \centering
    \includegraphics[width=\textwidth]{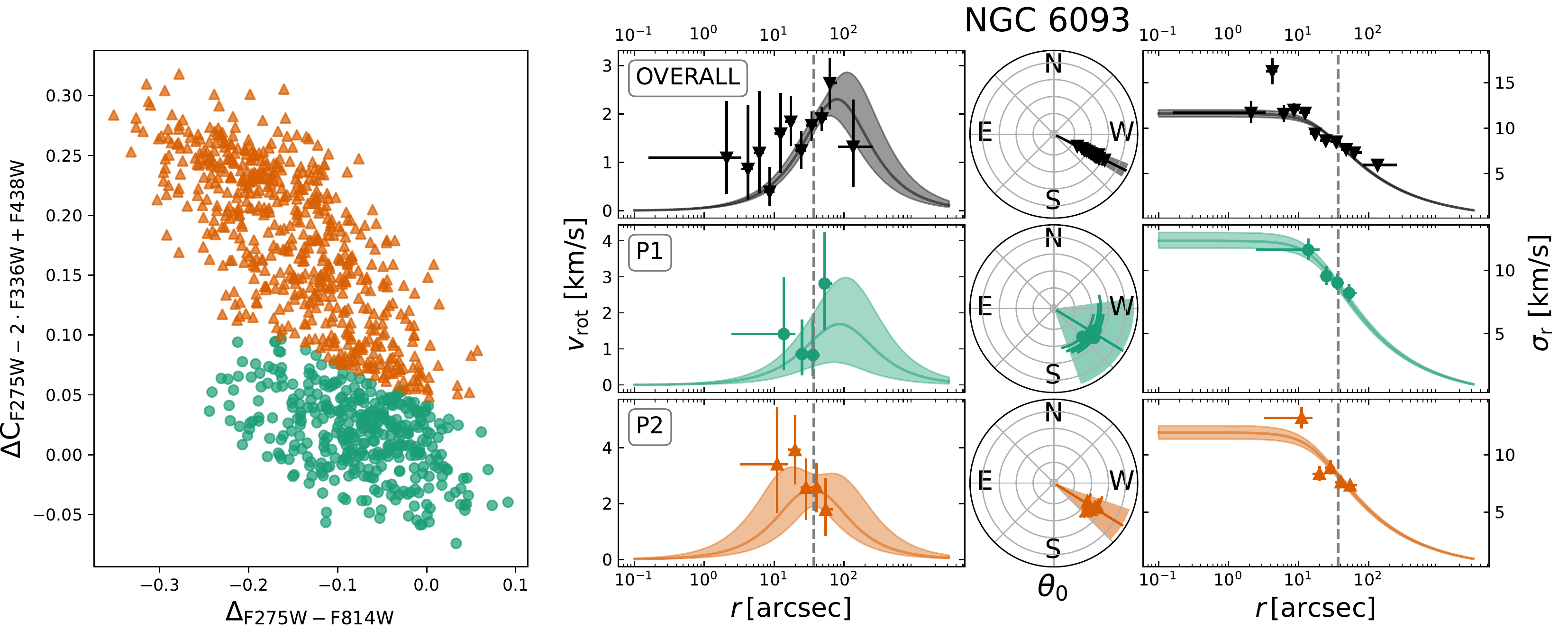}
    \caption{Continuation of Fig. \ref{fig:profiles_ngc104} for NGC~6093.}
    \label{fig:profiles_ngc6093}
\end{figure*}
\begin{figure*}[b]
    \centering
    \includegraphics[width=\textwidth]{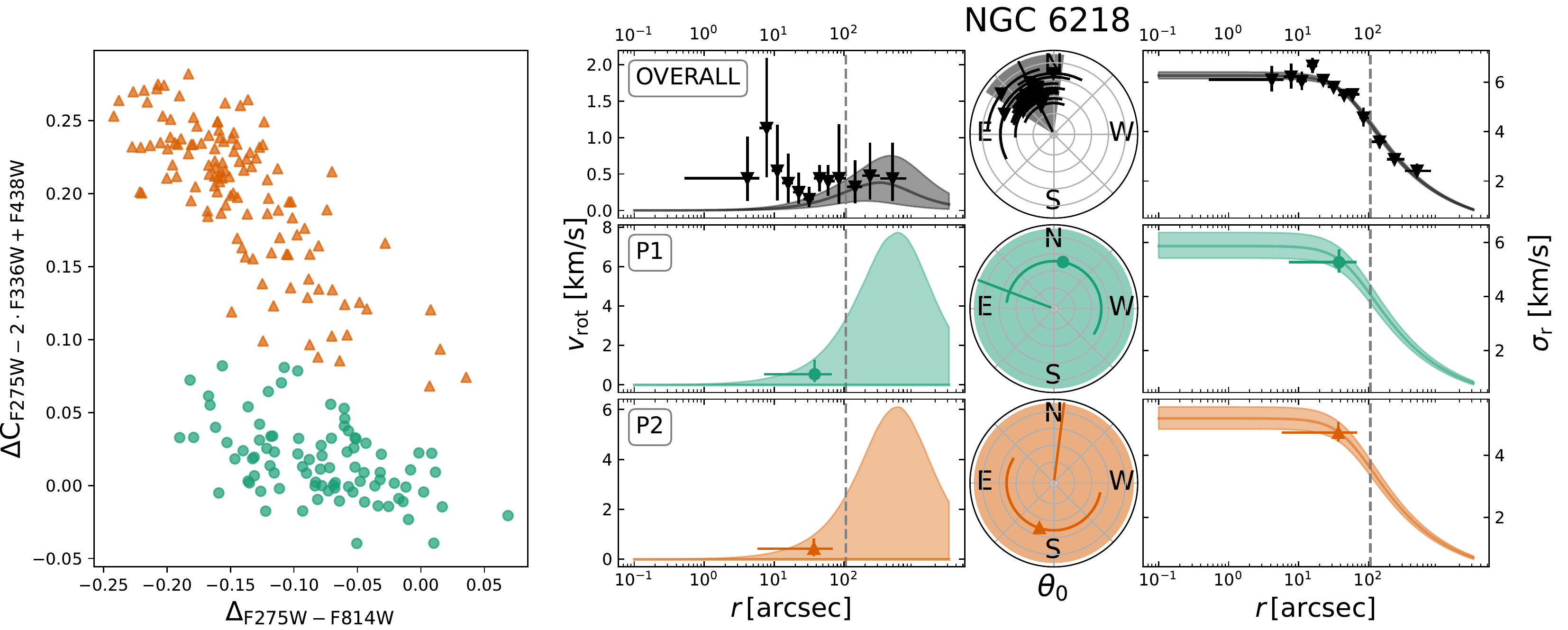}
    \caption{Continuation of Fig. \ref{fig:profiles_ngc104} for NGC~6218.}
    \label{fig:profiles_ngc6218}
\end{figure*}
\begin{figure*}[b]
    \centering
    \includegraphics[width=\textwidth]{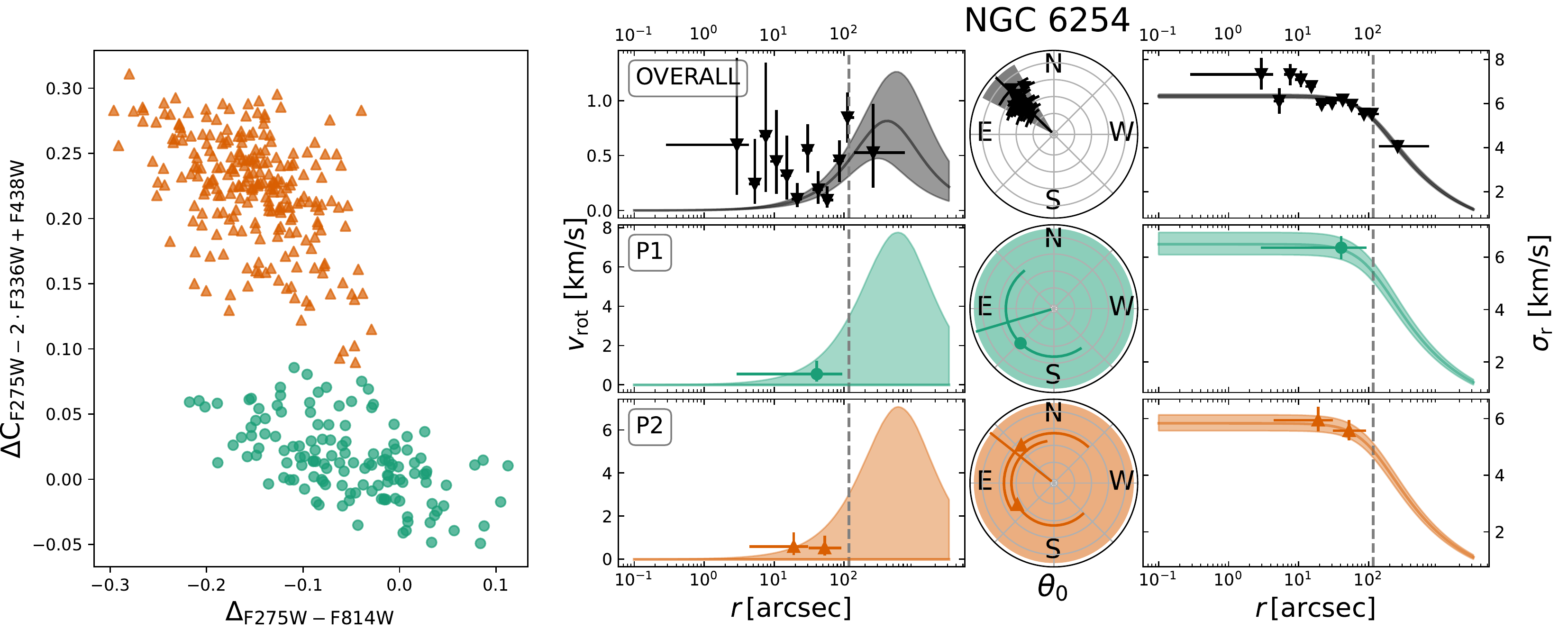}
    \caption{Continuation of Fig. \ref{fig:profiles_ngc104} for NGC~6254.}
    \label{fig:profiles_ngc6254}
\end{figure*}
\begin{figure*}[b]
    \centering
    \includegraphics[width=\textwidth]{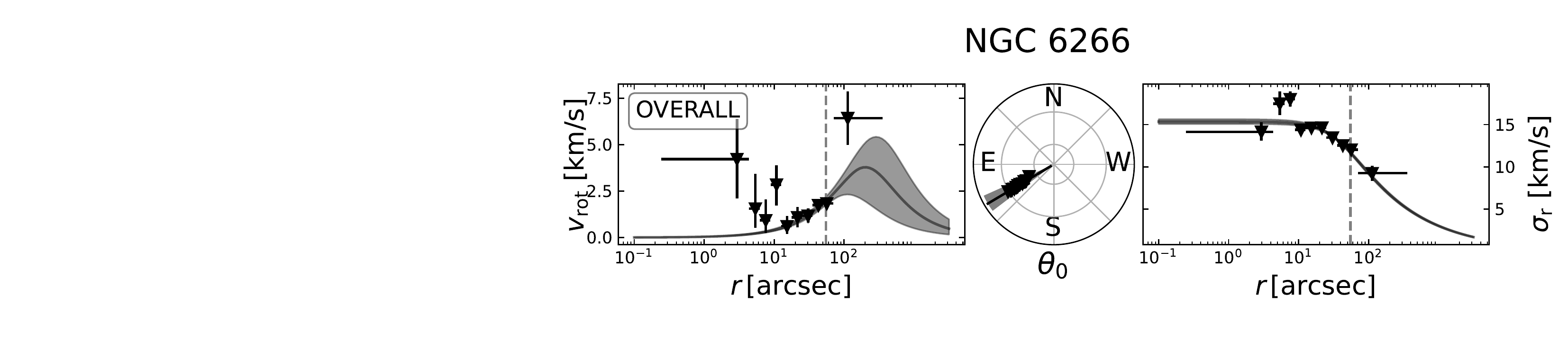}
    \caption{Continuation of Fig. \ref{fig:profiles_ngc104} for NGC~6266.}
    \label{fig:profiles_ngc6266}
\end{figure*}
\begin{figure*}[b]
    \centering
    \includegraphics[width=\textwidth]{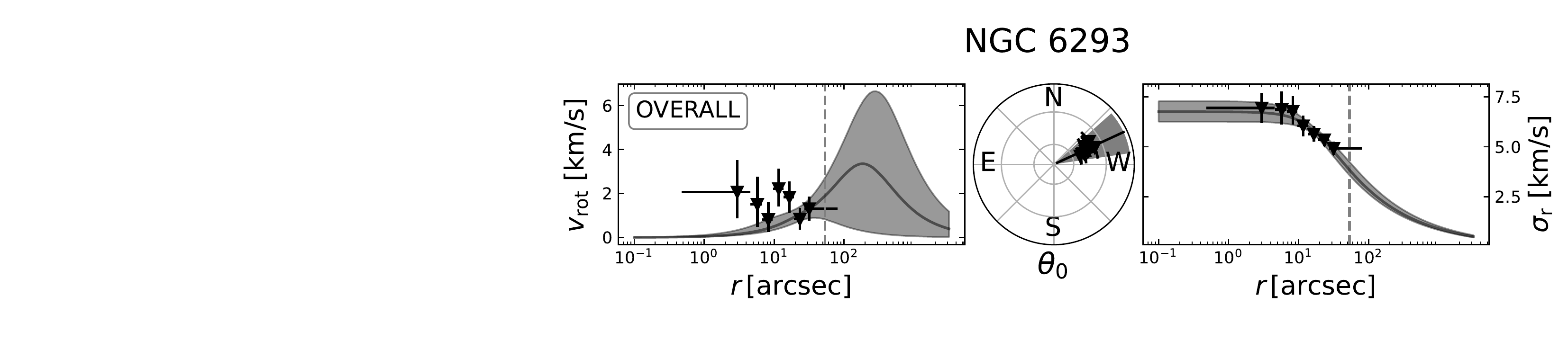}
    \caption{Continuation of Fig. \ref{fig:profiles_ngc104} for NGC~6293.}
    \label{fig:profiles_ngc6293}
\end{figure*}
\begin{figure*}[b]
    \centering
    \includegraphics[width=\textwidth]{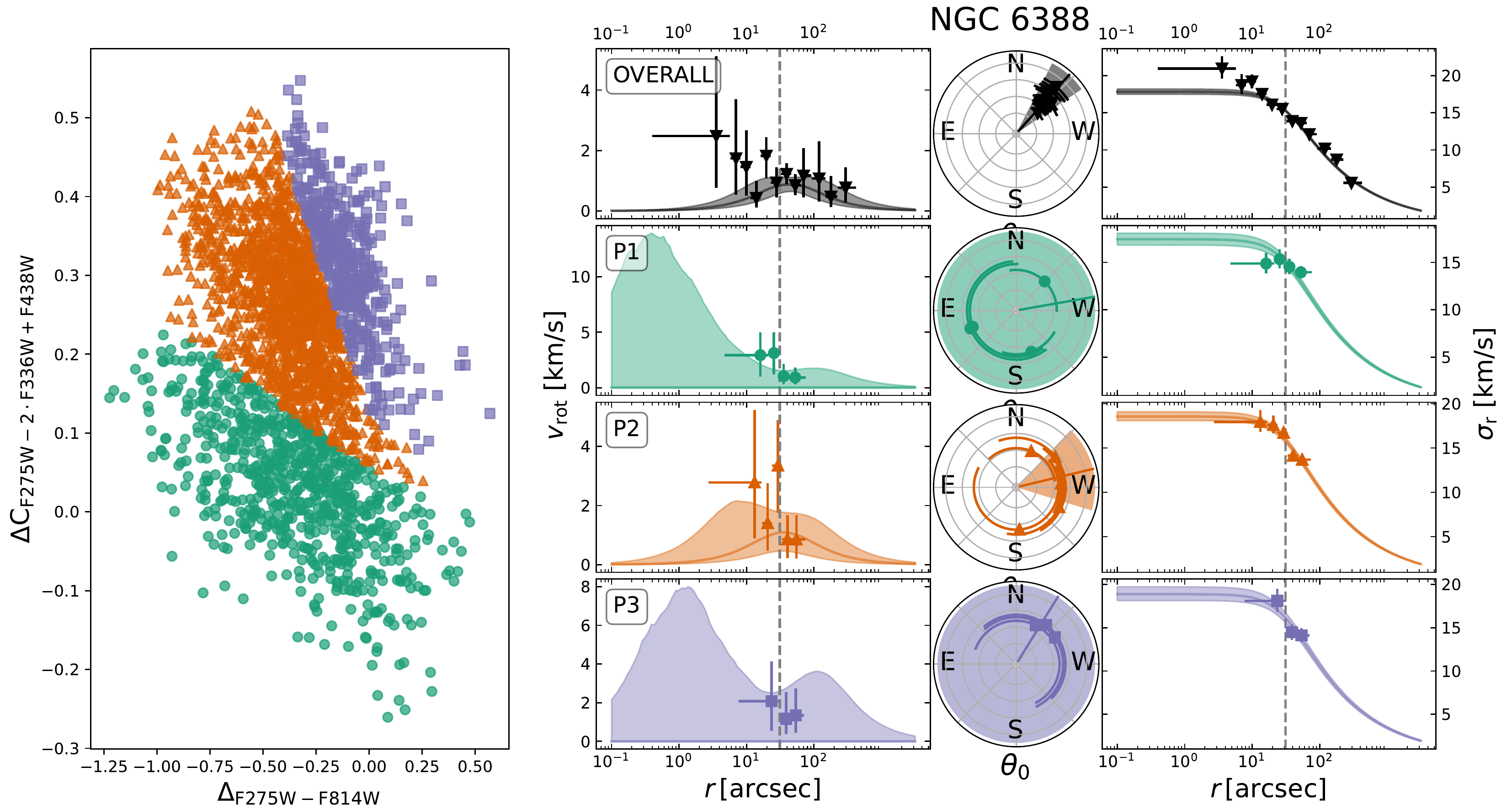}
    \caption{Continuation of Fig. \ref{fig:profiles_ngc104} for NGC~6388.}
    \label{fig:profiles_ngc6388}
\end{figure*}
\begin{figure*}[b]
    \centering
    \includegraphics[width=\textwidth]{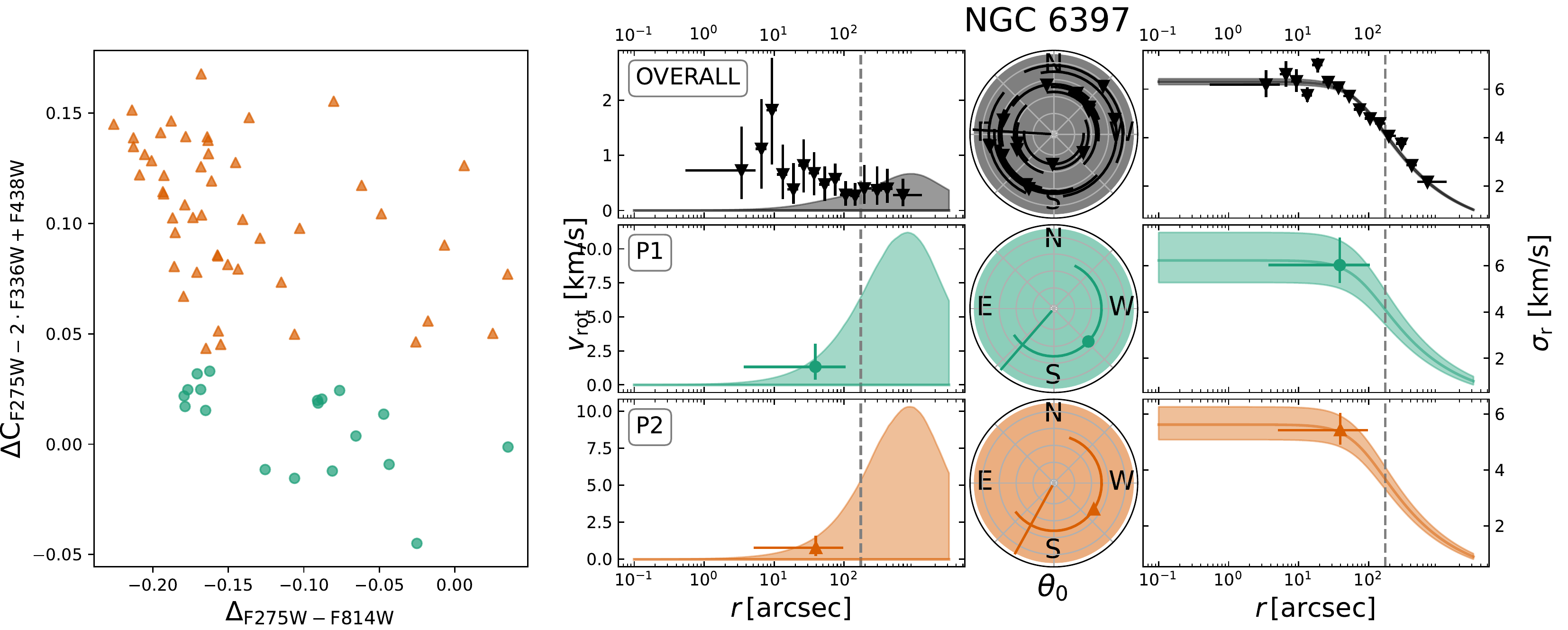}
    \caption{Continuation of Fig. \ref{fig:profiles_ngc104} for NGC~6397.}
    \label{fig:profiles_ngc6397}
\end{figure*}
\begin{figure*}[b]
    \centering
    \includegraphics[width=\textwidth]{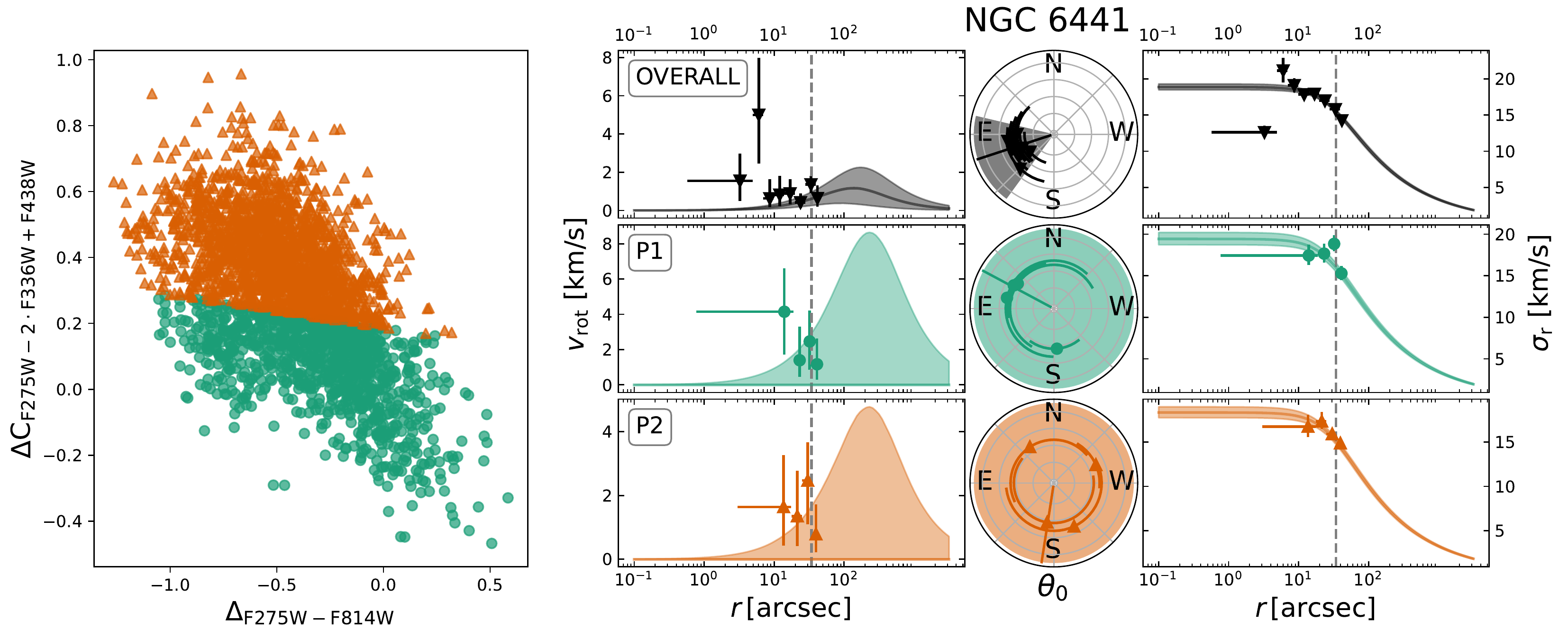}
    \caption{Continuation of Fig. \ref{fig:profiles_ngc104} for NGC~6441.}
    \label{fig:profiles_ngc6441}
\end{figure*}
\begin{figure*}[b]
    \centering
    \includegraphics[width=\textwidth]{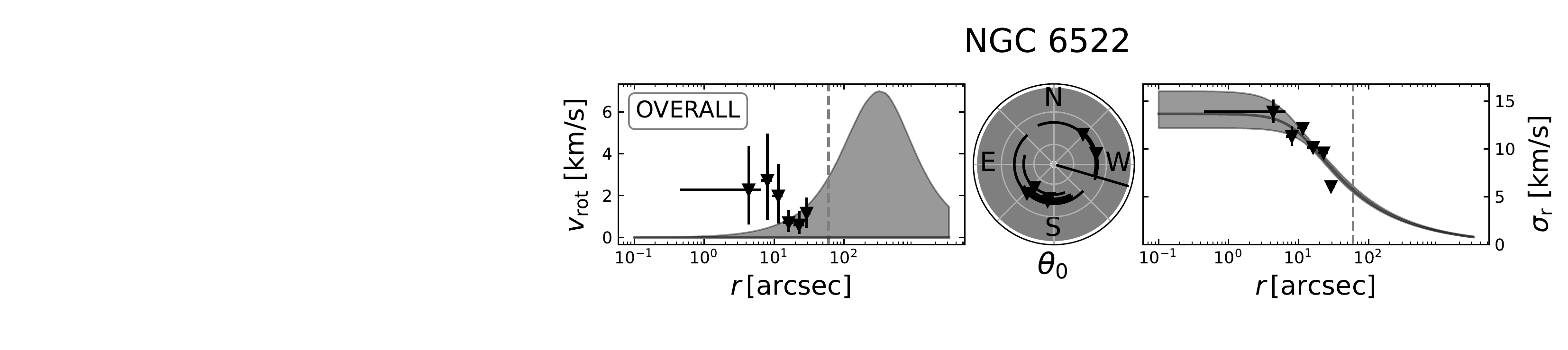}
    \caption{Continuation of Fig. \ref{fig:profiles_ngc104} for NGC~6522.}
    \label{fig:profiles_ngc6522}
\end{figure*}
\begin{figure*}[b]
    \centering
    \includegraphics[width=\textwidth]{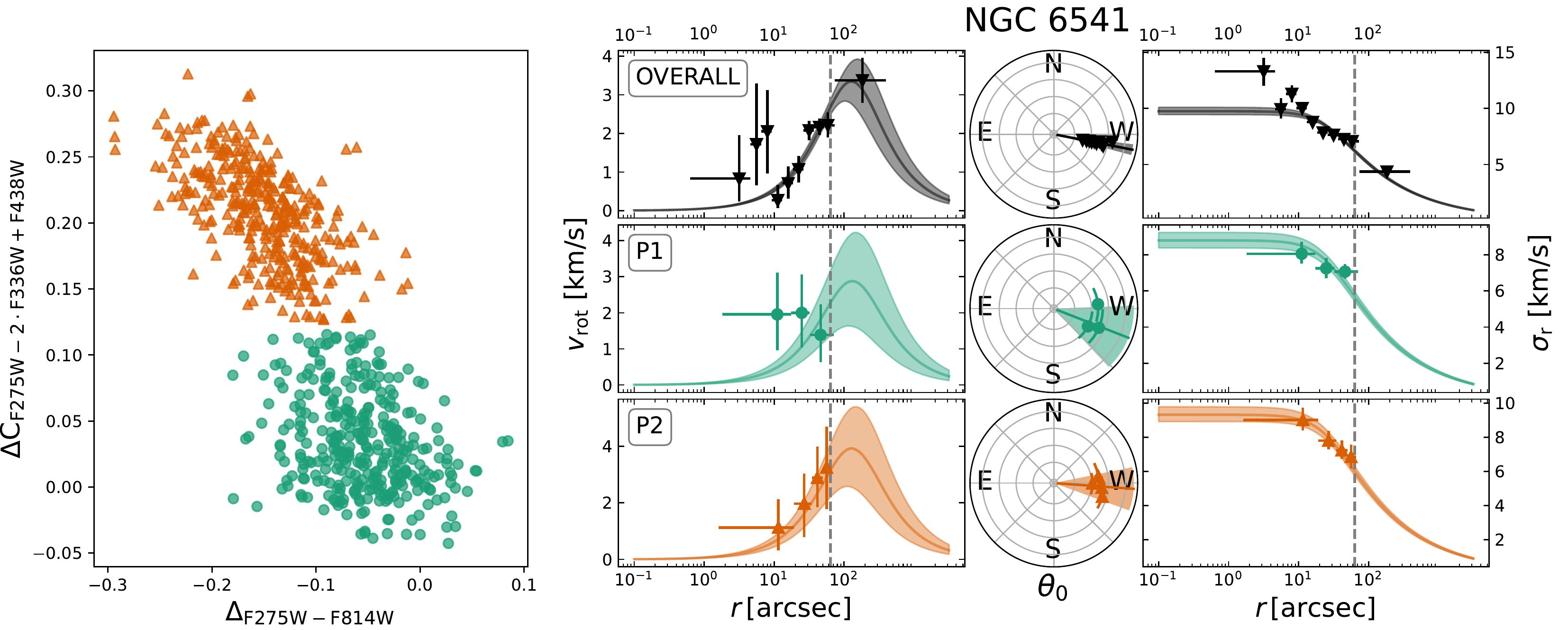}
    \caption{Continuation of Fig. \ref{fig:profiles_ngc104} for NGC~6541.}
    \label{fig:profiles_ngc6541}
\end{figure*}
\begin{figure*}[b]
    \centering
    \includegraphics[width=\textwidth]{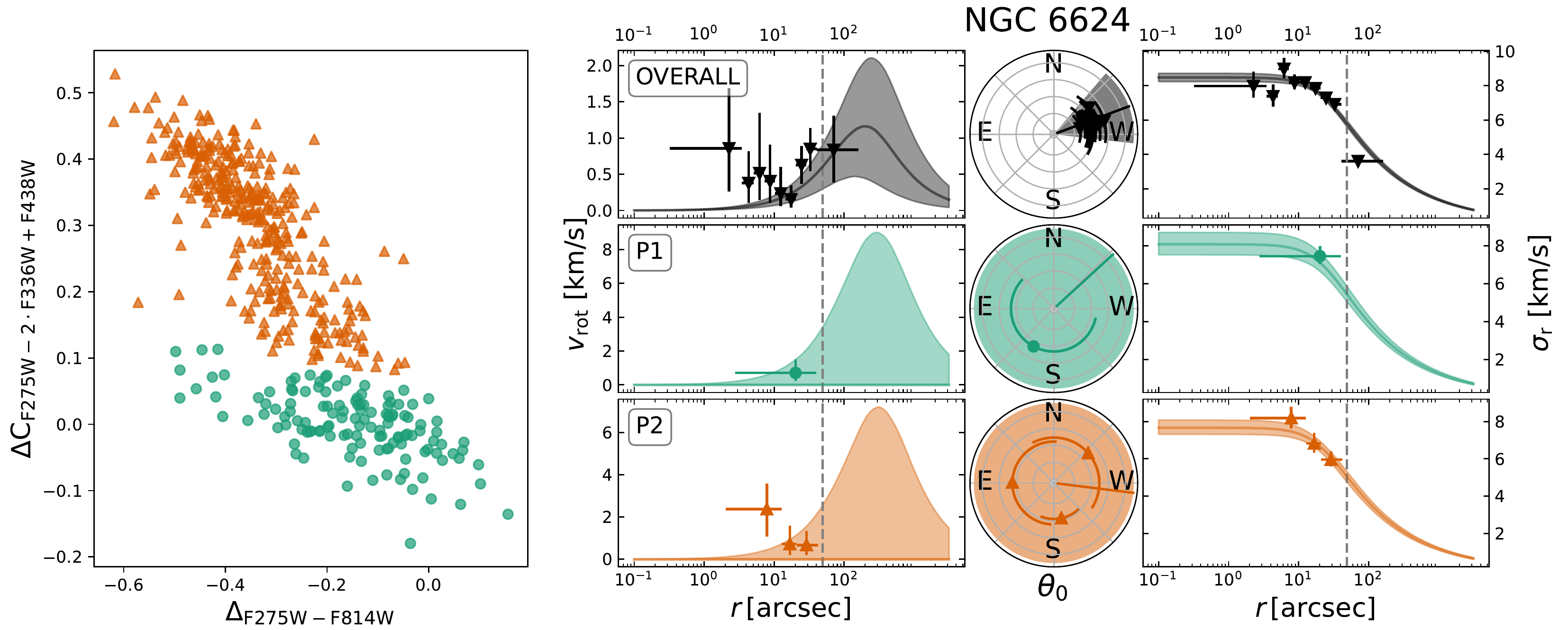}
    \caption{Continuation of Fig. \ref{fig:profiles_ngc104} for NGC~6624.}
    \label{fig:profiles_ngc6624}
\end{figure*}
\begin{figure*}[b]
    \centering
    \includegraphics[width=\textwidth]{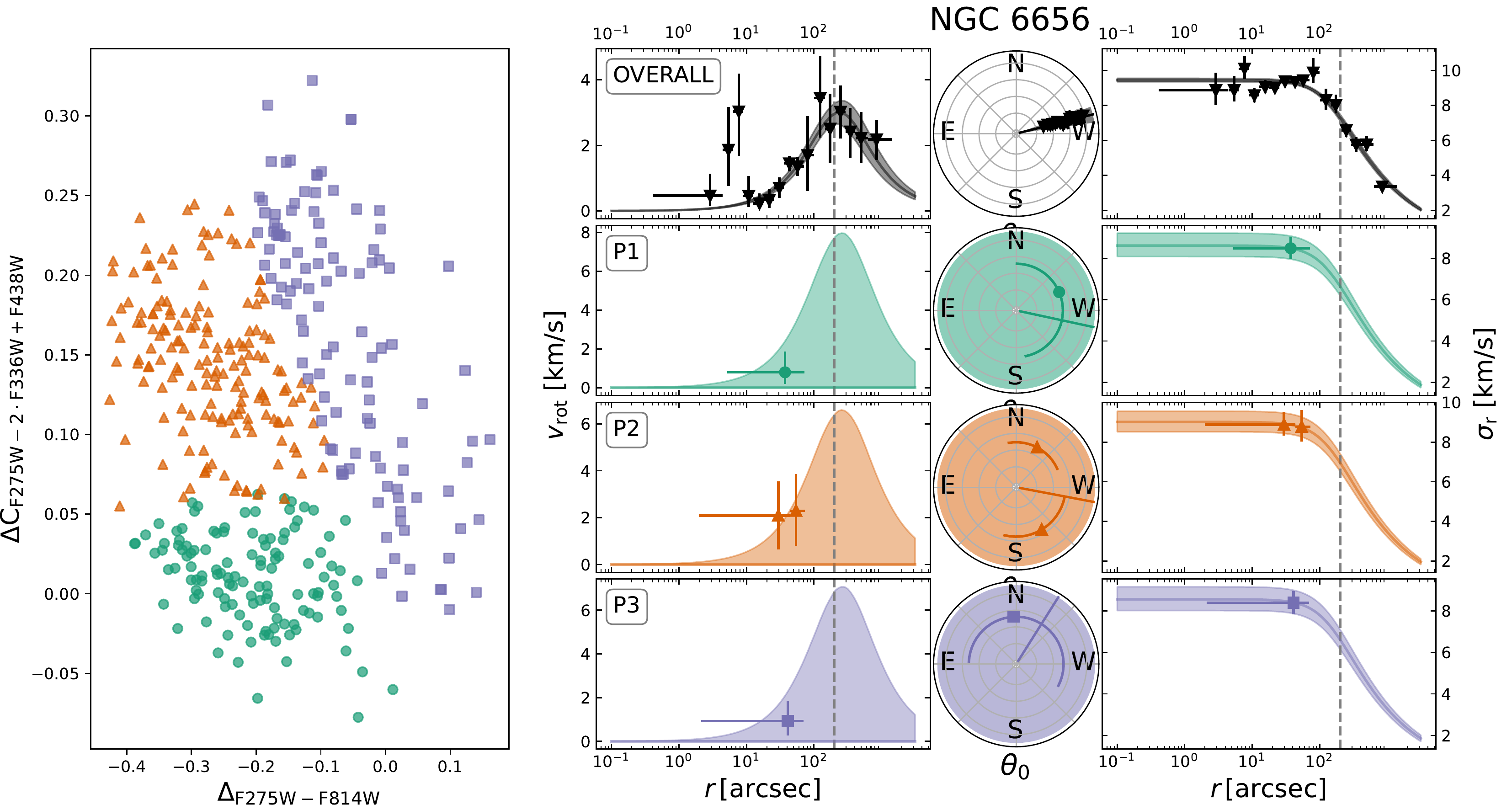}
    \caption{Continuation of Fig. \ref{fig:profiles_ngc104} for NGC~6656.}
    \label{fig:profiles_ngc6656}
\end{figure*}
\begin{figure*}[b]
    \centering
    \includegraphics[width=\textwidth]{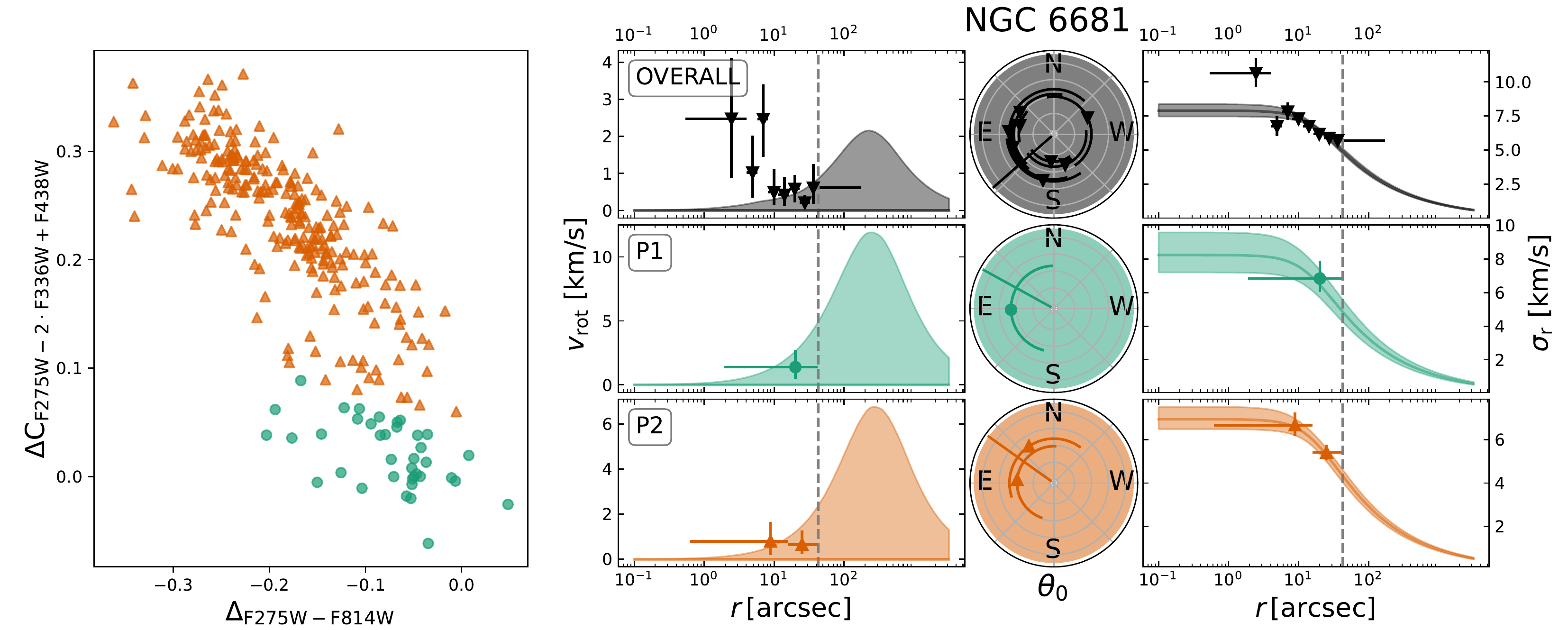}
    \caption{Continuation of Fig. \ref{fig:profiles_ngc104} for NGC~6681.}
    \label{fig:profiles_ngc6681}
\end{figure*}
\begin{figure*}[b]
    \centering
    \includegraphics[width=\textwidth]{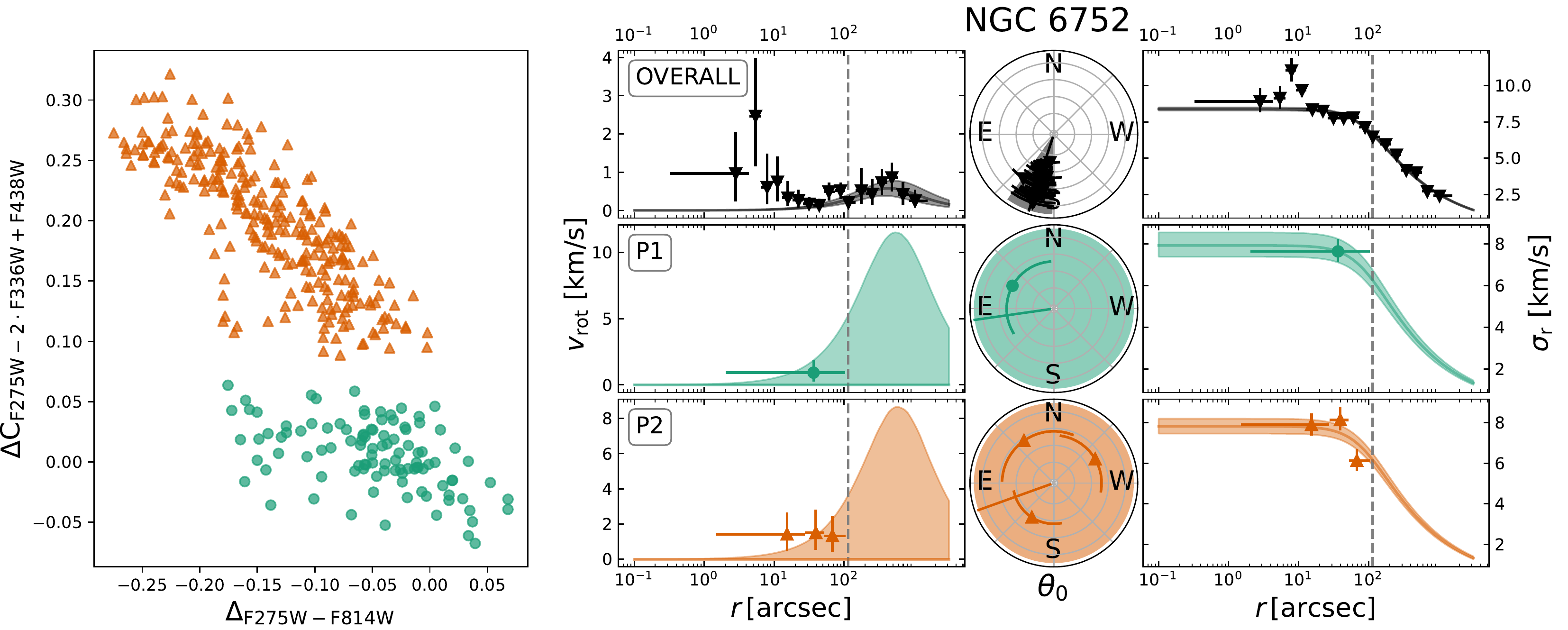}
    \caption{Continuation of Fig. \ref{fig:profiles_ngc104} for NGC~6752.}
    \label{fig:profiles_ngc6752}
\end{figure*}
\begin{figure*}[b]
    \centering
    \includegraphics[width=\textwidth]{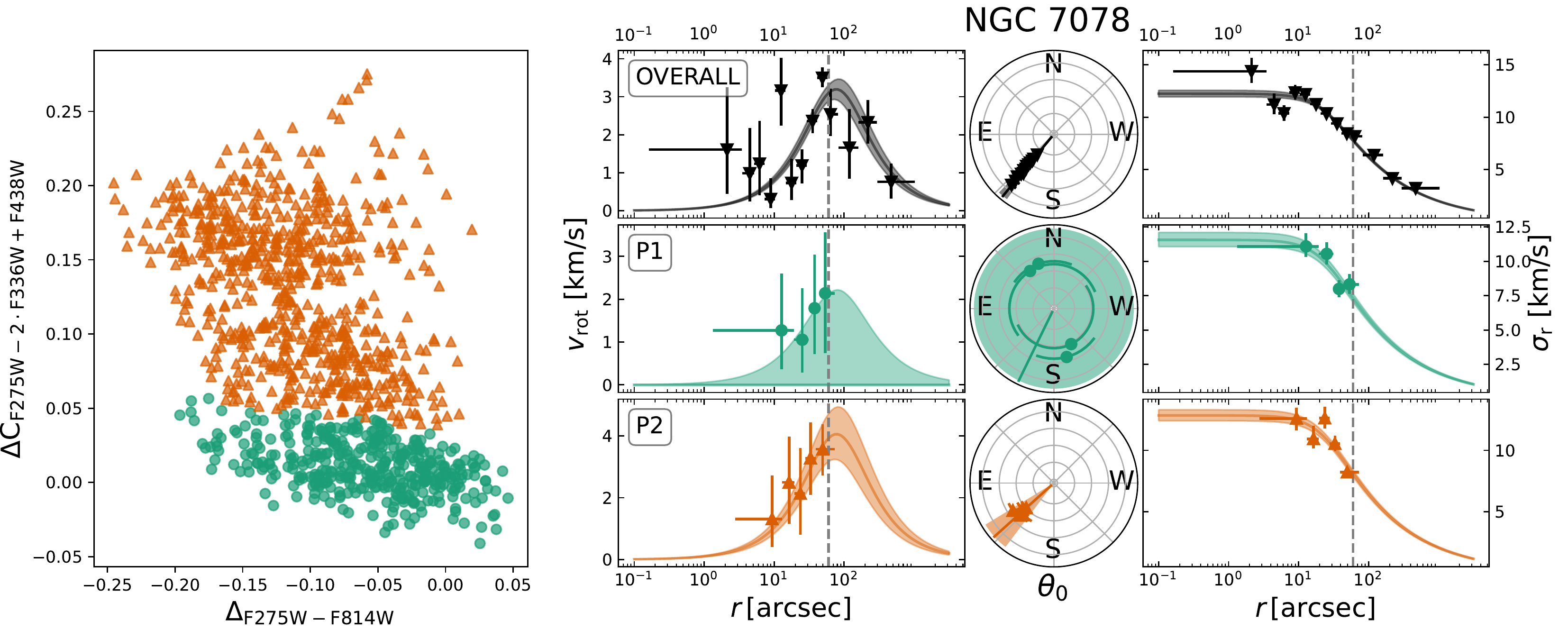}
    \caption{Continuation of Fig. \ref{fig:profiles_ngc104} for NGC~7078.}
    \label{fig:profiles_ngc7078}
\end{figure*}
\begin{figure*}[b]
    \centering
    \includegraphics[width=\textwidth]{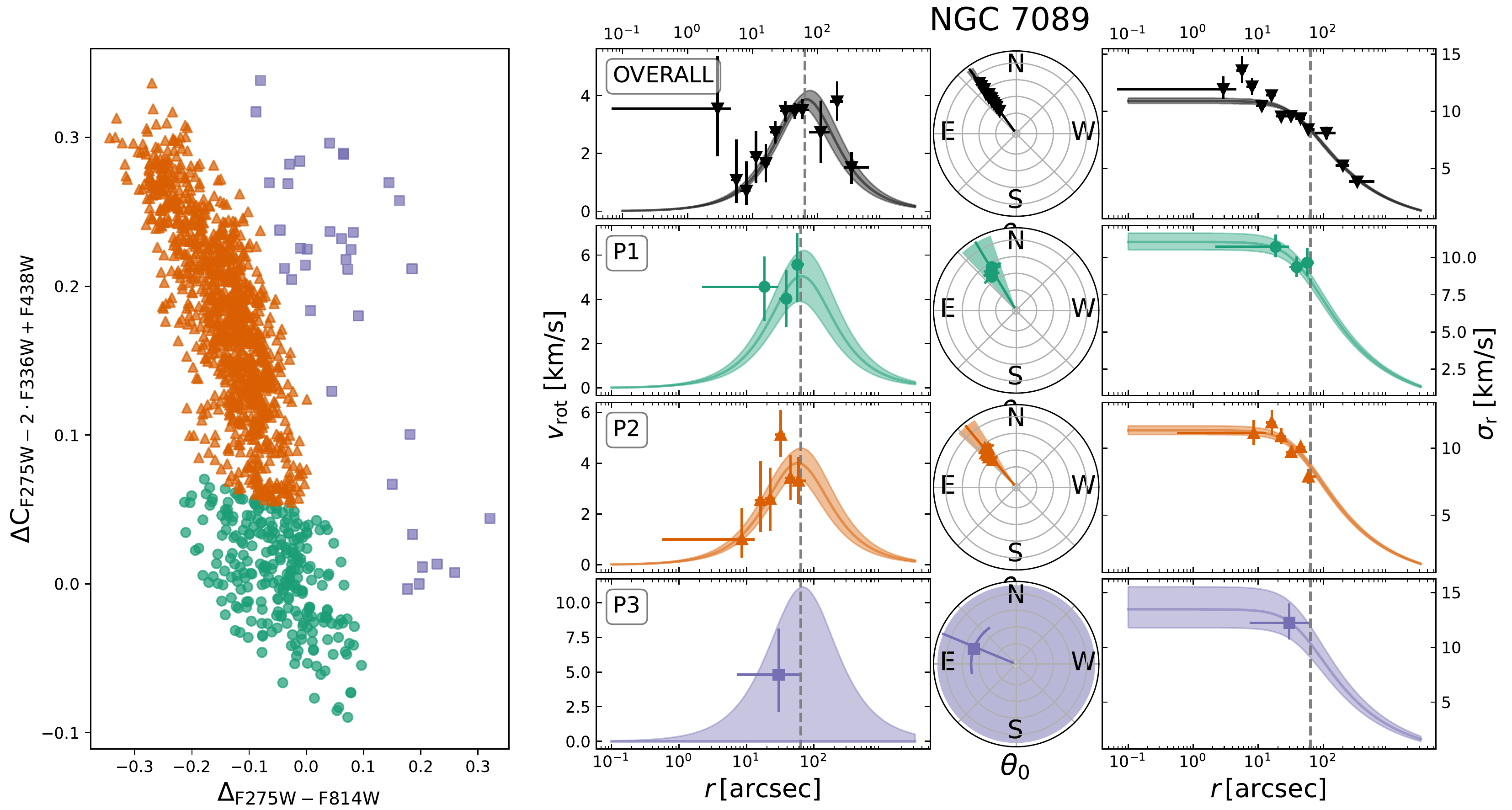}
    \caption{Continuation of Fig. \ref{fig:profiles_ngc104} for NGC~7089.}
    \label{fig:profiles_ngc7089}
\end{figure*}
\begin{figure*}[b]
    \centering
    \includegraphics[width=\textwidth]{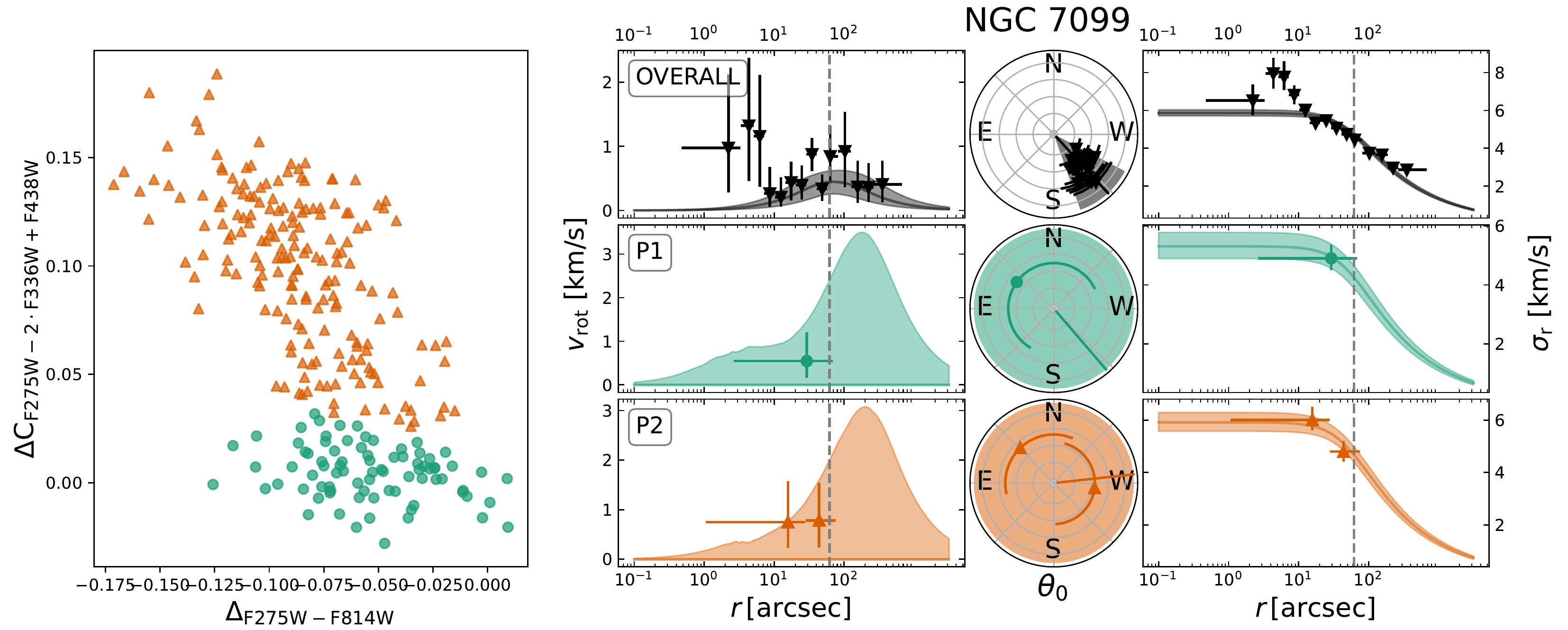}
    \caption{Continuation of Fig. \ref{fig:profiles_ngc104} for NGC~7099.}
    \label{fig:profiles_ngc7099}
\end{figure*}

\begin{figure*}[b]
    \centering
    \includegraphics[width=\textwidth]{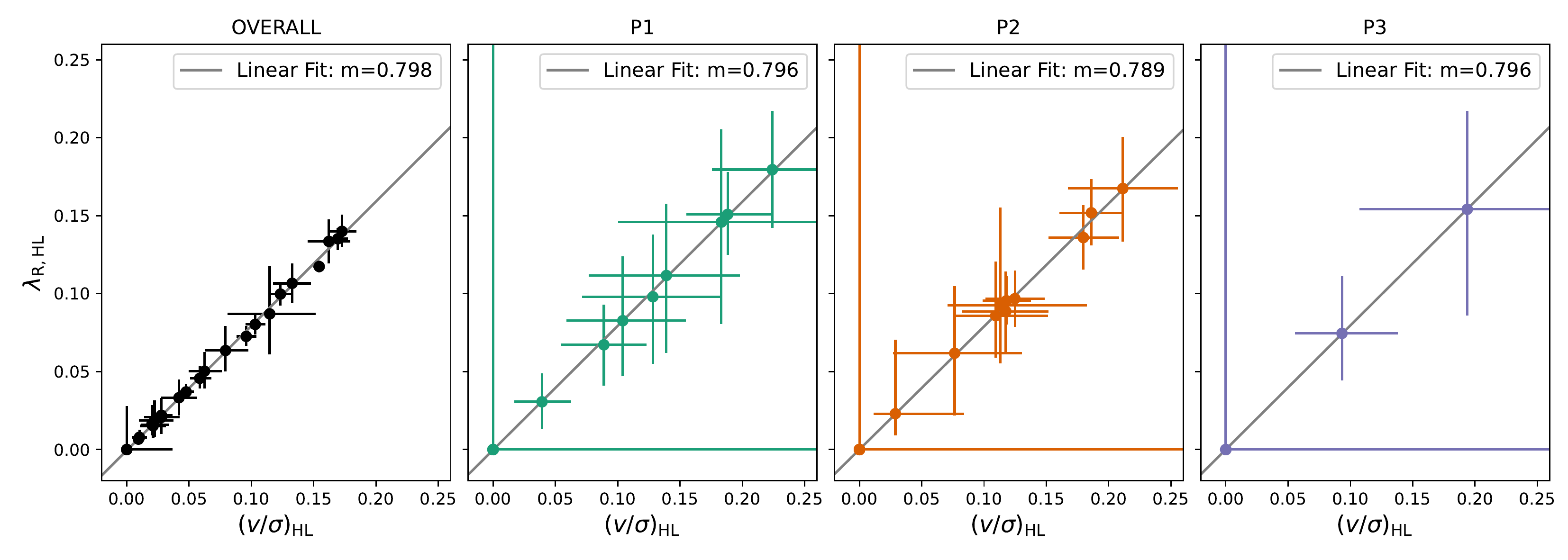}
    \caption{Derived values of $\lambda_\mathrm{R, HL}$ plotted against $\left(v/\sigma\right)_\mathrm{HL}$ for the overall cluster and each population with a linear fit.}
    \label{fig:vsigma_vs_lambdar}
\end{figure*}

\begin{figure*}[t]
    \centering
    \includegraphics[width=0.49\textwidth]{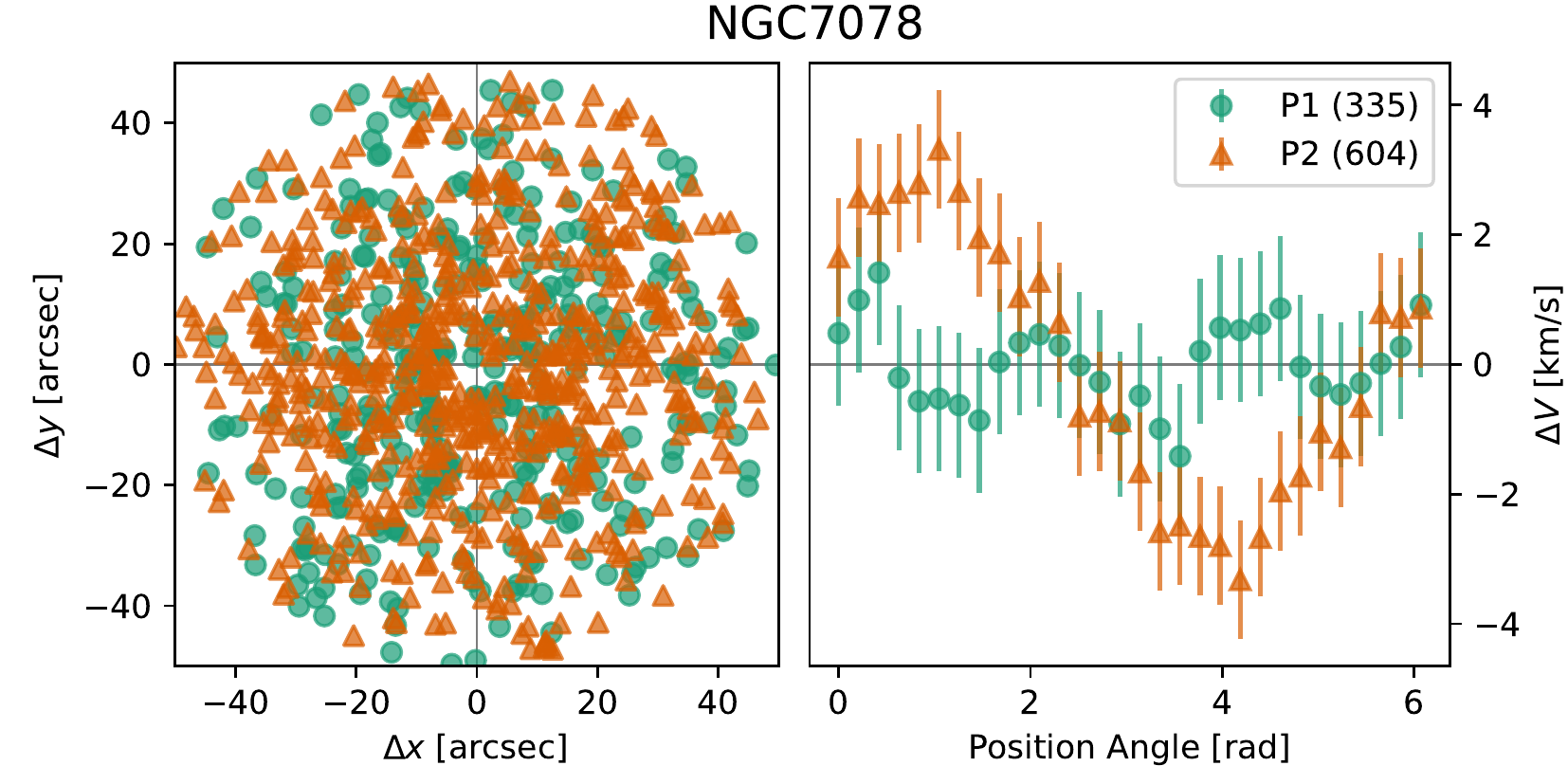}
    \caption{Differential rotation profile for P1 and P2 of NGC~7078, where the difference in mean radial velocity between the two subsets of stars is plotted against the position angle of their line of separation.}
    \label{fig:ngc7078_theta_binning_1}
\end{figure*}

\begin{figure*}[t]
    \centering
    \includegraphics[width=0.49\textwidth]{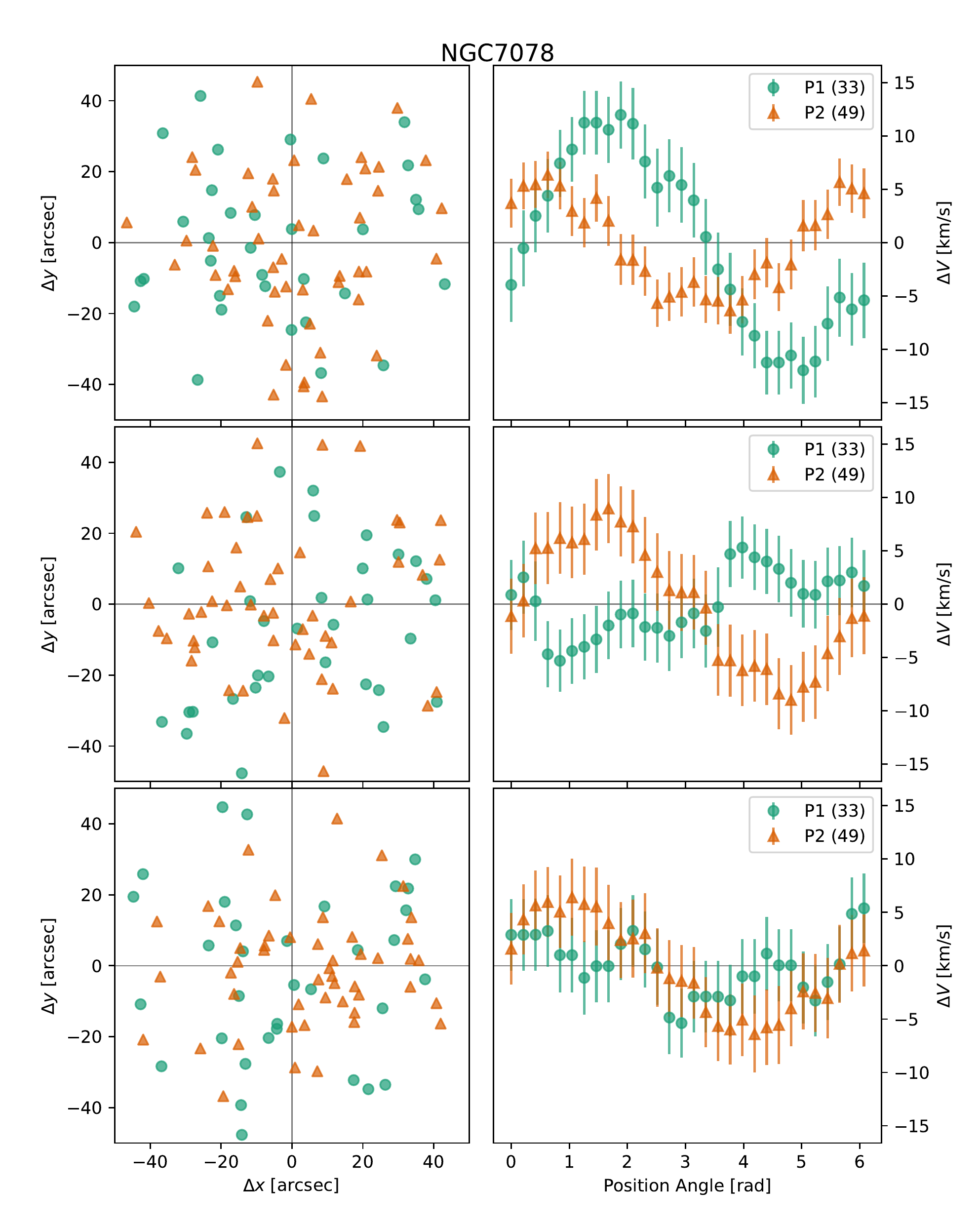}
    \caption{Differential rotation profile for randomly sampled stars of P1 and P2 for NGC~7078, where the difference in mean radial velocity between the two subsets of stars is plotted against the position angle of their line of separation.}
    \label{fig:ngc7078_theta_binning_3}
\end{figure*}

\end{appendix}

\end{document}